\documentclass[prb,aps,twocolumn,superscriptaddress]{revtex4}

\renewcommand{\mod}{\text{ mod }}
\theoremstyle{plain}
\newcommand{\pprl}[1]{Phys. Rev. Lett. \textbf{#1}}
\newcommand{\pprb}[1]{Phys. Rev. B \textbf{#1}}
\newcommand{\npb}[1]{Nucl. Phys. B \textbf{#1}}

\newcommand{\ksc}[1]{k^\text{sc}_{\gamma;#1}}

\newcommand{\hsc}[1]{h^\text{sc}_{#1}}
\newcommand{\thsc}[1]{\tilde{h}^\text{sc}_{#1}}

\newcommand{\dN}{\mathds N}
\newcommand{\dR}{\mathds R}
\newcommand{\dZ}{\mathds Z}
\renewcommand{\t}{\tilde}

\begin{document}

\title{Non-Abelian Quantum Hall States and their Quasiparticles:\\
from the Pattern of Zeros to Vertex Algebra}
\date{{\small Oct. 2009}}

\author{Yuan-Ming Lu}
\affiliation{Department of Physics, Boston College, Chestnut Hill,
MA 02467, USA}

\author{Xiao-Gang Wen}
\affiliation{ Department of Physics, Massachusetts Institute of
Technology, Cambridge, MA 02139}
\affiliation{ Perimeter Institute for Theoretical Physics, 31 Caroline
Street North, Waterloo, Ontario N2J 2Y5, Canada }

\author{Zhenghan Wang}
\affiliation{ Microsoft Station Q, CNSI Bldg. Rm 2237, University of
California, Santa Barbara, CA 93106}

\author{Ziqiang Wang}
\affiliation{Department of Physics, Boston College, Chestnut Hill,
MA 02467, USA}

\begin{abstract}

In the pattern-of-zeros approach to quantum Hall states, a set of data
$\{n;m;S_a|a=1,...,n; n,m,S_a\in \dN \}$ (called the pattern of zeros)
is introduced to characterize a quantum Hall wave function.  In this
paper we find sufficient conditions on the pattern of zeros so that
the data correspond to a valid wave function.  Some times, a set of
data $\{n;m;S_a\}$ corresponds to a unique quantum Hall state, while
other times, a set of data corresponds to several different quantum
Hall states.  So in the latter cases, the patterns of zeros alone does not
completely characterize the quantum Hall states.  In this paper, We
find that the following expanded set of data $\{n;m;S_a;c|a=1,...,n;
n,m,S_a\in \dN; c\in \dR \}$ provides a more complete characterization
of quantum Hall states.  Each expanded set of data completely
characterizes a unique quantum Hall state, at least for the examples
discussed in this paper.  The result is obtained by combining the
pattern of zeros and $Z_n$ simple-current vertex algebra which
describes a large class of Abelian and non-Abelian quantum Hall states
$\Phi_{Z_n}^\text{sc}$.  The more complete characterization in terms
of $\{n;m;S_a;c\}$ allows us to obtain more topological properties of
those states, which include the central charge $c$ of edge states, the
scaling dimensions and the statistics of quasiparticle excitations.

\end{abstract}

\maketitle

{\small \setcounter{tocdepth}{2} \tableofcontents }

\section{Introduction}

\renewcommand{\v}[1]{\boldsymbol{#1}}

Materials can have many different forms, which is partially due to
the very rich ways in which atoms and electrons can organize. The
different organizations correspond to different phases of matter (or
states of matter).  It is very important for physicists to
understand these different states of matter and the phase
transitions between them.  At zero-temperature, the phases are
described by the ground state wave functions, which are complex wave
functions $\Phi(\textbf{r}_1,\textbf{r}_2,\cdots,\textbf{r}_N)$ with
$N \to \infty$ variables.  So mathematically, to describe
zero-temperature phases, we need to characterize and classify the
ground state wave functions with $\infty$ variables, which is a very
challenging mathematical problem.

For a long time we believe that all states of matter and all phase
transitions between them are characterized by their broken
symmetries and the associated order parameters\cite{landau}.  A
general theory for phases and phase transitions is developed based
on this symmetry breaking picture.  So within the paradigm of
symmetry breaking, a many-body wave function is characterized by its
symmetry properties. Landau's symmetry breaking theory is a very
successful theory and has dominated the theory of phases and phase
transitions until the discovery of fractional quantum Hall (FQH)
effect\cite{fqhexp,laughlin}.

FQH states cannot be described by symmetry breaking since different
FQH states have exactly the same symmetry. So different FQH states
must contain a new kind of order. The new order is called topological
order\cite{wenniu,to1,to2} and the associated phase called topological
phase, because their characteristic universal properties (such as the
ground states degeneracy on a torus\cite{wenniu}) are invariant under
\emph{any} small perturbations of the system. Unlike symmetry-breaking
phases described by local order parameters, a topological phase is
characterized by a pattern of long-range quantum
entanglement\cite{entanglement1,entanglement2,entanglement3}. In
\Ref{Wrig}, the non-Abelian Berry phases for the degenerate ground
states are introduced to systematically characterize and classify
topological orders in FQH states (as well as other topologically
ordered states).  In this paper, we further develop another systematic
characterization of the topological orders in FQH states based on the
pattern of zeros approach.\cite{poz1,poz2}

In the strong magnetic field limit, a FQH wave function with filling
factor $\nu<1$ is an anti-symmetric holomorphic polynomial of
complex coordinates $\{z_i=x_i+\imth y_i\}$ (except for a common
factor that depends on geometry: say, a Gaussian factor
$\exp\Big(\sum_{i}\frac{|z_i|^2}{4}\Big)$ for a planar geometry).
After factoring out an anti-symmetric factor of
$\prod_{i<j}(z_i-z_j)$, we can describe a quantum Hall state by a
symmetric polynomial $\Phi(z_1,\cdots,z_N)$ in the $N\to\infty$
limit.\cite{poz1} So the characterization and classification of
long-range quantum entanglements in FQH states become a problem of
characterizing and classifying symmetric polynomials with infinite
variables.

In a recent series of work,\cite{poz1,poz2,poz3} the pattern of
zeros is introduced to characterize and classify symmetric
polynomials of infinite variables. The pattern of zeros is described
by a sequence of integers $\{S_a|a=1,2,...\}$, where $S_a$ is the
lowest order of zeros of the symmetric polynomial when we fuse $a$
different variables together.  The data $\{S_a|a=1,2,...\}$ can be
further compactified into a finite set $\{n;m;S_a|a=1,2,...,n;n,m\in
\dN \}$ for $n$-cluster quantum Hall states.  Here
$\dN=\{0,1,2,...\}$ is the set of non-negative integers.  It has
been shown\cite{poz1,poz2} that all known one-component Abelian and
non-Abelian quantum Hall states can be (partially) characterized by
pattern of zeros.  It is also shown\cite{poz1,poz2} that, for any
given pattern of zeros $\{S_a\}$, we can construct an ideal local
Hamiltonian\cite{pp0,pp1,GWW,RR99,Read06} $H_{\{S_a\}}$ such that
the FQH state with the pattern of zeros is a zero energy ground
state of the Hamiltonian.

We would like to point out that, strictly speaking, a FQH state must
be a state with a finite energy gap. But in this paper, we will use
the term more loosely. We will call one state a FQH state if it can
be an zero energy state of an ideal Hamiltonian.  So our FQH states
may not be gapped.

Due to the length of this paper, in the following, we are going to
summarize the issues that we are going to discuss in this paper.  We
will also summarize the main results that we obtain on those issues.

\subsection{Sufficient conditions on pattern of zeros}

Within the pattern-of-zero approach, two questions naturally arise:
(1) Does any pattern of zeros, \ie an arbitrary integer sequence
$\{n;m;S_a\}$ corresponds to a symmetric polynomial
$\Phi(z_1,\cdots,z_N)$? Are there any ``illegal'' patterns of zeros
that do not correspond to any symmetric polynomial? (2) Given a
``legal'' pattern of zeros, can we construct a corresponding FQH
many-body wave function? Is the FQH many-body wave function uniquely
determined by the pattern of zeros?

For question (1), it turns out that the pattern of zeros must
satisfy some consistent conditions\cite{poz1,poz2} in order to
describe an existing symmetric polynomial. In other words, some
sequences $\{n;m;S_a\}$ don't correspond to any symmetric
polynomials.  However, \Ref{poz1,poz2} only obtain some necessary
conditions on the pattern of zeros $\{n;m;S_a\}$.  We still do not
have a set of sufficient conditions on pattern of zeros that
guarantee a pattern of zeros to correspond to an existing symmetric
polynomial.

For the question (2), right now, we do not have an efficient way to
obtain corresponding FQH many-body wave function from a ``legal''
pattern of zeros. Further more, while some patterns of zeros can
uniquely determine the FQH wave function, it is known that some other
patterns of zeros cannot uniquely determine the FQH wave function: \ie
in those cases, two different FQH wave functions can have the same
pattern of zeros.\cite{poz1,S3} This means that, some patterns of
zeros do not provide complete information to fully characterize
FQH states. In this case it is important to expand the data of pattern of
zeros to obtain a more complete characterization of FQH states.

We see that the above two questions are actually closely related. In
this paper, we will try to address those questions. Motivated by the
conformal field theory (CFT) construction of FQH wave
functions,\cite{MR,wenwu,wenwuh,coset,nass} we will try to use the
patterns of zeros to define and construct vertex algebras (which are
CFTs). Since the correlation function of the electron operator in
the constructed vertex algebra gives us the FQH wave function, once
the vertex algebra is obtained from a pattern of zeros, we
effectively find the corresponding FQH wave function for the pattern
of zeros. In this way, we establish the connection between the
pattern of zeros and the FQH wave function through the vertex
algebra.

In order for the correlation of electron operators in the vertex
algebra to produce a single-valued electron wave function with
respect to electron variables $\{z_1,\cdots,z_N\}$, electron
operators need to satisfy a so-called ``simple-current'' property
(see \eqn{psiapsibG} and \eqn{psiapsib}). Also the vertex algebra
need to satisfy the generalized Jacobi identity (GJI) which
guarantees the associativity of the corresponding vertex
algebra.\cite{va0} We find that only a certain set of patterns of
zeros can give rise to simple-current vertex algebras that satisfy
the GJI.  So the GJIs in simple-current vertex algebras give us a
set of sufficient conditions on a pattern of zeros so that this
pattern of zeros does correspond to an existing symmetric
polynomial.

In this paper, we first try to use the pattern of zeros $\{n;m;S_a\}$
to define a $Z_n$ vertex algebra.  From some of the GJI of the $Z_n$
vertex algebra, we obtain more necessary conditions on the pattern of
zeros $\{n;m;S_a\}$ than those obtained in \Ref{poz1,poz2} (see
section \ref{FQHVA}).  It is not clear if those  conditions are
actually sufficient or not.

Then, we try to use the pattern of zeros $\{n;m;S_a\}$ to define a
$Z_n$ simple-current vertex algebra.  From the complete GJI of the
$Z_n$ simple-current vertex algebra, we obtain sufficient conditions
on the pattern of zeros $\{n;m;S_a\}$ (see section \ref{GJIVA}).

\subsection{How to expand the pattern-of-zeros data to completely
characterize the topological order}

If a pattern of zeros  $\{n;m;S_a\}$ can uniquely describe the
topological order in a quantum Hall ground state, then from such a
quantitative description, we should be able to calculate the
topological properties from the data $\{n;m;S_a\}$. Indeed, this can
be done.  First different types of quasiparticles can also be
quantitatively described and labeled by a set of sequences
$\{S_{\ga;a}\}$ that can be determined from the pattern-of-zeros
data $\{n;m;S_a\}$\cite{poz2}. Those quantitative characterizations
of the quantum Hall ground state and quasiparticles allow us to
calculate the number of different quasiparticle types, quasiparticle
charges, fusion algebra between the quasiparticles, and topological
ground state degeneracy on a Riemann surface of any
genus.\cite{poz2,poz3}

However, from the pattern-of-zeros data, $\{n;m;S_a\}$ and
$\{S_{\ga;a}\}$, we still do not know how to calculate the
quasiparticle statistics and scaling dimensions, as well as the
central charge $c$ of the edge states.  This difficulty is related to
the fact that some patterns of zeros do not uniquely characterize a
FQH state. Thus one cannot expect to calculate the topological
properties of FQH state from the pattern-of-zeros data alone in those
cases.

In this paper, we will try to solve this problem.  We first introduce
a more complete characterization for FQH states in terms of a expanded
data set: $\{n;m;S_a;c\}$. Then, we use the data set $\{n;m;S_a;c\}$
to define a so called $Z_n$ simple-current vertex algebra. The $Z_n$
simple-current vertex algebra contain a subalgebra, Virasoro algebra,
generated by the energy-momentum tensor $T$ and $c$ is the central
charge of the Virasoro algebra.  It contains only $n$ primary fields
$\psi_a$, $a=0,1,...,n-1$ of the Virasoro algebra, with a $Z_n$ fusion
rule $\psi_a\psi_b\sim \psi_{(a+b)\mod n}$, $\psi_n=\psi_0$. Those
$\psi_a$ are called simple currents.  The extra data $c$ is the one of
the structure constants of the $Z_n$ simple-current vertex algebra.
One may want to include all the structure constants $\{C_{ab}\}$ in
the data set to have a complete characterization.  But for the
examples discussed in this paper, we find that data set
$\{n;m;S_a;c\}$ already provides a complete characterization.  So in
this paper, we will use $\{n;m;S_a;c\}$ to characterize FQH states. If
later we find that $\{n;m;S_a;c\}$ is not sufficient, we can always
add additional data, such as $C_{ab}$.  Every $Z_n$ simple-current
vertex algebra uniquely define a FQH state, and the data
$\{n;m;S_a;c\}$ that defines a $Z_n$ simple-current vertex algebra
also completely characterizes a FQH state.

We would like to remark that although the data $\{n;m;S_a;c\}$
and the corresponding $Z_n$ simple-current vertex algebras describe
a large class of FQH states, they do not describe all FQH states.
For example let $\Phi_{\cA_i}$ be the FQH wave function described by
a $Z_{n_i}$ simple-current vertex algebra $\cA_i$, $i=1,2$. Then, in
general, the FQH state described by the product wave function $\Phi=
\Phi_{\cA_1} \Phi_{\cA_2}$ cannot be described by a simple-current
vertex algebra. Such a product state is described by the product
vertex algebra $\cA_1\otimes \cA_2$, which is in general no longer a
simple-current vertex algebra.  So a more general FQH state should
have a form
\begin{equation}
\label{cAi} \Phi=\prod_i \Phi_{\cA_i}.
\end{equation}
The study in \Ref{poz1,poz2,poz3} reveal that many FQH states
described by pattern of zeros have the following form
\begin{align}
\label{Zkana}
 \Psi(\{z_i\})=\prod_a \Phi_{Z_{n_a}^{(k_a)}}(\{z_i\})
\end{align}
where $\Phi_{Z_{n_a}^{(k_a)}}(\{z_i\})$ is the wave function
described by $Z_{n_a}^{(k_a)}$ parafermion vertex
algebra.\cite{poz3} The $Z_n$ simple-current vertex algebra
mentioned above is a natural generalization of the $Z_{n_a}^{(k_a)}$
parafermion vertex algebra, and \eqn{cAi} naturally generalizes
\eqn{Zkana}.  (Note that there are many $Z_n$ simple-current vertex
algebras even for a fixed $n$, so there are many different $Z_n$
simple-current states.)

For the subclass of FQH states described by $Z_n$ simple-current
vertex algebra (which includes Virasoro algebra as an essential
part), the quasiparticle statistics and scaling dimensions, as well
as the central charge $c$ of the edge states can be calculated from
the data $\{n;m;S_a;c\}$.  Certainly, we can also calculate the
number of different quasiparticle types, quasiparticle charges,
fusion algebra between the quasiparticles, and topological ground
state degeneracy on a Riemann surface of any genus.

Obviously, not every collection $\{n;m;S_a;c\}$ corresponds to a $Z_n$
simple-current vertex algebra and a FQH state.  GJIs of the $Z_n$
simple-current vertex algebra generate the consistent conditions on
the data set $\{n;m;S_a;c\}$.  Only those data sets $\{n;m;S_a;c\}$
that satisfy the GJIs can describe a $Z_n$ simple-current vertex
algebra and FQH states.

For some patterns of zeros $\{n;m;S_a\}$, we
find that they uniquely define the vertex algebras by completely
determining the structure constants $c$ and $C_{ab}$.  So those
patterns of zeros completely specify the corresponding FQH wave
functions.  While for other patterns of zeros, we find that they
cannot uniquely define the vertex algebras. For those patterns of
zeros, many different sets of structure constants can satisfy the GJIs
for the same set of pattern of zeros.  This corresponds to the
situation where there are many different FQH wave functions that share
the same pattern of zeros. In this case, the pattern of zeros does not
completely characterize FQH wave functions. We need additional data to
completely characterize quantum Hall wave functions.  Here we choose
to add the structure constant $c$ of the Virasoro algebra (which is
the central charge) and use $\{n;m;S_a;c|a=1,...,n; n,m,S_a\in \dN;
c\in \dR \}$ to characterize FQH states.  For all the examples that we
considered in this paper, the data $\{n;m;S_a;c\}$ completely
determine the simple-current vertex algebra.

\subsection{The organization of the paper}

This paper is organized as follows. In section \ref{POZ}, we review
and extend the pattern-of-zeros approach to quantum Hall states.  In
section \ref{FQHVA} we use the pattern of zeros to define $Z_n$
vertex algebra, and then use associativity conditions (\ie the GJIs)
of the vertex algebra to obtain extra conditions on the pattern of
zeros that describe generic FQH states.  In section \ref{NUM} we
list some numerical solutions of the pattern of zeros for the
generic FQH states that also satisfy those extra consistent
conditions found in section \ref{FQHVA}.  In section \ref{GJIVA} we
define and construct the $Z_n$ simple-current vertex algebra from
the pattern of zeros.  We list the consistent conditions obtained
from GJIs of $Z_n$ simple-current vertex algebra.  The detailed
derivations of those consistent conditions are discussed in appendix
\ref{app:GJI}, \ref{gVAhC} and \ref{app:subleading}.  The consistent
conditions on the patterns of zeros that describe a $Z_n$
simple-current vertex algebra are more restrictive than those for a
generic $Z_n$ vertex algebra. Some of the solutions of the $Z_n$
simple-current pattern of zeros are listed in section \ref{EXM}.  In
section \ref{qpva}, we discuss how to represent quasiparticles in
the $Z_n$ simple-current vertex algebra, and to calculate the
topological properties of quasiparticles from the $Z_n$
simple-current pattern of zeros. In section \ref{EXM}, we apply the
vertex-algebra approach developed here to study some simple (but
non-trivial) examples of FQH states, which include $Z_n$ parafermion
states (the Read-Rezayi states\cite{RR99}), the $Z_n$ simple-current
FQH states of $Z_n|Z_n$ type, a $Z_4$ simple-current FQH state of
$Z_4|Z_2$ type, etc.

\section{Pattern-of-zeros approach to generic FQH states}
\label{POZ}

In this section, we will review how to use the pattern of zeros to
characterize and classify different FQH states that have one
component.\cite{poz1,poz2,poz3}  A discussion on two-component FQH
states can be find in \Ref{BWmultpoz}.

\subsection{FQH wave functions and symmetric polynomials}

Generally speaking, to classify generic complex wave functions
$\Phi(\textbf{r}_1,\cdots,\textbf{r}_N)$ is not even a well-defined problem.
Fortunately, under a strong magnetic field, electrons are spin-polarized in the
lowest Landau level (LLL) when the electron filling fraction $\nu_e$ is less
than 1. The wave function of a single electron in LLL (we set magnetic length
$l_B=\sqrt{\hbar/eB}$ to be unity hereafter) is $\Psi_m(z)=z^m\e^{-|z|^2/4}$ in
a planar geometry. $m$ is the angular momentum of this single particle state.
Thus the many-body wave function of \emph{spin-polarized} electrons in the LLL
should be
\begin{equation}
\Psi_e(z_1,\cdots,z_N)=\tilde\Phi_e(z_1,\cdots,z_N)
\e^{-\sum_{i=1}^N\frac{|z_i|^2}{4}}
\end{equation}
where $\tilde\Phi_e(\{z_i\})$ is an anti-symmetric holomorphic
polynomial of electron coordinates $\{z_i=x_i+\imth y_i\}$. The
electron filling fraction $\nu_e$ is defined as:
\begin{equation}\label{electron filling fraction}
\nu_e=\lim_{N\rightarrow\infty}\frac{N}{N_{\phi}}=\lim_{N\rightarrow\infty}\frac{N^2}{2N_p}
\end{equation}
where $N_\phi$ is the total number of flux quanta piercing through
the sample, and $N_p$ is the total degree of polynomial
$\tilde\Phi_e(\{z_i\})$. For FQH states $\nu_e<1$, we can extract a
Jastraw factor $\prod_{i<j}(z_i-z_j)$ and the remaining part
\begin{equation}
\Phi(z_1,\cdots,z_N)=\frac{\tilde\Phi_e(z_1,\cdots,z_N)}{\prod_{i<j}(z_i-z_j)}
\end{equation}
would be a symmetric polynomial of $\{z_i\}$. We will concentrate on
this symmetric polynomial to characterize and classify FQH states.

For the symmetric polynomial $\Phi(\{z_i\})$ we can also define a
filling fraction $\nu$ in the same way as in \eqn{electron filling
fraction}, only $N_p$ replaced by the total degree of bosonic
polynomial $\Phi(\{z_i\})$. The electron filling fraction $\nu_e$
has the following relation with this \emph{bosonic filling fraction}
$\nu$:
\begin{align}
\nu_e=\frac{1}{1+\nu^{-1}}<1
\end{align}

\subsection{Fusion of $a$ variables: the Pattern of Zeros}

The pattern of zeros\cite{poz1,poz2} is introduced to describe
symmetric polynomials $\Phi(\{z_i\})$ through certain local
properties, i.e. fusion of $a$ different variables $z_1,\cdots,z_a$.
More specifically, we bring these $a$ variables together, viewing
$z_{a+1},\cdots,z_N$ as fixed coordinates. By writing the $a$
variables in the following manner
$z_i=\lambda\xi_i+z^{(a)}$, $i=1,\cdots,a$,
where $z^{(a)}=\frac{z_1+\cdots+z_a}{a}$ and $\sum_{i=1}^a\xi_a=0$,
we can bring these $a$ variables together by letting $\lambda$ tend
to zero.  Then we can expand the polynomial $\Phi(\{z_i\})$ in
powers of $\lambda$:
\begin{align}\label{derived polynomial}
&\ \ \ \lim_{\lambda\rightarrow0^+}
\Phi(\lambda\xi_1+z^{(a)},\cdots,\lambda\xi_a+z^{(a)};z_{a+1},\cdots,z_N)
\\
&=\lambda^{S_a}P_{S_a}[z^{(a)},(\xi_1,\cdots,\xi_a);z_{a+1},\cdots,z_N]
+O(\lambda^{S_a+1}) \nonumber
\end{align}
In other words, $\{S_a\}$ is the lowest order of zeros when we fuse $a$
variables together. The pattern of zeros, by definition, is this sequence of
integers $\{S_a\}$.  In this paper, we will only consider the polynomials that
satisfy a unique fusion condition: the fusion of $a$ variables is unique, i.e.
$P_{S_a}$ in \eqn{derived polynomial} has the same form except for an overall
factor no matter how $\{\xi_i\}$ are chosen.

There are other equivalent descriptions of the pattern of zeros. One
of them is the orbital description:
\begin{align}
l_a=S_a-S_{a-1}\ \ \ a=1,2,\cdots
\end{align}
where $\{l_a\}$ labels the orbital angular momentum of the
single-particle state occupied by the $a$-th particle.  Another is
the occupation description in terms of a sequence of integers
$\{n_l\}$\cite{seidel,bergholtz,BH0802,BH0802a}, denoting the number of
particles occupying the orbital with angular momentum $l$.

\subsection{Consistent conditions for the Pattern of Zeros}

In this section, we will review and summarize the consistent
conditions on $\{S_a\}$ derived in \Ref{poz1,poz2}.

\subsubsection{Translational invariance}

A translational invariant wave function $\Phi(z_1,\cdots,z_N)=
\Phi(z_1-z,\cdots,z_N-z) $ satisfies $\Phi(0,z_2,\cdots,z_N)\neq0$.  As a
result we have $S_1=0$.

\subsubsection{Symmetry condition}

After we fuse $a$ variables together to form an $a$-particle cluster
($a$-cluster), it is natural to ask: what happens when we fuse an
$a$-cluster and another $b$-cluster together? Let $D_{a,b}$ be the
order of zeros obtained by fusing an $a$-cluster and another
$b$-cluster together. It satisfies $D_{a,b}=D_{b,a}\geq0$. Since the
final state is the same as fusing $a+b$ variables together, we find
an one-to-one relation between the two sets of data $D_{a,b}$ and
$S_a$\cite{poz1}
\begin{align}
D_{a,b}=S_{a+b}-S_a-S_b
\nonumber \\
S_a=\sum_{i=1}^{a-1}D_{i,1}
\end{align}

Since $\Phi(\{z_i\})$ is a symmetric polynomial, it describes a
state of bosonic particles seated at coordinates $\{z_i\}$. Thus the
$a$-cluster seated at $z_i^{(a)}$ can also be regarded as a bosonic
particle. The derived polynomial (see $P_{S_a}$ in \eqn{derived
polynomial} as an example) should be symmetric with respect to
interchange of two identical bosons seated at $z_1^{(a)}$ and
$z_2^{(a)}$. When we fuse such two identical bosonic clusters
$z_1^{(a)}$ and $z_2^{(a)}$ together, we have
\begin{align}
&\ \ \ \lim_{z_1^{(a)}\rightarrow
z_2^{(b)}}P(z_1^{(a)},z_2^{(a)},\cdots)
\nonumber\\
&=(z_1^{(a)}-z_2^{(a)})^{D_{a,a}}\tilde
P(\frac{z_1^{(a)}+z_2^{(a)}}{2},\cdots)
\nonumber\\
&\ \ \ \ \ \ \ \ \ +O((z_1^{(a)}-z_2^{(a)})^{D_{a,a}+1})
\end{align}
This leads to the symmetry condition
\begin{align}\label{symmetric}
D_{a,a}=\text{even}\Leftrightarrow S_{2a}=\text{even} .
\end{align}

\subsubsection{Concave conditions}

The 1st concave condition is the non-negativity of $D_{a,b}$
\begin{align}\label{concave1}
D_{a,b}\geq0\Leftrightarrow S_{a+b}\geq S_a+S_b
\end{align}
It comes naturally from the fusion of two clusters.

When we fuse three clusters together, we find the total order of
``off-particle'' zeros to be
\begin{align}
\Delta_3(a,b,c)=D_{a,b+c}-D_{a,b}-D_{a,c}\geq0
\end{align}
This gives the 2nd concave condition:
\begin{align}\label{concave2}
&\ \ \ \Delta_3(a,b,c)=
\\
&S_{a+b+c}+S_a+S_b+S_c-S_{a+b}-S_{a+c}-S_{b+c}\geq0 \nonumber
\end{align}

\subsubsection{$n$-cluster condition}

The above conditions \eqn{symmetric}, \eqn{concave1}, and \eqn{concave2} have
many solutions $\{S_a\}$.  Many of those solutions has a ``periodic'' structure
that the whole sequence $\{S_a\}$ can be determined from first a few terms:
\begin{align}\label{n cluster1}
S_{a+kn}=S_a+kS_n+mn\frac{k(k-1)}{2}+kma,
\end{align}
where
\begin{align}
m\equiv D_{n,1} .
\end{align}
We will call such a pattern of zeros the one that satisfies an $n$-cluster
condition.  We see that, for an $n$-cluster sequence, only the first $n$ terms,
$S_2$, ..., $S_{n+1}$, are independent, and the whole sequence is determined by
the first $n$ terms.

To understand the physical meaning of the $n$-cluster condition, we note
that \eqn{n cluster1} is equivalent to the following condition
\begin{align}
\Delta_3(kn,b,c)=0  \ \ \text{ for any } k.
\end{align}
This means that a symmetric polynomial that satisfies the $n$-cluster condition
has the following defining property: as a function of the $n$-cluster
coordinate $z^{(n)}$, the derived polynomial $P(z^{(n)},z_1^{(a)},\cdots)$
has no off-particle zeros.

Under the $n$-cluster condition, we see that
\begin{align}
\label{mneven}
 D_{n,n}=nm=\text{even}
\end{align}
We also note that the filling fraction is given by
\begin{align}
\label{numn}
 \nu=\frac{n}{m}
\end{align}
since $S_a \to \frac12 \frac{m}{n} a^2$ as $a\to \infty$.

We like to mention that the cluster condition plays a very important role in
the Jack polynomial approach to FQH.\cite{BH0802,BH0802a,ES0969} However, in
the pattern of zeros approach, the $n$-cluster condition only play a role of
grouping and tabulating solutions of the consistent conditions.  The
solutions with larger $n$ correspond to more complex wave functions which 
usually correspond to less stable FQH states.  Later, we will discuss the
relation between the pattern of zeros and CFT. We find that the solutions that
do not satisfy the $n$-cluster condition (\ie with $n=\infty$) correspond to
irrational CFT, which may always correspond to gapless FQH states.  The Jack
polynomial approach and the pattern-of-zeros approach have some close
relations.  The Jack polynomials are special cases of the polynomials
characterized by pattern of zeros.  

\subsubsection{Summary}

To summarize, we see that the pattern of zeros for an $n$-cluster
polynomial is described by a set of positive integers
$\{n;m;S_2,...,S_n\}$. Introducing $S_1=0$ and
\begin{align}\label{ncluster}
S_{a+kn}=S_a+kS_n+mn\frac{k(k-1)}{2}+kma
\end{align}
which define $S_{n+1}$, $S_{n+2}$, ..., we find that the data
$\{n;m;S_2,...,S_n\}$ must satisfy
\begin{align}\label{Dcon}
D_{a,b} &=S_{a+b}-S_a-S_b\geq0
\nonumber\\
D_{a,a} &=\text{even}
\end{align}
\begin{align}\label{del3con}
&\ \ \ \Delta_3(a,b,c)
\\
&=S_{a+b+c}+S_a+S_b+S_c-S_{a+b}-S_{a+c}-S_{b+c}\geq0 \nonumber
\end{align}
for all $a,b,c=1,2,3\cdots$.

The conditions \eq{Dcon} and \eq{del3con} are necessary conditions
for a pattern of zeros to represent a symmetric polynomial. Although
\eqn{Dcon} and \eqn{del3con} are very simple, they are quite
restrictive and are quite close to be sufficient conditions. In fact
if we add an additional condition
\begin{equation}
\label{del3even}
 \Delta_3(a,b,c) = \text{even}
\end{equation}
the three conditions \eq{Dcon}, \eq{del3con}, and \eq{del3even} may
even become sufficient conditions for a pattern of zeros to
represent a symmetric polynomial.\cite{poz1,poz2} However, this
condition is too strong to include many valid symmetric polynomials
such as Gaffnian,\cite{Gaffnian} a nontrivial $Z_4$ state discussed
in detail in section \ref{EXM}. We will obtain some additional
conditions in section \ref{subsect:generic cond from GJI}, which
combined with \eq{Dcon} and \eq{del3con} form a set of necessary and
(potentially) sufficient conditions for a valid pattern of zeros.

\subsection{Label the pattern of zeros by $\hsc{a}$}

In this section, we will introduce a new labeling scheme of the
pattern of zeros.
We can label the pattern of zeros in terms of
\begin{align}
\label{hscaSa} h^\text{sc}_a & = S_a-
\frac{aS_n}{n}+\frac{am}{2}-\frac{a^2m}{2n}.
\end{align}
This labeling scheme is intimately connected to the vertex algebra
approach that we will discuss later.

The $n$-cluster condition \eq{ncluster} of $S_a$ implies that
$\hsc{a}$ is periodic
\begin{equation}
\label{hscperi} h^\text{sc}_0=0,\ \ \ \ \ \ \
h^\text{sc}_a=h^\text{sc}_{a+n}
\end{equation}
The two sets of data $\{n;m;S_2,...,S_{n} \}$ and
$\{n;m;\hsc{1},...,\hsc{n-1} \}$ has a one-to-one correspondence,
since
\begin{equation}
\label{Sahsca}
 S_a=h^\text{sc}_a-ah^\text{sc}_1+\frac{a(a-1)m}{2n} .
\end{equation}

We can translate the conditions on $\{m;S_a\}$ to the equivalent
conditions on $\{m;\hsc{a} \}$.  First, we have
\begin{align}
\label{hcond1} & 2nS_a = 2nh^\text{sc}_a-2nah^\text{sc}_1+a(a-1)m=0
\mod 2n
\nonumber\\
& nS_{2a} = nh^\text{sc}_{2a}-2nah^\text{sc}_1+a(2a-1)m=0 \mod 2n
\nonumber\\
& m>0,\ \ \ \ mn=\text{even}
\end{align}

$n S_{2n}=0\mod2n$ in \eqn{hcond1} leads to $2n h^\text{sc}_1 + m =
0 \text{ mod } 2$, from which we see that $2n h^\text{sc}_1$ is an
integer. From $2n h^\text{sc}_a-a (2n h^\text{sc}_1)+a(a-1)m=$ even
integer, we see that $2n h^\text{sc}_a$ are always integers. Also
$2nh^\text{sc}_{2a}$ are always even integers, and
$2nh^\text{sc}_{2a+1}$ are either all even or all odd. Since
$h^\text{sc}_n=0$, thus when $n=$ odd, $2n h^\text{sc}_a$ are all
even.  Only when $n=$ even, can $2n h^\text{sc}_{2a+1}$ either be
all even or all odd. When $m=$even, $2nh^\text{sc}_{2a+1}$ are all
even. When $m=$odd, $2nh^\text{sc}_{2a+1}$ are all odd.

The two concave conditions become
\begin{align}
\label{hcond2}
&\ \ \ h^\text{sc}_{a+b}-h^\text{sc}_a-h^\text{sc}_b + \frac{abm}{n}
= D_{ab} =\text{integer} \geq 0
\\
\label{hcond3} &\ \ \  h^\text{sc}_{a+b+c} -h^\text{sc}_{a+b}
-h^\text{sc}_{b+c} -h^\text{sc}_{a+c} +h^\text{sc}_a +h^\text{sc}_b
+h^\text{sc}_c
\nonumber\\
&= \Del_3(a,b,c) = \text{integer} \geq 0
\end{align}
The valid data $\{n;m;\hsc{1},...,\hsc{n-1}\}$ can be obtained by
solving \eqn{hscperi}, \eqn{hcond1}, \eqn{hcond2}, and \eqn{hcond3}.

Choosing $1\leq a,b<a+b\leq n$ in \eqn{hcond3}, we have
\begin{align}
0& \leq\Delta_3(a,b,n-a-b)
\nonumber\\
&=(\hsc{n-a-b}-\hsc{a+b})-(\hsc{n-a}-\hsc{a})-(\hsc{n-b}-\hsc{b})
\nonumber\\
&=-\Delta_3(n-a,n-b,a+b)\leq0 \nonumber
\end{align}
which implies the following \emph{reflection condition} on
$\{\hsc{a}\}$:
\begin{align}\label{symmetric4scaling dimension}
\hsc{a}-\hsc{n-a}=a(\hsc{1}-\hsc{n-1})=0
\end{align}

{}From (\ref{symmetric4scaling dimension}) we see that partially
solving conditions (\ref{hcond3}) reduces the number of independent
variables characterizing a pattern of zeros from $n-1$ in
$\{S_2,\cdots,S_n\}$ to $[\frac{n}{2}]$ in
$\{\hsc{1},\cdots,\hsc{[\frac{n}{2}]}\}$. However, being a sequence
of fractions rather than integers, $\{\hsc{a}\}$ labeling scheme
imposes some difficulty in numerically solving conditions
(\ref{hscperi}), (\ref{hcond1}), (\ref{hcond2}), and (\ref{hcond3}).
In Appendix (\ref{label:a_j}) and (\ref{pMk}) we will further use
consistent conditions (\ref{hcond3}) to introduce two schemes
labeling the pattern of zeros with a sequence of non-negative
integers or half-integers. They turn out to be quite efficient for
numerical studies, since consistent conditions (\ref{hscperi}),
(\ref{hcond1}), (\ref{hcond2}), and (\ref{hcond3}) can be reduced to
a much smaller set after introducing a new labeling scheme
$\{M_k;p;m\}$ as in Appendix \ref{pMk}. In particular, this
$\{M_k;p;m\}$ labeling scheme is the same one as adopted in the
literature of parafermion vertex algebra.\cite{ZF0}\\

\section{Constructing FQH wave functions from $Z_n$ vertex algebras}
\label{FQHVA}

If we use $\{n;m;\hsc{a}\}$ to characterize $n$-cluster symmetric
polynomial $\Phi(\{z_i\})$, the conditions \eq{hcond1}, \eq{hcond2},
and \eq{hcond3} are required by the single-valueness of the
symmetric polynomial.  Or more precisely, eqns. \eq{hcond1},
\eq{hcond2}, and \eq{hcond3} come from a simple requirement that the
zeros in $\Phi(\{z_i\})$ all have integer orders.  However, the
conditions \eq{hcond1}, \eq{hcond2}, and \eq{hcond3} are incomplete
in the sense that some patterns of zeros $\{n;m;\hsc{a}\}$ can
satisfy those conditions but still do not correspond to any valid
polynomial.

\subsection{FQH wave function as a correlation function
in $Z_n$ vertex algebra}

To find more consistent conditions, in the rest of this paper, we
will introduce a new requirement for the symmetric polynomial.  We
require that the symmetric polynomial can be expressed as a
correlation function in a vertex algebra.
More specifically, we
have\cite{MR,wenwu,wenwuh}
\begin{align}
\label{PhiV}
\Phi(\{z_i\})=\lim_{z_\infty\rightarrow\infty}z^{2h_N}_\infty\langle
V(z_\infty)\prod_{i=1}^NV_e(z_i)\rangle
\end{align}
where $V_e(z)$ is an electron operator and $V(\infty)$ represents a
positive background to guarantee the charge neutral condition. This
new requirement, or more precisely, the associativity of the vertex
algebra, leads to new conditions on $\hsc{a}$.

The electron operator has the following form
\begin{align}
V_e(z)=\psi(z):\e^{\imth \phi(z)/\sqrt\nu}:
\end{align}
where $:\e^{\imth \phi(z)/\sqrt\nu}:$ (:: stands for normal
ordering, which is implicitly understood hereafter) is a vertex
operator in a Gaussian model. It has a scaling dimension of
$\frac{1}{2\nu}$ and the following operator product expansion
(OPE)\cite{cft-bible}
\begin{align}
\e^{\imth a\phi(z)}\e^{\imth b\phi(w)}=(z-w)^{ab} \e^{\imth
(a+b)\phi(w)}+O\Big((z-w)^{ab+1}\Big)
\end{align}
The operator $\psi$ is a primary field of Virasoro algebra obeys an
quasi-Abelian fusion rule
\begin{align}
\label{psiapsibG} \psi_a\psi_b\sim\psi_{a+b} + ...,\ \ \ \ \ \ \
\psi_a\equiv(\psi)^a .
\end{align}
where $...$ represent other primary fields of Virasoro
algebra whose scaling dimensions are higher than that of
$\psi_{a+b}$ by some integer values. We believe that the integral
difference of the scaling dimensions is necessary to produce a
single-valued correlation function(see \eqn{PhiV}).

Let $\thsc{a}$ be the scaling dimension of the simple current
$\psi_a$. Therefore the $a$-cluster operator
\begin{align}
V_a\equiv(V_e)^a=\psi_a(z)\e^{\imth a\phi(z)/\sqrt\nu}
\end{align}
has a scaling dimension of
\begin{align}
\label{hahsca} h_a=\thsc{a}+\frac{a^2}{2\nu}
\end{align}

The vertex algebra is defined through the following OPE of the
$a$-cluster operators
\begin{align}
\label{VaVbOPE} V_a(z)V_b(w)
&=C^V_{a,b}\frac{V_{a,b}(w)}{(z-w)^{h_a+h_b-h_{a+b}}}
\nonumber\\
&\ \ \ \ \ \ \ \ +O\Big((z-w)^{h_{a+b}-h_a-h_b+1}\Big).
\end{align}
where $C^V_{a,b}$ are the structure constants. However, the above
OPE is not quite enough.  To fully define the vertex algebra, we
also need to define the relation between $V_a(z)V_b(w)$ and $V_b(w)
V_a(z)$.

The correlation functions is calculated through the expectation
value of radial-ordered operator
product.\cite{cft-ginsparg,cft-bible,va0} The radial-ordered
operator product is defined through
\begin{align}
\label{rorder} &\ \ \ (z-w)^{\al_{V_aV_b}} R[V_a(z)V_b(w)]
\nonumber\\
&=
\begin{cases}
\ \ \ \ \ \ (z-w)^{\al_{V_aV_b}} V_a(z)V_b(w),&  |z|>|w| \\
\mu_{ab}    (w-z)^{\al_{V_aV_b}} V_b(w)V_a(z),&  |z|<|w| \\
\end{cases}
\end{align}
where
\begin{align}
\al_{V_aV_b}=h_a+h_b-h_{a+b}.
\end{align}
Note that the extra complex factor $\mu_{ab}$ is introduced in the
above definition of radial order.  In the case of standard
conformal algebras, where $\al_{V_aV_b}\in \dZ$, we choose
$\mu_{ab}=-\e^{\imth\pi\al_{V_aV_b}}$ if both $V_a$ and $V_b$ are
fermionic and $\mu_{ab}=\e^{\imth\pi\al_{V_aV_b}}$ if at least one
of them is bosonic. But in general, the commutation factor can be
different from $\pm 1$ and can be chosen more arbitrarily.

To gain an intuitive understanding of the above definition of radial
order, let us consider the Gaussian model and choose $V_a=\e^{\imth
a\phi}$ and $V_b=\e^{\imth  b\phi}$.  The scaling dimension of $V_a$
and $V_b$ are $h_a=\frac{a^2}{2}$ and $h_b=\frac{b^2}{2}$.
$\al_{V_aV_b}=h_a+h_b-h_{a+b}$. We see that $\al_{V_aV_b}\in \dZ$ if
$a,b\in \dZ$ and such a Gaussian model is an example of standard
conformal algebras.  If both $a$ and $b$ are odd, then $h_a$ and
$h_b$ are half integers and $V_a$ and $V_b$ are fermionic operators.
In this case $\al_{V_aV_b}=-ab=$ odd.  So under the standard choice
$\mu_{ab}=-\e^{\imth\pi\al_{V_aV_b}}$, we have $\mu_{ab}=1$.  If one
of $a$ and $b$ is even, $\al_{V_aV_b}=-ab=$ even and one of $V_a$
and $V_b$ is bosonic operators.  Under the standard choice
$\mu_{ab}=\e^{\imth\al_{V_aV_b}}$, we have again $\mu_{ab}=1$.  Even
when $a$ and $b$ are not integers, in the Gaussian model, the radial
order of $V_a=\e^{\imth a\phi}$ and $V_b=\e^{\imth  b\phi}$ is still
defined with a choice $\mu_{ab}=1$. This is a part of the definition
of the Gaussian model.  In this paper, we will choose a more general
definition of radial order where $\mu_{ab}$ are assumed to be
generic complex phases $|\mu_{ab}|=1$.\\

The vertex algebra generated by $\psi$ have a form
\begin{align}
\label{psiapsibOPE} \psi_a(z)\psi_b(w)
&=C_{a,b}\frac{\psi_{a+b}(w)}{(z-w)^{\t h^\text{sc}_a+\t
h^\text{sc}_b-\t h^\text{sc}_{a+b}}}
\nonumber\\
&\ \ \ \ \ \ \ \ +O\Big((z-w)^{\t h^\text{sc}_{a+b}-\t
h^\text{sc}_a-\t h^\text{sc}_b+1}\Big).
\end{align}
where
\begin{align}
\label{Cnonzero} C_{a,b}\neq 0.
\end{align}
When combined with the $U(1)$ Gaussian model, the above vertex
algebra can produce the wave function for a FQH state (see
\eqn{PhiV}).

We will also limit ourselves to the vertex algebra that satisfies
the $n$-cluster condition:
\begin{equation}
\label{psin1}
 \psi_n=1
\end{equation}
where $1$ stands for the identity operator defined in Appendix
\ref{app:muABcon}. Those vertex algebras are in some sense ``finite'' and
correspond to rational conformal field theory.  We will call such vertex
algebra $Z_n$ vertex algebra.  We see that in general, a FQH state can be
described by the direct product of a $U(1)$ Gaussian model and a $Z_n$ vertex
algebra.  Some exmaples of  $Z_n$ vertex algebra are studied in
\Ref{DGZ0297,DGZ0249}.

Note that the $Z_n$ vertex algebras are different from the $Z_n$
simple-current vertex algebras that will be defined in section
\ref{GJIVA}. The $Z_n$ simple-current vertex algebras are special
cases of the $Z_n$ vertex algebras.  In this and the next sections,
we will consider $Z_n$ vertex algebras. We will further limit
ourselves to $Z_n$ simple-current vertex algebras in section
\ref{GJIVA} and later.

As a result
\begin{align}
 \thsc{a}&=\thsc{a+n},\ \ \ \ \ \thsc{n}=0
\nonumber\\
 \mu_{ab}&= \mu_{a+n,b}= \mu_{a,b+n}, \ \ \ \ \mu_{n,a}=\mu_{a,n}=1
\nonumber\\
 C_{a,b}&= C_{a+n,b}= C_{a,b+n}, \ \ \ \ C_{n,a}=C_{a,n}=1\nonumber\\
 C_{a,b}&=\mu_{a,b}C_{b,a}
\end{align}
By choosing proper normalizations for the operators $\psi_a$, we can
have
\begin{align}\label{Cstr:normalization}
\nonumber C_{a,-a}=\left\{\begin{aligned}1,~~~&a\mod n\leq n/2\\
\mu_{a,-a},~~~&a\mod n>n/2\end{aligned}\right.\\
C_{a,b}=1,~~~\text{if}~a~\text{or}~b=0\mod n
\end{align}

To summarize, we see that the $Z_n$ vertex algebras (whose correlation
functions give rise to electron wave functions) are characterized by
the following set of data $\{n;m;\thsc{a};C_{a,b},...|a,b=1,...,n\}$,
where $m=n/\nu$.  Here the $...$ represent other structure constants
in the subleading terms.  The commutation factors $\mu_{ab}$ are not
included in the above data because they can be expressed in terms of
$\thsc{a}$ and are not independent (see \eqn{muij}). Since the $Z_n$
vertex algebra encodes the many-body wave function of electrons, we
can say that the data $\{n;m;\thsc{a};C_{a,b},...|a,b=1,...,n\}$ also
characterize the electron wave function.  We can study all the
properties of electron wave functions by studying the data
$\{n;m;\thsc{a};C_{a,b},...|a,b=1,...,n\}$. In the pattern-of-zero
approach, we use data $\{n;m;\hsc{a}\}$ to characterize the wave
functions.  We will see that the
$\{n;m;\thsc{a};C_{a,b},...|a,b=1,...,n\}$ characterization is more
complete, which allows us to obtain some new results.

\subsection{Relation between $\thsc{a}$ and $\hsc{a}$}

What is the relation between the two characterizations:
$\{n;m;\hsc{a}|a=1,...,n\}$ and
$\{n;m;\thsc{a};C_{ab}|a,b=1,...,n\}$? The single-valueness of the
correlation function $\Phi(\{z_i\})$ requires that the zeros in
$\Phi(\{z_i\})$ all have integer orders. In this section, we derive
conditions on the scaling dimension $\thsc{a}$, just from this
integral-zero condition.  This allows us to find a simple relation
between $\{n;m;\hsc{a}|a=1,...,n\}$ and
$\{n;m;\thsc{a};C_{ab}|a,b=1,...,n\}$.

{}From the definition of $D_{ab}$ and the OPE \eq{VaVbOPE}, we see
that
\begin{align}
\label{Dabha} D_{a,b} & \equiv S_{a+b}-S_a-S_b =h_{a+b}-h_a-h_b
\nonumber\\
& =\thsc{a+b}-\thsc{a}-\thsc{b}+\frac{ab}{\nu}=D_{b,a}
\end{align}
We see that $D_{1,n}=\frac{n}{\nu}$. So $\frac{n}{\nu}$ is an
positive integer which is called $m$.

{}From \eqn{Dabha}, we can show that\cite{poz1,poz2}
\begin{align}\label{S-hsc}
S_a=\sum_{i=1}^{a-1}D_{i,1}=h_a-ah_1=\thsc{a}-a\thsc{1}+\frac{a(a-1)}{2\nu}
\end{align}
and
\begin{align}
\label{thscaSa} \t h^\text{sc}_a & = S_a-
\frac{aS_n}{n}+\frac{am}{2}-\frac{a^2m}{2n}.
\end{align}
Therefore, the $\hsc{a}$ introduced before is nothing but the
scaling dimensions $\thsc{a}$ of the simple currents $\psi_a$ (see
\eqn{hscaSa}).  In the following, we will use $\hsc{a}$ to describe
the scaling dimensions of $\psi_a$.  Thus the data
$\{n;m;\thsc{a};C_{a,b}|a,b=1,...,n\}$ can be rewritten as
$\{n;m;\hsc{a};C_{a,b}|a,b=1,...,n\}$. Those $\hsc{a}$ satisfy eqns.
\eq{hcond1}, \eq{hcond2}, and \eq{hcond3}.

As emphasized in \Ref{poz1,poz2}, the conditions \eq{hcond1},
\eq{hcond2}, and \eq{hcond3}, although necessary, are not
sufficient. In the following, we will try to find more conditions
from the vertex algebra.

\subsection{Conditions on $\hsc{a}$ and $C_{a,b}$
from the associativity of vertex algebra }\label{subsect:generic
cond from GJI}

The multi-point correlation of a $Z_n$ vertex algebra can be
obtained by fusing operators together, thus reducing the original
problem to calculating a correlation of fewer points.\cite{ZF0} It
is the associativity of this vertex algebra that guarantees any
different ways of fusing operators would yield the same correlation
in the end,\cite{va0} so that the electron wave function would be
single-valued. The associativity of a $Z_n$ vertex algebra requires
$\hsc{a}$ and $C_{a,b}$ to satisfy many consistent conditions. Those
are the extra consistent conditions we are looking for. The
consistent conditions come from two sources.  The first source is
the consistent conditions on the commutation factors $\mu_{a,b}$ as
discussed in appendix \ref{app:muABcon}. When applied to our vertex
algebra \eq{psiapsibOPE}, we find that some consistent conditions on
$\mu_{a,b}$ allow us to express $\mu_{a,b}$ in terms of
$h^\text{sc}_a$. Then other consistent conditions on $\mu_{a,b}$
will become consistent conditions on $h^\text{sc}_a$ (see appendix
\ref{nconcom}).  The second source is GJI for the vertex algebra
\eq{psiapsibOPE} as discussed in appendix \ref{nconGJI}.  We like to
stress that the discussions so far are very general. The consistent
conditions that we have obtained for generic $Z_n$ vertex algebra
are necessary conditions for any FQH states.

A detailed derivation of those conditions on $\hsc{a}$ and $C_{a,b}$
is given in appendix \ref{gVAhC}.  Here we just summarize the new
and old conditions in a compact form.  The consistent conditions can
be divided in two classes.  The first type of consistent conditions
act only on the pattern of zeros $\{n;m;h^\text{sc}_a\}$ (see eqns.
\eq{hcond1}, \eq{hcond2}, \eq{hcond3}, \eq{alijint}, \eq{alijeven},
\eq{cchsc1}, \eq{cchsc2}, and \eq{pozGJI}):
\begin{align}
\label{gVAcon1}
& nh^\text{sc}_{2a}-2nah^\text{sc}_1+a(2a-1)m=0 \mod 2n ,
\nonumber\\
& m>0,\ \ \ \ mn=\text{even} ,
\nonumber \\
&h^\text{sc}_{a+b}-h^\text{sc}_a-h^\text{sc}_b + \frac{abm}{n} \in
\dN ,
\nonumber \\
& h^\text{sc}_{a+b+c} -h^\text{sc}_{a+b} -h^\text{sc}_{b+c}
-h^\text{sc}_{a+c} +h^\text{sc}_a +h^\text{sc}_b +h^\text{sc}_c \in
\dN ,
\nonumber\\
& {n\alpha_{1,1}=\text{even}} ,
\nonumber\\
&{a^2\alpha_{1,1}-\alpha_{a,a}=\text{even}\ \ \ \forall a=1,2,\cdots
n-1},
\nonumber\\
&\Delta_3(\frac n2,\frac n2,\frac n2)=4\hsc{\frac n2}\neq1, ~\text{
if }n=~\text{even},
\end{align}
where $h^\text{sc}_a=h^\text{sc}_{a+n}$ and $\al_{a,b}=
h^\text{sc}_a +h^\text{sc}_b -h^\text{sc}_{a+b}$.

The second type of consistent conditions act on the structure
constants (see eqns.  \eq{GJIC1}, \eq{GJIC2}, \eq{GJIC3}, and
\eq{GJIC4}): For any $a,b,c$
\begin{align}
\label{gVAcon2}
C_{a,b}C_{a+b,c}&=C_{b,c}C_{a,b+c}=\mu_{a,b}C_{a,c}C_{b,a+c}
\nonumber\\
&\ \ \ \  \text{
if } \Del_3(a,b,c)=0,
\nonumber\\
C_{a,b}C_{a+b,c}&=C_{b,c}C_{a,b+c}+\mu_{a,b}C_{a,c}C_{b,a+c}
\nonumber\\
&\ \ \ \   \text{
if } \Del_3(a,b,c)=1 ,
\end{align}
where $\mu_{a,b}$ is a function of the pattern of zeros
$\{h^\text{sc}_a\}$:
\begin{equation*}
\mu_{ij}=(-1)^{ij\alpha_{1,1}-\alpha_{i,j}}=\pm1 .
\end{equation*}
 For any $a\neq n/2$
\begin{align}
\label{gVAcon3} & C_{a,-a}=C_{a,a} C_{2a,-a}=1 &&\text{ if }
\Del_3(a,a,-a)=0,
\nonumber\\
& 2C_{a,-a}=C_{a,a} C_{2a,-a} &&\text{ if } \Del_3(a,a,-a)=1.
\end{align}
Here $C_{a,b}$ satisfies the normalization condition
\eq{Cstr:normalization}. There may be additional conditions when
$\Del_3(a,b,c)\neq 0,1$. But we do not know how to derive those
conditions systematically at this time.

\section{Examples of generic FQH states
described by the $Z_n$ vertex algebra} \label{NUM}

To obtain the examples of generic FQH states, we have numerical
solved the conditions \eq{gVAcon1}. (We don't require \eqn{cchsc3}
to be satisfied, in order to include some valid interesting
solutions, like Gaffnian which violates \eqn{cchsc3}.)
In this section, we list some of those solutions in terms of
$\{n;m;\hsc{a}|_{a=1,...,n-1}\}$. First we note that, for two
$n$-cluster symmetric polynomial $\Phi_1$ and $\Phi_2$ described by
$\{n;m_1;\hsc{1,a}\}$ and $\{n;m_2;\hsc{2,a}\}$, the product
$\Phi=\Phi_1\Phi_2$ is also an $n$-cluster symmetric polynomial.
$\Phi$ is described by the pattern of zeros
\begin{equation}
\label{nmhcomp} \{n;m;\hsc{a}\}= \{n;m_1+m_2;\hsc{1,a}+\hsc{2,a}\} .
\end{equation}
Most of the solutions can be decomposed according to \eqn{nmhcomp}.
We will call the solutions that cannot be decomposed primitive
solutions. We will only list those primitive solutions.  We only
searched solutions with a filling fraction $\nu \geq 1/4$. We can
see that most solutions shown also satisfy condition \eq{cchsc3},
which means they obey OPE \eq{OPE:i,n-i} and correspond to special
$Z_n$ vertex algebras. However, some solutions such as a $4$-cluster
state called Gaffnian, explicitly violates condition \eq{cchsc3} and
their OPE's take the more general form \eq{OPE:i,n-i:i<=n/2} and
\eq{OPE:i,n-i:i>n/2}. They are described by generic $Z_n$ vertex
algebras.

\subsection{$n=1$ case}

There is only one $n=1$ primitive solution:
\begin{align}
 &n =1:\ \ \ c=0
\nonumber\\
 &\{m;\hsc{1}..\hsc{n-1}\}=\{2;\}
\nonumber\\
 &\{p;M_{1}..M_{n-1}\}=\{0; \}
\nonumber\\
 &\ \ \ \ \{n_0..n_{m-1}\} = \{1\ 0 \} .
\end{align}
It is $\nu=1/2$ Laughlin state.  Note that $\hsc{a}=0$, indicating
that the simple-current part of vertex algebra is trivial and has a
zero central charge $c=0$.  The vertex algebra contains only the
$U(1)$ Gaussian part.

\subsection{$n=2$ case}

We note that the $n=1$ primitive solution also appears as a $n=2$
primitive solution. We find only one new $n=2$ primitive solution:
\begin{align}
 &n =2:\ \ \ c=1/2\ \ \ (Z_2\text{ parafermion state})
\nonumber\\
 &\{m;\hsc{1}..\hsc{n-1}\}=\{2;\frac12 \}
\nonumber\\
 &\{p;M_{1}..M_{n-1}\}=\{1; 0\}
\nonumber\\
 &\ \ \ \ \{n_0..n_{m-1}\} = \{2\ 0 \} .
\end{align}
It is the $\nu=1$ Pfaffian state $\Phi_{Z_2}$.  The simple current
part of the vertex algebra is a $Z_2$ parafermion CFT.  If we only
use the conditions \eq{hcond1}, \eq{hcond2}, and \eq{hcond3} obtained
in \Ref{poz1,poz2}, then
\begin{align}
 &n =2:
\nonumber\\
 &\{m;\hsc{1}..\hsc{n-1}\}=\{2;\frac14 \}
\nonumber\\
 &\{p;M_{1}..M_{n-1}\}=\{\frac12; 0\}
\nonumber\\
 &\ \ \ \ \{n_0..n_{m-1}\} = \{1\ 0 \} .
\end{align}
will be a solution.  Such a solution does not correspond to any
symmetric polynomial, indicating that the conditions \eq{hcond1},
\eq{hcond2}, and \eq{hcond3} are incomplete.  An extra condition
\eq{cchsc1} from commutation factors remove such an incorrect
solution.

\subsection{$n=3$ case}

Apart from the $n=1$ primitive solution, we find only one new $n=3$
primitive solution

\begin{align}
 &n =3:\ \ \ c=4/5\ \ \ (Z_3\text{ parafermion state})
\nonumber\\
 &\{m;\hsc{1}..\hsc{n-1}\}=\{2;\frac23\ \frac23 \}
\nonumber\\
 &\{p;M_{1}..M_{n-1}\}=\{2; 0\ 0\}
\nonumber\\
 &\ \ \ \ \{n_0..n_{m-1}\} = \{3\ 0  \} .
\end{align}
It is the $Z_3$ parafermion state $\Phi_{Z_3}$.

\subsection{$n=4$ case}

Apart from the $n=1$ primitive solutions, we find only two new $n=4$
primitive solutions using conditions \eq{hcond1}, \eq{hcond2},
\eq{hcond3}, \eq{alijeven}, \eq{cchsc1}, and \eq{cchsc2}:
\begin{align}
 &n =4: \ \ \ c=1\ \ \ (Z_4\text{ parafermion state})
\nonumber\\
 &\{m;\hsc{1}..\hsc{n-1}\}=\{2;\frac34\ 1\ \frac34 \}
\nonumber\\
 &\{p;M_{1}..M_{n-1}\}=\{2; 0\ 0\ 0\}
\nonumber\\
 &\ \ \ \ \{n_0..n_{m-1}\} = \{4\ 0 \} .
\end{align}
which is the $Z_4$ parafermion state $\Phi_{Z_4}$, and
\begin{align}
 &n =4:
\nonumber\\
 &\{m;\hsc{1}..\hsc{n-1}\}=\{2;\frac14\ 0\ \frac14 \}
\nonumber\\
 &\{p;M_{1}..M_{n-1}\}=\{1; \frac12\ 1\ \frac12\}
\nonumber\\
 &\ \ \ \ \{n_0..n_{m-1}\} = \{1\ 0\ 1\ 0 \} .
\end{align}

We like to point out that a non primitive solution $
\{m;\hsc{1},..,\hsc{n-1}\}= 2\times\{2;\frac14,0,\frac14 \}
=\{4;\frac12,0,\frac12 \} $ is the $Z_2$ parafermion state (the
Pfaffian state). Consistent conditions from a study of useful GJI's
show that it has central charge $c=1/2$ (the same as $Z_2$ Pfaffian
state) and $\mu_{2,a}=1,~~\partial\psi_2=0$, indicating that
$\psi_2=1$ is the identity operator here. In other words, this $Z_4$
simple-current vertex algebra is generated by a $Z_2$ simple
current.

Another non primitive solution $ \{m;\hsc{1},..,\hsc{n-1}\}=
3\times\{2;\frac14,0,\frac14 \} =\{6;\frac34,0,\frac34 \} $ is the
Gaffnian state\cite{Gaffnian}. Gaffnian vertex algebra is a $Z_4$
simple-current vertex algebra with
$\mu_{1,3}=\mu_{1,2}=\mu_{2,3}=-1$ and $\partial\psi_2=0$. In
comparison with $Z_4$ Pfaffian, this $Z_4$ Gaffnian vertex algebra
cannot be generated by any $Z_2$ simple current. This example will
be analyzed in detail in section \ref{EXM}.

\subsection{Including conditions \eq{gVAcon2} and \eq{gVAcon3}}
\label{gVAcon23}

In the above, we only considered the conditions \eq{gVAcon1}.  Those
patterns of zeros that satisfy \eqn{gVAcon1} may not satisfy the
conditions \eq{gVAcon2} and \eq{gVAcon3}, \ie one may not be able to
find $C_{a,b}$ that satisfy eqns. \eq{gVAcon2} and \eq{gVAcon3}.
However, we do not know how to check the conditions \eq{gVAcon2} and
\eq{gVAcon3} systematically.  We have to check them on a case by
case basis.

For the $Z_2$ and $Z_3$ parafermion states, we find that eqns.
\eq{gVAcon2} and \eq{gVAcon3} reduce to trivial identities after
using \eqn{Cstr:normalization}.  So the non-trivial $C_{1,1}$ and
$C_{2,2}$ for the $Z_3$ parafermion vertex algebra cannot be
determined from eqns. \eq{gVAcon2} and \eq{gVAcon3}, which means
that the conditions \eq{gVAcon2} and \eq{gVAcon3} can be satisfied
by any choices of $C_{a,b}$ that are consistent with
\eqn{Cstr:normalization}.

For the state with pattern of zeros
$\{n;m;h^\text{sc}_a\}=\{4;2;\frac14\ 0\ \frac14\}$, we find that by
choosing $(a,b,c)=(1,2,3)$ and $(1,3,3)$ in \eq{gVAcon2}, we can
obtain the following equations
\begin{align}
C_{1,2}C_{3,3}&=C_{2,3}C_{1,1}=-1 ,
\nonumber\\
1&=C_{3,3}C_{1,2}-1  .
\end{align}
Clearly no $\{C_{a,b}\}$ can satisfy the above two equations. Thus
the $n=4$ pattern of zeros $\{m;h^\text{sc}_a\}=\{2;\frac14\ 0\
\frac14\}$ do not correspond to any valid symmetric polynomial. It's
interesting to note that the $n=4$ pattern of zeros
$\{m;h^\text{sc}_a\}=2\times \{2;\frac14\ 0\ \frac14\}=\{4;\frac12\
0\ \frac12\}$ correspond to the $Z_2$ parafermion state and the
$n=4$ pattern of zeros $\{m;h^\text{sc}_a\}=3\times \{2;\frac14\ 0\
\frac14\}=\{6;\frac34\ 0\ \frac34\}$ correspond to the Gaffnian
state, both being valid symmetric polynomials.

For the state with pattern of zeros $\{n;m;h^\text{sc}_a\}=\{4;4;1\
1\ 1\}$, we find that by choosing $(a,b,c)=(1,1,1)$, $(1,1,2)$, and
$(1,2,0)$ in \eq{gVAcon2}, we can obtain the following equations
\begin{align}
C_{1,1}C_{2,1}&=C_{1,1}C_{1,2}+C_{1,1}C_{1,2}
\nonumber\\
C_{1,1}&=C_{1,2}=C_{1,2}
\nonumber\\
C_{1,2}&=C_{1,2}=-C_{2,1}
\end{align}
which can be reduced to $-C_{1,1}^2=2C_{1,1}^2$.  We see that the
only solution is $C_{1,1}=C_{1,2}=C_{2,1}=0$, which is not allowed
by \eqn{Cnonzero}.  Thus the $n=4$ pattern of zeros
$\{m;h^\text{sc}_a\}=\{4;1\ 1\ 1\}$ do not correspond to any valid
symmetric polynomial.

\subsection{Summary}

In \Ref{poz1,poz2}, we have seen that the conditions \eq{hcond1},
\eq{hcond2}, and \eq{hcond3} are not enough since they allow the
following pattern of zeros $\{n;m;h^\text{sc}_a\}=\{2;1;\frac14\}$.
Such a pattern of zeros does not correspond to any valid polynomial.
The conditions \eq{gVAcon1} obtained in this paper rule out the
above invalid solution.  So the conditions \eq{gVAcon1} is more
complete than the conditions \eq{hcond1}, \eq{hcond2}, and
\eq{hcond3}. However, the conditions \eq{gVAcon1} is still
incomplete, since they allow the invalid patterns of zeros such as
$\{n;m;h^\text{sc}_a\} = \{4;2;\frac14\ 0\ \frac14\}$ and $\{4;4;1\
1\ 1\}$.  Both of them can be ruled out by the conditions
\eq{gVAcon2} and \eq{gVAcon3}.

The conditions \eq{gVAcon1}, \eq{gVAcon2}, and \eq{gVAcon3} are
the consistent conditions that we can find from some of GJI, based on the
most general form of OPE (\ref{psiapsibOPE}).
So those conditions are necessary, but may not be sufficient.

The correspondence between the patterns of zeros $\{n;m;\hsc{a}\}$ and
FQH states is not one-to-one. There can be many polynomials that have
the same pattern of zeros.  This is not surprising since the pattern
of zeros only fixes the highest-order zeros in electron wave functions
(symmetric polynomials), while different patterns of lower-order zeros
could lead to different polynomials in principle. In other words, the
leading-order OPE (\ref{psiapsibOPE}) alone might not suffice to
uniquely determine the correlation function of the vertex algebra. The
examples studied in this section support such a belief.  Explicit
calculations for some examples suggest that the pattern of zeros
\emph{together with} the central charge $c$ and simple current
condition would uniquely determine the FQH state. This is a reason why
we introduce $Z_n$ simple-current vertex algebra in the next section.

\section{$Z_n$ simple-current vertex algebra}
\label{GJIVA}

In the last section, we discuss ``legal'' patterns of zeros that
satisfy the consistent conditions \eq{gVAcon1}, \eq{gVAcon2}, and
\eq{gVAcon3} and describe existing FQH states.  If we believe that a
``legal'' pattern of zeros $\{n;m;\hsc{a}\}$, or more precisely the
data $\{n;m;\hsc{a};c\}$, can completely describe a FQH state,
then we should be able to calculate all the topological properties
of the FQH states.  But so far, from the pattern of zeros
$\{n;m;\hsc{a}\}$, we can only calculate the number of different
quasiparticle types, quasiparticle charges, and the fusion algebra
between the quasiparticles.\cite{poz2,poz3} Even with the more
complete data $\{n;m;\hsc{a};c\}$, we still do not know, at
this time, how to calculate the quasiparticle statistics and scaling
dimensions.

One idea to calculate more topological properties from the data
$\{n;m;\hsc{a};c\}$ is to use the data to define and construct
the corresponding $Z_n$ vertex algebra, and then use the $Z_n$
vertex algebra to calculate the quasiparticle scaling dimensions and
the central charge $c$. However, so far we do not know how to use
the data $\{n;m;\hsc{a};c\}$ to completely construct a $Z_n$
vertex algebra in a systematic manner.

Starting from this section, we will concentrate on a subset of
``legal'' patterns of zeros that correspond to a subset of $Z_n$
vertex algebra. Such a subset is called $Z_n$ simple-current vertex
algebras.  The FQH states described by those $Z_n$ simple-current
vertex algebras are called $Z_n$ simple-current states.  We will
show that in many cases the quasiparticle scaling dimensions and the
central charge $c$ can be calculated from the data
$\{n;m;\hsc{a};c\}$ for those $Z_n$ simple-current states.

\subsection{OPE's of $Z_n$ simple-current vertex algebra}

The $Z_n$ simple-current vertex algebra is defined through an
Abelian fusion rule with cyclic $Z_n$ symmetry for primary fields
$\{\psi_a\}$ of Virasoro algebra\cite{ZF0,GepnerQiu}
\begin{align}
\label{psiapsib} \psi_a\psi_b\sim\psi_{a+b},\ \ \ \ \ \ \
\psi_a\equiv(\psi)^a .
\end{align}
Compared to \eqn{psiapsibG}, here we require that $\psi_a$ and
$\psi_b$ fuse into a single primary field of Virasoro algebra
$\psi_{a+b}$.  Such operators are called simple currents.  The $Z_n$
simple-current vertex algebra is defined by the following OPE of
$\psi_a$\cite{ZF0,GepnerQiu}:
\begin{align}
\label{OPE-Zn-parafermion}
&\psi_a(z)\psi_b(w)=C_{a,b}\frac{\psi_{a+b}(w)}{(z-w)^{\alpha_{a,b}}}
+O\Big((z-w)^{1-\alpha_{a,b}}\Big)
\end{align}
\begin{align}
\label{OPE-Zn-parafermionT}
&\psi_a(z)\psi_{-a}(w)=\frac{1+\frac{2\hsc
a}{c}(z-w)^2T(w)}{(z-w)^{2\hsc a}}+O\Big((z-w)^{3-2\hsc a}\Big)
\end{align}
where we define
\begin{align}
\alpha_{a,b}\equiv \hsc{a}+\hsc{b} -\hsc{a+b}
\end{align}
$\psi_{-a}\equiv\psi_{n-a}$ and $\psi_a=\psi_{n+a}$ is understood
due to the $Z_n$ symmetry. In the context the subscript $a$ of $Z_n$
simple currents is always defined as $a\mod n$.

We like to point out here that the form of the OPE
\eq{OPE-Zn-parafermionT} is a special property of the $Z_n$
simple-current vertex algebra. For a more general $Z_n$ vertex algebra
that describes a generic FQH state, the correspond OPE has a more
general form
\begin{align}
\psi_a(z)\psi_{-a}(w) &=\frac{1+\frac{2\hsc
a}{c}(z-w)^2T(w)}{(z-w)^{2\hsc a}}
\\
&\ \ \ \ \ \ +\frac{T'_a}{(z-w)^{2\hsc a-2}} +O\Big((z-w)^{3-2\hsc
a}\Big) \nonumber
\end{align}
where $T'_a$ are dimension-2 primary fields of Virasoro algebra
($\{T_a^\prime,~a=1,\cdots,[\frac{n}{2}]\}$ may not be linearly
independent though). Also, for the $Z_n$ simple-current vertex
algebra, the subleading terms in \eq{OPE-Zn-parafermion} are also
determined. For more details, see appendix \ref{app:subleading}.

$\{C_{a,b}\}$ are the structure constants of this vertex algebra. We
also have conformal symmetry
\begin{align}
T(z)\psi_a(w)=\frac{\hsc
a}{(z-w)^2}\psi_a(w)+\frac{1}{z-w}\partial\psi_a(w)+O(1)
\end{align}
and Virasoro algebra
\begin{align}
T(z)T(w)=\frac{c/2}{(z-w)^4}+\frac{2T(w)}{(z-w)^2}+\frac{\partial
T(w)}{z-w}+O(1)
\end{align}
where $T(z)$ represents the energy-momentum tensor, which has a
scaling dimension of $2$. $c$ stands for the central charge as usual,
which is also a structure constant.

Using the notation of generalized vertex algebra\cite{va0} (see
Appendix \ref{app:muABcon}), we have
\begin{align}
\label{OPE:i,j} &[\psi_i\psi_j]_{\alpha_{i,j}}=C_{i,j}\psi_{i+j},\ \
\ i+j\neq0\mod n
\end{align}
\begin{align}
\label{OPE:i,n-i} &[\psi_i\psi_{-i}]_{\alpha_{i,-i}}=1,\ \ \
[\psi_i\psi_{-i}]_{\alpha_{i,-i}-1}=0,
\nonumber\\
& [\psi_i\psi_{-i}]_{\alpha_{i,-i}-2}=\frac{2\hsc{i}}{c}T
\end{align}
\begin{align}
\label{OPE:T,i}
&[T\psi_i]_2=\hsc{i}\psi_i=[\psi_iT]_2,\ \ \ [T\psi_i]_1=\partial\psi_i\\
\nonumber&[\psi_iT]_1=(\hsc{i}-1)\partial\psi_i
\end{align}
\begin{align}
\label{OPE:T,T} &[TT]_4=\frac{c}{2},\ \ \ [TT]_3=0,\ \ \ [TT]_2=2T,\
\ \ [TT]_1=\partial T
\end{align}
with $\alpha_{T,\psi_i}=2,~\alpha_{T,T}=4$. We call it a
\emph{special $Z_n$ simple-current vertex algebra} if it satisfies
OPE's \eq{OPE:i,j}-\eq{OPE:T,T}. For example, the $Z_n$ parafermion
states\cite{RR99} correspond to a series of special $Z_n$
simple-current vertex algebras.

The commutation factor $\mu_{AB}$ equals unity if either $A$ or $B$ is
the energy-momentum tensor $T$:
$\mu_{T,\psi_i}=\mu_{\psi_i,T}=\mu_{T,T}=1$. Similarly we have
$\mu_{A,1}=\mu_{1,A}=1$ for the identity operator $1$ and any operator
$A$. However, $\mu_{i,j}\equiv\mu_{\psi_i,\psi_j}$ given in \eqn{muij}
can be $\pm1$ in general.  In deriving OPE \eq{OPE:i,n-i} we have
assumed that $\mu_{i,-i}=1,~~\forall i$, which is not necessary.  For
example, the $Z_4$ Gaffnian does not satisfy $\mu_{i,-i}=1,~~\forall i$.
So, we will adopt the more general OPE
\eq{OPE:i,n-i:i<=n/2} and \eq{OPE:i,n-i:i>n/2} instead of
\eqn{OPE:i,n-i} to include examples like Gaffnian which do give a
FQH wave function. OPE (\ref{OPE:i,n-i}) is for a special $Z_n$
simple-current vertex algebra that
satisfies $\mu_{i,-i}=1,~~\forall i$. For a more general $Z_n$
simple-current vertex algebra, they become
\begin{align}\label{OPE:i,n-i:general}
 &[\psi_i\psi_{-i}]_{\alpha_{i,-i}}=C_{i,-i},\ \ \
[\psi_i\psi_{-i}]_{\alpha_{i,-i}-1}=0,
\nonumber\\
& [\psi_i\psi_{-i}]_{\alpha_{i,-i}-2}=\frac{2C_{i,-i}\hsc{i}}{c}T\\
&\nonumber C_{i,-i}=\left\{\begin{aligned}1,~~~i\leq n/2\mod
n\\\mu_{i,-i},~~~i>n/2\mod n\end{aligned}\right.
\end{align}
so that we always have $C_{a,b}=\mu_{a,b}C_{b,a}$ for any subscripts
$a$ and $b$ in such an associative vertex algebra.

The OPE's \eq{OPE:i,j}, \eq{OPE:i,n-i:general}, \eq{OPE:T,i},
\eq{OPE:T,T}, \eq{OPE:psi,sigma} and \eq{OPE:T,sigma} define the
\emph{generalized $Z_n$ simple-current vertex algebra}, or simply
$Z_n$ simple-current vertex algebra.  The Gaffnian state corresponds
to a generalized $Z_4$ simple-current vertex algebra with
$\mu_{a,-a}\neq 1$. When $\mu_{a,-a}=1$, we have a special $Z_n$
simple-current vertex algebra.

What kind of pattern of zeros $\{n;m;\hsc{a}\}$, or more precisely
what kind of data $\{n;m;\hsc{a};c,C_{ab}\}$, can produce a $Z_n$
simple-current vertex algebra? Since the $Z_n$ simple-current vertex
algebras are special cases of  $Z_n$ vertex algebras, the data
$\{n;m;\hsc{a};c,C_{ab}\}$ must satisfy the conditions \eq{gVAcon1},
\eq{gVAcon2}, and \eq{gVAcon3}.  However, the data
$\{n;m;\hsc{a};c,C_{ab}\}$ for $Z_n$ simple-current vertex algebras
should satisfy more conditions.  Those conditions can be obtained
from the GJI of $Z_n$ simple-current vertex algebras. In Appendix
\ref{nconGJI}, we derived all those extra consistent conditions for
a generic $Z_n$ vertex algebra, from the useful GJI's based on OPE
(\ref{psiapsibOPE}). Now based on OPE's summarized in this section,
we can similarly derive a set of extra consistent conditions for a
$Z_n$ simple-current vertex algebra. These conditions are summarized
in section \ref{concon}. For those valid data that satisfy all the
consistent conditions, the full properties of simple-current vertex
algebra can be obtained. This in turn allows us to calculate the
physical topological properties of the FQH states associated with
those valid patterns of zeros.

We like to point out that many examples of $Z_n$ simple-current vertex algebra
have been studied in detail.  They include the simplest $Z_n$ simple-current
vertex algebra -- the $Z_n$ parafermion algebra.\cite{ZF0,ZF1,GepnerQiu} More
general exmaples that have been studied are the higher generations of $Z_n$
parafermion algebra\cite{DJS0359,DJS0377,DJS0464,DJS0486,DS0589} and graded
parafermion algebra.\cite{N0959,JM0233,JM0524} In those exapmles, the $Z_n$
simple-current algebras are studied by embedding the algebras into some known
CFT, such as coset models of Kac-Moody current algebras and/or Coulomb gas
models.  However, in this paper, we will not assume such kind of embeding.  We
will try to calculate the properties of $Z_n$ simple-current vertex algebra
directly from the data $\{n;m;\hsc{a},c,...\}$ without assuming any embedding.

\subsection{Consistent conditions from useful GJI's}
\label{concon}

In Appendix \ref{app:list of useful GJIs:example}, we show how to
obtain the consistent conditions on the data
$\{n;m;\hsc{a};c,C_{ab}\}$ characterizing a generic $Z_n$ vertex
algebra from a set of useful GJI's as described in Appendix
\ref{app:GJI}, requiring that OPE (\ref{OPE:i,j:generic}) is obeyed.
Here for a $Z_n$ simple-current vertex algebra, requiring that OPE's
\eq{OPE:i,j}, \eq{OPE:i,n-i:general}, \eq{OPE:T,i} and \eq{OPE:T,T} 
\eq{OPE:psi,sigma} and \eq{OPE:T,sigma} are obeyed, we can derive a
larger set of consistent conditions on the data
$\{n;m;\hsc{a};c,C_{ab}\}$.  For the examples studied in this paper,
we find that $C_{ab}$ can be uniquely determined from
$\{n;m;\hsc{a};c\}$ using those consistent conditions.  Thus, for
those states, $C_{ab}$ are not independent and can be dropped.

Since $\Delta_3(A,B,C)=\Delta_3(A,C,B)=\Delta_3(C,A,B)$ for an
associative vertex algebra, we can combine the consistent conditions
obtained from GJI's of all possible 6 permutations of 3 operators
$(A,B,C)$ together. In this section we summarize the consistent
conditions obtained from useful GJI's (just like in Appendix
\ref{app:list of useful GJIs:example}) and list them in a compact
manner. These extra consistent conditions, together with conditions
\eq{gVAcon1} should form a complete set of consistent conditions,
which allows us to obtain a valid pattern of zeros and construct the
associated simple-current vertex algebra and FQH wave function.

\subsubsection{$\{A,B,C\}=\{\psi_a,\psi_b,\psi_c\}$,
$a+b,b+c,a+c\neq0\mod n$} \label{consistency:a,b,c}

For $\Delta_3(a,b,c)=0$, we have the following consistent
conditions:
\begin{align}
C_{a,b}C_{a+b,c}=C_{b,c}C_{a,b+c}=\mu_{a,b}C_{a,c}C_{b,a+c}
\end{align}
Notice that the consistent conditions obtained from useful GJI's of
$(\psi_a,\psi_b,\psi_c)$ and of $(\psi_b,\psi_a,\psi_c)$ only differ
by a factor of $\mu_{a,b}$ since $C_{a,b}=\mu_{a,b}C_{b,a}$. Thus
they are not independent conditions. Similarly it's easy to show
that other permutations yield consistent conditions linearly
dependent with the above condition, using the fact that
$\mu_{a,b}\mu_{a,c}=\mu_{a,b+c}$ here since $\Delta_3(a,b,c)=0$.

For $\Delta_3(a,b,c)=1$, we have the following consistent
conditions:
\begin{align}
C_{a,b}C_{a+b,c}=C_{b,c}C_{a,b+c}+\mu_{a,b}C_{a,c}C_{b,a+c}
\end{align}

For $\Delta_3(a,b,c)\geq2$ there are no extra consistent conditions.

\subsubsection{$\{A,B,C\}=\{\psi_a,\psi_{b},\psi_{-b}\}$, $a\pm b\neq0\mod
n$} \label{consistency:a,b,-b}

For $\Delta_3(a,b,-b)=0$ we have the following independent
consistent conditions
\begin{align}
\label{pozCabb0} \hsc{a}\hsc{b}=\al_{a,\pm
b}=0,~~~\hsc{b}\partial\psi_a\equiv0
\nonumber \\
C_{a,b}C_{a+b,-b}=\mu_{a,b}C_{a,-b}C_{b,a-b}=C_{b,-b}
\end{align}
since we know \eqn{cf1} and $\mu_{a,0}=1$.

For $\Delta_3(a,b,-b)=1$ the independent consistent conditions are
\begin{align}
\label{pozCabb1} c&=\frac{4\hsc{a}\hsc{b}}{\al_{a,b}(1-\al_{a,b})},
\nonumber \\
C_{a,b}C_{a+b.-b}&=(1-\al_{a,b})C_{b,-b}
\nonumber \\
\mu_{a,b}C_{a,-b}C_{b,a-b}&=-\al_{a,b}C_{b,-b}
\end{align}

For $\Delta_3(a,b,-b)=2$ the independent consistent conditions are
\begin{align}
\mu_{a,b}C_{a,-b}C_{b,a-b}&=[\frac{\al_{a,b}(\al_{a,b}-1)}{2}+\frac{2\hsc{a}\hsc{b}}{c}]C_{b,-b}
\nonumber \\
C_{a,b}C_{a+b,-b}&=[\frac{(\al_{a,b}-1)(\al_{a,b}-2)}{2}+\frac{2\hsc{a}\hsc{b}}{c}]C_{b,-b}
\end{align}

For $\Delta_3(a,b,-b)=3$ the independent consistent conditions are
\begin{align}
&C_{a,b}C_{a+b,-b}-\mu_{a,b}C_{a,-b}C_{b,a-b}
\nonumber \\
&=[\frac{(\al_{a,b}-1)(\al_{a,b}-2)}{2}+\frac{2\hsc{a}\hsc{b}}{c}]C_{b,-b}
\end{align}

For $\Delta_3(a,b,-b)\geq4$ there are no extra consistent conditions
from useful GJI's.

\subsubsection{$\{A,B,C\}=\{\psi_a,\psi_{a},\psi_{-a}\}$, $a\neq n/2\mod n$}\label{consistency:a,a,-a}

For $\Delta_3(a,a,-a)=\al_{a,a}+2\hsc{a}=0$ the consistent
conditions are summarized as:
\begin{align}
\label{pozCaaa0} \hsc{a}=\al_{a,a}=0,~~~\partial\psi_a\equiv0
\nonumber \\
C_{a,a}C_{2a,-a}=C_{a,-a}=C_{-a,a}=\mu_{a,-a}=1
\end{align}

For $\Delta_3(a,a,-a)=\al_{a,a}+2\hsc{a}=1$ the corresponding
consistent conditions are:
\begin{align}
\label{pozCaaa1} \al_{a,a}=-1,~~\hsc{a}=1,~~c=-2,~~\mu_{a,-a}=-1
\nonumber \\
C_{a,a}C_{2a,-a}=2C_{a,-a},~~~C_{a,-a}=-C_{-a,a}
\end{align}

For $\Delta_3{(a,a,-a)}=2$ the independent consistent conditions are
\begin{align}
\label{pozCaaa2} & c=\frac{2\hsc{a}}{3-2\hsc{a}},
\nonumber \\
& C_{a,a}C_{2a,-a}=2\hsc{a},~~C_{a,-a}=C_{-a,a}=\mu_{a,-a}=1
\end{align}
since we have $C_{2a,-a}=C_{-a,2a}$ here.

For $\Delta_3(a,a,-a)=\al_{a,a}+2\hsc{a}=3$ the extra consistent
conditions are
\begin{align}
\label{pozCaaa3}
& \mu_{a,-a}=-1 \nonumber \\
&
c=\frac{-2(\hsc{a})^2}{(2\hsc{a}-3)(\hsc{a}-2)},~~~C_{-a,a}=-C_{a,-a}
\nonumber \\
& C_{a,a}C_{2a,-a}=4(\hsc{a}-1)C_{a,-a}
\end{align}

For $\Delta_3(a,a,-a)=4$ the independent consistent conditions are
\begin{align}
\label{pozCaaa4} &C_{a,-a}=C_{-a,a}=\mu_{a,-a}=1
\nonumber \\
&C_{a,a}C_{2a,-a}=\hsc{a}(2\hsc{a}-3)+\frac{2(\hsc{a})^2}{c}
\end{align}

For $\Delta_3(a,a,-a)=\al_{a,a}+2\hsc{a}=5$ there is only 1 useful
GJI and the consistent conditions is:
\begin{align}
\label{pozCaaa5} &\mu_{a,-a}=-1,~~C_{a,-a}=-C_{-a,a}
\nonumber \\
&C_{a,a}C_{2a,-a}=2[(\hsc{a}-2)(2\hsc{a}-3)+\frac{2(\hsc{a})^2}{c}]C_{a,-a}
\nonumber
\end{align}

For $\Delta_3(a,a,-a)=\al_{a,a}+2\hsc{a}\geq6$ there are no useful
GJI's and no extra consistent conditions.

\subsubsection{$\{A,B,C\}=\{\psi_{n/2},\psi_{n/2},\psi_{n/2}\}$, $n=$~even}\label{consistency:n/2,n/2,n/2}

{}Just like shown in Appendix \ref{app:list of useful GJIs:example},
we require that
\begin{equation}
\label{pozCnnn} \Delta_3(n/2,n/2,n/2)\neq1,3,5.
\end{equation}
otherwise the useful GJI's would yield a contradiction
$\psi_{n/2}=0$.

For $\Delta_3(n/2,n/2,n/2)=0$ the extra consistent conditions are
\begin{align}
\label{pozCnnn0} \hsc{n/2}=0,~~~\partial\psi_{n/2}\equiv0
\end{align}

For $\Delta_3(n/2,n/2,n/2)=2$ the extra consistent conditions are
\begin{align}
\label{pozCnnn2} c=\hsc{n/2}=1/2
\end{align}

For $\Delta_3(n/2,n/2,n/2)=4$ the extra consistent conditions are
\begin{align}
\label{pozCnnn4} c=\hsc{n/2}=1
\end{align}

For $\Delta_3(n/2,n/2,n/2)=6$ there are no extra consistent conditions. 

For $\Delta_3(n/2,n/2,n/2)=7$ the extra consistent conditions are
\begin{align}
\label{pozCnnn7} c=49,~~\hsc{n/2}=7/4
\end{align}

For $\Delta_3(n/2,n/2,n/2)\geq8$ there are no extra consistent
conditions.

\section{Representing quasiparticles in $Z_n$ simple-current vertex algebra}
\label{qpva}

Since the $Z_n$ simple-current vertex algebras completely determine
the FQH states and their topological orders, we should be able to
calculate all the topological properties from the vertex algebras.
In this section, we will discuss how to represent quasiparticles and
how to calculate quasiparticle properties from the vertex algebras.

\subsection{The pattern of zeros for quasiparticles and
its consistent conditions}

First, let us review the pattern of zeros description for
quasiparticles in FQH states.\cite{poz2,poz3}

\subsubsection{Definition and consistent conditions}

The pattern of zeros for the ground state wave function can be easily
generalized to describe the wave functions with quasiparticle
excitations. If a symmetric polynomial $\Phi(\{z_i\})$ has a
quasiparticle at $z=0$, $\Phi(\{z_i\})$ will have a different
pattern of zeros $\{S_{\ga;a}\}$ as
$z_1=\la\xi_1,\cdots,z_a=\la\xi_a$ approach $0$:
\begin{align}
\nonumber\lim_{\la\rightarrow0^+}\Phi(\{z_i\})=\la^{S_{\ga;a}}P_\ga(\xi_1,\cdots,\xi_a;z_{a+1},\cdots)
\nonumber \\
+O(\la^{S_{\ga;a}+1}).
\end{align}
Thus we can use the sequence of non-negative integers $\{S_{\ga;a}\}$
to quantitatively characterize quasiparticles.

It was shown\cite{poz2,poz3} that there are similar consistent
conditions on the quasiparticle pattern of zeros $\{S_{\ga;a}\}$:\\
First concave condition
\begin{align}\label{concave1qp}
D_{\ga+a,b}\equiv S_{\ga;a+b}-S_{\ga;a}-S_b\geq0
\end{align}
Second concave condition
\begin{align}\label{concave2qp}
\nonumber\Delta_3(\ga+a;b,c)\equiv S_{\ga;a+b+c}+S_{\ga;a}+S_b+S_c
\nonumber \\
-S_{\ga;a+b}-S_{\ga;a+c}-S_{b+c}\geq0
\end{align}
$n$-cluster condition
\begin{align}\label{ncluster qp}
S_{\ga;a+kn}=S_{\ga;a}+k(S_{\ga;n}+ma)+mn\frac{k(k-1)}{2}
\end{align}

$\{S_{\ga;a}\}$ is a quantitative way to label all types of the
quasiparticles in the FQH state described by $\{S_{a}\}$.  The
question is that is $\{S_{\ga;a}\}$ an one-to-one label of the
quasiparticles? Can two different quasiparticles share the same
pattern of zeros? The answer is yes and no.  For certain FQH states
(such as all the generalized and composite parafermion FQH states),
$\{S_{\ga;a}\}$ is an one-to-one label of all the quasiparticles.
While for other FQH states, such as $Z_2|Z_2$ and $Z_3|Z_3$ in
section \ref{EXM}, $\{S_{\ga;a}\}$ is not an one-to-one label and
two different quasiparticles can have the same pattern of zeros.

If we assume $\{S_{\ga;a}\}$ to be an one-to-one label of all  the
quasiparticles, then by solving the above consistent conditions, we
can obtain the number of quasiparticle types, which happens to equal
the ground state degeneracy of the FQH state on a torus.  We can also
calculate other physical properties of quasiparticles from
$\{S_{\ga;a}\}$. For example, the quasiparticle charge $Q_\ga$ can be
obtained from the pattern of zeros as\cite{poz2}
\begin{align}\label{qp charge}
Q_\ga=\frac{S_{\ga;n}-S_n}{m}
\end{align}
(The above formula is valid even when $\{S_{\ga;a}\}$ is not
an one-to-one label.)

\subsubsection{Label quasiparticle pattern of zeros by $\{\ksc{a};Q_\ga\}$}

Another way to label the quasiparticle pattern of zeros can be
obtained by introducing the $\{\ksc{a}\}$ vector (which is denoted
by $\t l^\text{sc}_{\ga;a}$ in \Ref{poz3}):
\begin{align}\label{tlsc-def}
\ksc{a}\equiv S_{\ga;a}-S_{\ga;a-1}+\hsc1-\frac{m(Q_\ga+a-1)}{n}
\end{align}
Conversely we have
\begin{align}\label{tlsc-S}
S_{\ga;a}=\sum_{i=1}^a\ksc{i}+a(\frac{mQ_\ga}{n}-\hsc1)+\frac{ma(a-1)}{2n}
\end{align}
The $n$-cluster condition \eq{ncluster qp} of $\{S_{\ga;a}\}$
sequence results in the periodic property of $\{\ksc{a}\}$
\begin{align}\label{tlsc-periodic}
\ksc{n+a}=\ksc{a}
\end{align}
Therefore we can use the set of data
$\{\ksc{1},...,\ksc{n};Q_\ga\}$.  to describe quasiparticles.

Let $a=n$ in \eqn{tlsc-S} and use \eqn{qp charge} we can see that
\begin{align}\label{tlsc-sum=0}
\sum_{i=1}^n\ksc{i}=0
\end{align}
The two concave conditions \eq{concave1qp} and \eq{concave2qp} for
this set of data now becomes
\begin{align}\label{concave1qp tlsc}
D_{\ga+a,b}=\sum_{i=1}^b\ksc{a+i}-\hsc{b}+b\frac{mQ_\ga}{n}+\frac{mab}{n}\in\mathds{N}
\end{align}
\begin{align}\label{concave2qp tlsc}
\nonumber\Delta_3(\ga+a,b,c)=\sum_{i=1}^c(\ksc{a+b+i}-\ksc{a+i})
\nonumber \\
+\hsc{b}+\hsc{c}-\hsc{b+c}\in\mathds{N}
\end{align}
A set of $\{\ksc{1},...,\ksc{n};Q_\ga\}$ satisfying the above two
conditions and $S_{\ga;a}\geq 0$ can generate a valid quasiparticle
pattern of zeros, which corresponds to quasiparticle above a ground
state with the pattern of zeros $\{\hsc{a}\}$.

We note that $\ga+1$ corresponds to a bound state between a
$\ga$-quasiparticle and a hole (the absence of an electron).  The
$(\ga+1)$-quasiparticle is labeled by
\begin{equation*}
\{k_{{\ga+1};1}^\text{sc},\cdots,
k_{{\ga+1};n}^\text{sc};Q_{\ga+1}\} = \{k_{\ga;2}^\text{sc},\cdots,
k_{\ga;n}^\text{sc}, k_{\ga;1}^\text{sc};Q_\ga+1\}.
\end{equation*}
Since two quasiparticles that differ by an electron are regarded as
equivalent, we can use the above equivalence relation to pick an
equivalent label that has the minimal charge and satisfies
$S_{\ga;a}\geq 0$.  For each equivalence class, there exists only
one such label. In this paper, we will use such a label to label
inequivalent quasiparticles.

\subsection{Quasiparticle wave functions and quasiparticle operators}

Just like the ground state wave function \eq{PhiV}, the wave function
with a quasiparticle can also be written as a R-ordered correlation
function between electron operators and quasiparticle operators in the
vertex algebra
\begin{align}
\label{Phiga}
\Phi_\ga(w;\{z_i\})=\lim_{z_\infty\rightarrow\infty}z^{2h_N}_\infty
\langle V(z_\infty)\Big[\prod_{i}V_e(z_i)\Big]V_\ga(w)\rangle ,
\end{align}
where $w$ is the location of the quasiparticle and $V_\ga(w)$ is the
quasiparticle operator.  By definition, a quasiparticle operator can
be any operator that is mutually local respect to the electron
operators $V_e(z)$.

In our simple-current$\times U(1)$ vertex algebra, the quasiparticle
operator $V_\ga$ has the following form
\begin{align}
V_\ga(z)=\si_\ga(z):\e^{\imth\phi(z)Q_\ga/\sqrt\nu}:
\end{align}
where $\si_\ga$ is a disorder operator\cite{ZF1} that generates a
representation of the simple current part of the vertex algebra.
When $a$ electrons and one quasiparticle are fused together we have
\begin{align}
& V_{\ga+a}(z)\propto V_aV_\ga
=\si_{\ga+a}(z):\e^{\imth\phi(z)(Q_\ga+a)/\sqrt\nu}:
\nonumber\\
& \si_{\ga+a}\propto \psi_a\si .
\end{align}
The OPE between the quasiparticle operator and the electron operator
can be written as
\begin{align}
V_e(z)V_{\ga+a}(w)\propto (z-w)^{l_{\ga;a+1}}V_{\ga+a+1}(w)+\cdots
\end{align}
The mutual locality between the quasiparticle operator and the
electron operator requires $l_{\ga;a}$ to be integers.  In order for
the quasiparticle wave function $ \Phi_\ga(w;\{z_i\})$ to contain no
poles, we also require that $l_{\ga;a} \geq 0$.

In fact, the sequence $l_{\ga;a}$, $a=1,2,...$, provides a
quantitative way to label the quasiparticles (and quasiparticle
operators).  We have introduced another quantitative label of the
quasiparticles in terms of $S_{\ga;a}$, $a=1,2,...$ The two labeling
schemes are related by\cite{poz2,poz3}
\begin{align}
\label{lgaSga}
S_{\ga;a}=\sum_{i=1}^al_{\ga;i},\ \ \ \ \
 l_{\ga;a}=S_{\ga;a}-S_{\ga;a-1}.
\end{align}
We can also convert the orbital sequence $l_{\ga;a}$ into an
occupation sequence $n_{\ga;l}$.  If we view $l_{\ga;a}$ as the
index of the orbital occupied by the $a$-th particle, then
$n_{\ga;l}$ is simply the number of particles occupying the
$l^\text{th}$ orbital.

Let us denote the scaling dimension of disorder operators
$\si_{\ga}$ as $\hsc{\ga}$.  Can we calculate those scaling
dimensions from the data $l_{\ga;a}$ that characterize the
quasiparticle?  From the OPE of the quasiparticle operators, we find
the following relations
\begin{align}
l_{\ga;a+1}=\hsc{\ga+a+1}-\hsc{\ga+a}-\hsc{1}+\frac{m(Q_\ga+a)}{n}
\end{align}
and
\begin{align}
S_{\ga;a}&=\sum_{i=1}^al_{\ga;i}
\\
&=\hsc{\ga+a}-\hsc\ga+a(\frac{mQ_\ga}{n}-\hsc1)+\frac{ma(a-1)}{2n} ,
\nonumber
\end{align}

Making use of \eqn{tlsc-def} we immediately obtain the relations
between $\{\ksc{a}\}$ and $\{\hsc{\ga+b}\}$
\begin{align}
\ksc{a}=\hsc{\ga+a}-\hsc{\ga+a-1}
\end{align}
which implies that
\begin{align}
\label{hgaahga} \hsc{\ga+a}=\hsc\ga+\sum_{i=1}^a\ksc{i} .
\end{align}
Moreover, \eqn{tlsc-periodic} and \eqn{tlsc-sum=0} lead to the
periodic condition on $\hsc{\ga+a}$
\begin{align}
\hsc{\ga+a+n}=\hsc{\ga+a}
\end{align}
which is implied by the fusion rule $\psi_n\si=\si$ since
$\psi_n=1$.

We know that we can use $\{\ksc{1},...,\ksc{n};Q_\ga\}$ that
satisfies the two concave conditions \eq{concave1qp tlsc} and
\eq{concave2qp tlsc} to describe (or label) a quasiparticle operator
$V_\ga$ (or a quasiparticle $\ga$). The above result \eq{hgaahga}
only allows us to determine the scaling dimensions $\{\hsc{\ga+a}\}$
of the associated disorder operators up to a
constant.
That is if we know the scaling dimension $\hsc{\ga}$ of a disorder
operator $\si_\ga$, then the scaling dimensions of a family of
disorder operators $\si_{\ga+a}$ can be determined.  However, the
scaling dimension $\hsc{\ga}$ cannot be determined from the
considerations discussed here.  Can we do a better job by fully
using the structure of the vertex algebra? In section
\ref{ConsistentConditionsQP} and \ref{EXM} we will show how to
extract the scaling dimension $\hsc{\ga+a}$ from useful GJI's
defined in Appendix \ref{app:GJI}.

\subsection{A more complete characterization of quasiparticles}

Through a study of $Z_n$ vertex algebra, we have realized that the
pattern-of-zero data $\{n;m;\hsc{a}\}$ does not fully describe a
symmetric polynomial (\ie a quantum Hall wave function).  We need to
at least expand $\{n;m;\hsc{a}\}$ to $\{n;m;\hsc{a};c\}$ to
characterize a quantum Hall wave function.  Similarly, the data
$\{\ksc{a};Q_\ga\}$ does not fully describe a quasiparticle either,
\ie some times, different quasiparticles can have the same pattern of
zeros $\{\ksc{a};Q_\ga\}$.

To see how to extend $\{\ksc{a};Q_\ga\}$, we note that a generic OPE
between $\si_{\ga+b}$ and $\psi_a$ has a form
\begin{align}
\psi_a(z)\si_{\ga+b}(w) = \frac{C_{a,\ga+b}}{
(z-w)^{\al_{a,\ga+b}}}\si_{\ga+a+b}(w)+\cdots
\end{align}
where
\begin{equation}
 \al_{a,\ga+b}=\hsc{a}+\hsc{\ga+b}-\hsc{\ga+a+b}.
\end{equation}
We also need to introduce the commutation factor $\mu_{a,\ga+b}$:
\begin{align}
\label{muagab} &\ \ \
(z-w)^{\alpha_{a,\ga+b}}\psi_a(z)\si_{\ga+b}(w)
\nonumber\\
&= \mu_{a,\ga+b}(w-z)^{\alpha_{a,\ga+b}}\si_{\ga+b}(w)\psi_a(z),
\end{align}
to describe the commutation relation between $\si_{\ga+b}$ and
$\psi_a$.  We see that in vertex algebra, we need additional data,
$C_{a,\ga+b}$, $C_{\ga+b,a}$, $\mu_{a,\ga+b}$, and $\mu_{\ga+b,a}$, to
describe the quasiparticle $\ga$.  (In appendix \ref{mugakga}, we give
a discussion about the relation between the quasiparticle commutation
factor $\mu_{a,\ga}$ and quasiparticle pattern of zeros
$\{\ksc{a};Q_\ga\}$.)

However, if we put the quasiparticle at $w=0$ (see \eqn{Phiga}), then
we do not need to use commutation factor $\mu_{a,\ga}$ when we
calculate the R-ordered correlation function \eq{Phiga}.  Thus, the
electron wave function with a quasiparticle do not depend on the
commutation factor $\mu_{a,\ga}$.  Similarly, the R-ordered
correlation function only depend on $C_{a,\ga+b}$.  Therefore, we only
need to add  $C_{a,\ga+b}$ to describe the quasiparticle $\ga$ more
completely.

Therefore, within
the simple-current vertex algebra, we can use the following more
complete data
\begin{equation}
\label{tlQkC} \{\ksc{a};Q_\ga;C_{a,\ga+b}\}
\end{equation}
to describe a quasiparticle.
By considering the full structure of the vertex algebra (see next
section \ref{ConsistentConditionsQP}), we can obtain many
self-consistent conditions on the data
$\{\ksc{a};Q_\ga;C_{a,\ga+b}\}$.  In particular, we can calculate
the scaling dimension $\hsc{\ga}$ of $\si_\ga$ from the data
$\{\ksc{a};Q_\ga;C_{a,\ga+b}\}$.

Once we find the scaling dimension $\hsc{\ga}$ of a disorder
operator $\si_\ga$, the scaling dimension $h_{\ga}$ of the
associated quasiparticle operator $V_\ga$ can be determined from
\begin{equation}
\label{hgahschga} h_\ga= \hsc{\ga}+h^\text{ga}_\ga =  \hsc{\ga} +
\frac{m Q_\ga^2}{2n} .
\end{equation}
where $h^\text{ga}_\ga$ is the scaling dimension of the $U(1)$ part
$\e^{\imth \phi Q_\ga/\sqrt\nu}$ of the quasiparticle operator.
$h_\ga$ is the intrinsic spin of the quasiparticle which is closely
related to the statistics of the quasiparticle.  (Note that in 2+1D
the intrinsic spin is not quantized as half integer.)

\subsection{Consistent conditions for quasiparticles from useful GJI's
}\label{ConsistentConditionsQP}

\subsubsection{Complete vertex algebra with quasiparticle operators}

To find more consistent conditions on the quasiparticle data
$\{\ksc{a};Q_\ga;C_{a,\ga+b}\}$, we need to write down the
complete OPE between the disorder operators $\si_{\ga+b}$ and the
simple currents
\begin{align}
& \psi_a(z)\si_{\ga+b}(w)=C_{a,\ga+b}
\frac{\si_{\ga+a+b}(w)}{(z-w)^{\alpha_{a,\ga+b}}}
+O((z-w)^{1-\alpha_{a,\ga+b}})
\nonumber \\
& T(z)\si_{\ga+a}(w)=\frac{\hsc{\ga+a}}{(z-w)^2}
\si_{\ga+a}(w)+\frac{1}{z-w}\partial\si_{\ga+a}(w)+O(1)
\end{align}
where we define
\begin{align}
\label{alpha a,gamma+b}
\alpha_{a,\ga+b}\equiv\hsc{a}-(\hsc{\ga+a+b}-\hsc{\ga+b}) =\hsc
a-\sum_{i=1}^a\ksc{b+i} ,
\end{align}
In other words we have
\begin{align}
&\label{OPE:psi,sigma}[\psi_a\si_{\ga+b}]_{\alpha_{a,\ga+b}}=C_{a,\ga+b}
\si_{\ga+a+b}\\
&\nonumber[\si_{\ga+b}\psi_a]_{\al_{a,\ga+b}}=\mu_{\ga+b,a}C_{a,\ga+b}\si_{\ga+a+b}
\end{align}
\begin{align}
&\label{OPE:T,sigma}[T\si_{\ga+a}]_2=\hsc{\ga+a}\si_{\ga+a}=[\si_{\ga+a}T]_2,
\\
\nonumber&[T\si_{\ga+a}]_1=\partial\si_{\ga+a},~~~[\si_{\ga+a}T]_1=(\hsc{\ga+a}-1)\partial\si_{\ga+a}
\end{align}
with $\alpha_{T,\si_{\ga+a}}=2$.  We set $C_{a,\ga}=1$ as the
definition of disorder operators $\si_{\ga+a},~a\neq0\mod n$, which
possess $Z_n$ symmetry.  Note that \eqn{OPE:T,sigma} can be used in
GJI's to determine the scaling dimension $\hsc{\ga+a}$ of disorder
operators, as will be shown in examples.

The consistent conditions on the quasiparticle data,
$\{\ksc{a};Q_\ga;C_{a,\ga+b}\}$ or
$\{\hsc{\ga+a};Q_\ga;C_{a,\ga+b}\}$, can also be obtained from useful
GJI's with respect to the OPE's \eq{OPE:psi,sigma} and
\eq{OPE:T,sigma}, just as we did in Appendix \ref{app:list of useful
GJIs:example} for simple currents of a generic $Z_n$ vertex algebra.
In the following, we'll list the obtained consistent conditions from
GJI's.

\subsubsection{Consistent conditions: $\{A,B,C\}=\{\psi_{a},\psi_{b},\si_{\ga+c}\}$,
$a+b\neq0\mod n$} \label{consistency:a,b,ga+c}

Apply the GJI to the quasiparticle algebra \eq{OPE:psi,sigma} and
\eq{OPE:T,sigma}, we can obtain many new consistent conditions. 

For $\Delta_3(a,b,\ga+c)=0$ the independent consistent conditions
are
\begin{align}
\nonumber
\mu_{a,b}C_{a,\ga+c}C_{b,\ga+a+c}&=C_{a,b}C_{a+b,\ga+c}\\
&=C_{b,\ga+c}C_{a,\ga+b+c}
\end{align}

For $\Delta_3(a,b,\ga+c)=1$ the only independent consistent
condition is:
\begin{align}
&\ \ \ \mu_{a,b}C_{a,\ga+c}C_{b,\ga+a+c}\\
\nonumber &=C_{a,b}C_{a+b,\ga+c}-C_{b,\ga+c}C_{a,\ga+b+c}
\end{align}

For $\Delta_3(a,b,\ga+c)\geq2$ there are no extra consistent
conditions.

\subsubsection{Consistent conditions: $\{A,B,C\}=\{\psi_{a},\psi_{-a},\si_{\ga+c}\}$}
\label{consistency:a,-a,ga+b}

For $\Delta_3(a,-a,\ga+b)=0$ the independent consistent conditions
are
\begin{align}
\hsc{a}\partial\si_{\ga+b}& \equiv0,~~~\al_{\pm
a,\ga+b}=\hsc{a}\hsc{\ga+b}=0,
\nonumber \\
C_{a,\ga+b}C_{-a,\ga+a+b} &=C_{-a,\ga+b}C_{a,\ga+b-a}=C_{a,-a}.
\end{align}
since $\mu_{\ga+b,0}=1$.

For $\Delta_3(a,-a,\ga+b)=1$ the independent consistent conditions
are
\begin{align}
\frac{\hsc{a}}{c}\hsc{\ga+b}&=\frac{\al_{a,\ga+b}(1-\al_{a,\ga+b})}{4},
\nonumber \\
C_{-a,\ga+b}C_{a,\ga+b-a}&=C_{a,-a}\al_{a,\ga+b},
\nonumber \\
C_{a,\ga+b}C_{-a,\ga+a+b}&=C_{-a,a}\al_{-a,\ga+b}.
\end{align}
Notice here the quasiparticle scaling dimension $\hsc\ga$ is
determined through useful GJI's.

For $\Delta_3(a,-a,\ga+b)=2$ the independent consistent conditions
are
\begin{align}
&C_{-a,\ga+b}C_{a,\ga+b-a}=
\nonumber \\
&\nonumber[\frac{2\hsc{a}\hsc{\ga+b}}{c}+\frac{\al_{a,\ga+b}(\al_{a,\ga+b}-1)}{2}]C_{a,-a},
\nonumber \\
&C_{a,\ga+b}C_{-a,\ga+a+b}=\\
&\nonumber[\frac{2\hsc{a}\hsc{\ga+b}}{c}+\frac{(\al_{a,\ga+b}-2)(\al_{a,\ga+b}-1)}{2}]C_{-a,a}.
\end{align}

For $\Delta_3(a,-a,\ga+b)=3$ the independent consistent condition is
\begin{align}
&\mu_{a,-a}C_{a,\ga+b}C_{-a,\ga+a+b}+C_{-a,\ga+b}C_{a,\ga+b-a}\nonumber\\
&=[\frac{2\hsc{a}\hsc{\ga+b}}{c}+\frac{(\al_{a,\ga+b}-2)(\al_{a,\ga+b}-1)}{2}]C_{a,-a}.
\end{align}

For $\Delta_3(a,-a,\ga+b)\geq4$ there are no extra consistent
conditions from useful GJI's.

\section{Examples of FQH states
described by $Z_n$ simple-current vertex algebras} \label{EXM}

In this section, we will examine some examples of FQH states that can
be described by $Z_n$ simple-current vertex algebra.

\subsection{Pattern of zeros for $Z_n$ simple-current vertex algebra}
\label{NUMSCVA}

When we consider FQH states described by $Z_n$ simple-current vertex
algebra, the patterns of zero for those FQH states satisfy many
additional conditions on top of the conditions \eq{gVAcon1},
\eq{gVAcon2}, and \eq{gVAcon3} for generic FQH states.  In section
\ref{concon}, we list those additional consistent conditions
obtained from GJI.  Many conditions do not contain the structure
constants $C_{ab}$, and those conditions become the extra conditions
on the pattern of zeros.  We have numerically solved all those
conditions on the pattern of zeros. In this section, we list some of
the numerical solutions.

We like to point out that the patterns of zeros for FQH states
described by simple-current vertex algebra do not have the additive
property.  This is because given two FQH wave functions described by
simple-current vertex algebra, their product in general cannot be
described any simple-current vertex algebra.  The direct product of
two simple-current vertex algebra, in general, contains at least one
dimension-2 primary field of Virasoro algebra that violates the
Abelian fusion algebra.  Thus the direct product of two
simple-current vertex algebra is not a simple-current vertex algebra
in general.

Among many solutions of the consistent conditions are the $Z_n$
parafermion algebras, which are the simplest simple-current vertex
algebra.  The $Z_n$ parafermion algebras give rise to $Z_n$
parafermion wave functions $\Phi_{Z_n}$.  As an example of no
additive property, the pattern of zeros for the product wave
function $\Phi_{Z_2\otimes Z_3}\equiv \Phi_{Z_2}\Phi_{Z_3}$ does not
satisfy the consistent conditions for the simple-current vertex
algebra, indicating that the direct product of $Z_2$ and $Z_3$
parafermion vertex algebras is not a simple-current vertex algebra. In
the following, we only list some solutions that are not $Z_n$
parafermion algebras.

$Z_2$ simple-current vertex algebra:
\begin{align}
 &n =2:\ \ \ c=1\ \ \ (Z_2|Z_2\text{ state})
\nonumber\\
 &\{m;\hsc{1}..\hsc{n-1}\}=\{4;1\}
\nonumber\\
 &\{p;M_{1}..M_{n-1}\}=\{2; 0\}
\nonumber\\
 &\ \ \ \ \{n_0..n_{m-1}\} = \{2\ 0\ 0\ 0  \} .
\end{align}
\begin{align}
  &n =2:\ \ \ \ \ \ \ (Z_2|Z_2|Z_2\text{ state})
\nonumber\\
 &\{m;\hsc{1}..\hsc{n-1}\}=\{6;\frac32\}
\nonumber\\
 &\{p;M_{1}..M_{n-1}\}=\{3; 0\}
\nonumber\\
 &\ \ \ \ \{n_0..n_{m-1}\} = \{2\ 0\ 0\ 0\ 0\ 0  \} .
\end{align}

$Z_3$ simple-current vertex algebra:
\begin{align}
 &n =3:\ \ \ (Z_3|Z_3\text{ state})
\nonumber\\
 &\{m;\hsc{1}..\hsc{n-1}\}=\{4;\frac43\ \frac43\}
\nonumber\\
 &\{p;M_{1}..M_{n-1}\}=\{2; 0\ 0\}
\nonumber\\
 &\ \ \ \ \{n_0..n_{m-1}\} = \{3\ 0\ 0\ 0  \} .
\end{align}

$Z_4$ simple-current vertex algebra:
\begin{align}
 &n =4:\ \ \ c=1\ \ \ (C_4\text{ state})
\nonumber\\
 &\{m;\hsc{1}..\hsc{n-1}\}=\{4;1\ 1\ 1 \}
\nonumber\\
 &\{p;M_{1}..M_{n-1}\}=\{2; \frac12\ 1\ \frac12\}
\nonumber\\
 &\ \ \ \ \{n_0..n_{m-1}\} = \{2\ 2\ 0\ 0  \} .
\end{align}
\begin{align}
 &n =4: \ \ \ c=1 \ \ \ (Z_4|Z_2\text{ state})
\nonumber\\
 &\{m;\hsc{1}..\hsc{n-1}\}=\{6;\frac54\ 1\ \frac54 \}
\nonumber\\
 &\{p;M_{1}..M_{n-1}\}=\{3; 1\ 2\ 1\}
\nonumber\\
 &\ \ \ \ \{n_0..n_{m-1}\} = \{2\ 0\ 2\ 0\ 0\ 0\} .
\end{align}
\begin{align}
 &n =4: (\text{Gaffnian state})
\nonumber\\
 &\{m;\hsc{1}..\hsc{n-1}\}=\{6;\frac34\ 0\ \frac34 \}
\nonumber\\
 &\{p;M_{1}..M_{n-1}\}=\{3; \frac32\ 3\ \frac32\}
\nonumber\\
 &\ \ \ \ \{n_0..n_{m-1}\} = \{2\ 0\ 0\ 2\ 0\ 0  \} .
\end{align}
\begin{align}
\label{C42}
 &n =4:  \ \ \ (C_4|C_4\text{ state})
\nonumber\\
 &\{m;\hsc{1}..\hsc{n-1}\}=\{8;2\ 2\ 2 \}
\nonumber\\
 &\{p;M_{1}..M_{n-1}\}=\{4; 1\ 2\ 1\}
\nonumber\\
 &\ \ \ \ \{n_0..n_{m-1}\} = \{2\ 0\ 2\ 0\ 0\ 0\ 0\ 0 \} .
\end{align}

$Z_6$ simple-current vertex algebra:
\begin{align}
 &n =6:
\nonumber\\
 &\{m;\hsc{1}..\hsc{n-1}\}=\{6;\frac32\ 2\ \frac52\ 2\ \frac32 \}
\nonumber\\
 &\{p;M_{1}..M_{n-1}\}=\{3; 1\ 2\ 2\ 2\ 1\}
\nonumber\\
 &\ \ \ \ \{n_0..n_{m-1}\} = \{2\ 2\ 2\ 0\ 0\ 0  \} .
\end{align}
\begin{align}
 &n =6: \ \ \ c=1
\nonumber\\
 &\{m;\hsc{1}..\hsc{n-1}\}=\{8;\frac43\ \frac43\ 1\ \frac43\ \frac43 \}
\nonumber\\
 &\{p;M_{1}..M_{n-1}\}=\{3; 2\ 4\ 5\ 4\ 2\}
\nonumber\\
 &\ \ \ \ \{n_0..n_{m-1}\} = \{2\ 1\ 0\ 1\ 2\ 0\ 0\ 0  \} .
\end{align}

We like to stress that the above pattern of zeros are only checked
to satisfy the consistent conditions that do not contain structure
constants $C_{a,b}$.  It remains to be shown that there exist
$C_{a,b}$ for those patterns of zeros that satisfy all the
consistent conditions for structure constants (from GJI's).  When we
check those additional conditions for $C_{a,b}$, we find that the
$C_4$ pattern of zero $\{n;m;\hsc{a}\}=\{4;4;1\ 1\ 1 \}$ does not
correspond to any symmetric polynomial as discussed in section
\ref{gVAcon23}.

We will discuss some other patterns of zeros in detail later.  We
will show how the central charge $c$, the structure constants
$C_{a,b}$ and the quasiparticle scaling dimension $\hsc{\ga+a}$ of
the corresponding vertex algebra can be determined from the pattern
of zeros $\{n,m;\hsc{a}\}$, through the consistent conditions in
section \ref{concon}, \ref{ConsistentConditionsQP} and in Appendix
\ref{app:subleading}. Those consistent conditions are generated by
useful GJI's: \eq{mul:gJacob} or \eq{GJI} with \eqn{GJI:useful}.

To calculate the central charge and the quasiparticle scaling
dimensions $\{c;C_{a,b};\hsc{\ga+a}\}$, in the first step we will
try to determine them from conditions in section \ref{concon}, \ie
we don't specify subleading order term \eq{OPE:i,j:1st} in OPE. If
these conditions don't give enough information, then we will resort
to more conditions in Appendix \ref{app:subleading}, which is based
on the subleading OPE term \eq{OPE:i,j:1st}.

We note that some pattern of zeros can directly fix the central
charge, and we list the central charge for those patterns of zeros as
in above.  The $Z_n$ parafermion patterns of zeros are examples in
this class.  While for other patterns of zeros, the central charges
depend on the structure constants $C_{a,b}$. We will calculate those
central charges below.  There are even patterns of zeros that do not
completely determine the simple-current vertex algebra.  We need to
include additional information $C_{ab}$ to determine the corresponding
simple-current vertex algebra.  The $Z_3|Z_3$, $Z_4|Z_2$ states \etc
are examples in this class of pattern of zeros.

We also give names for some patterns of zeros. For example, the
$C_n|C_n$ pattern of zeros $\{m;\hsc{1}..\hsc{n-1}\}=\{2n;2\ 2\
...2 \}$ is the sum of two $C_n$  pattern of zeros
$\{m;\hsc{1}..\hsc{n-1}\}=\{n;1\ 1\ ...1 \}$.  Also, the $Z_4|Z_2$
pattern of zeros is described by
$\{m;\hsc{1}..\hsc{n-1}\}=\{6;\frac54\ 1\ \frac54 \}$ which is a sum
of $\{m;\hsc{1}..\hsc{2n-1}\}=\{4;\frac12\ 0\ \frac12 \}$ for the
$Z_2$ parafermion state and $\{m;\hsc{1}..\hsc{n-1}\}=\{2;\frac34\
1\ \frac34 \}$ for the $Z_4$ parafermion state. (Note that the $Z_2$
parafermion state is also described by
$\{m;\hsc{1}..\hsc{n-1}\}=\{2;\frac12\}$.\cite{poz1}) However, the
wave function of such a $Z_4|Z_2$ state is different from the
product of a $Z_2$ parafermion wave function and a $Z_4$ parafermion
wave function.  The product wave function called the $Z_2\otimes
Z_4$ state, is described by a $Z_4$ vertex algebra given by the
direct product of the $Z_2$ parafermion algebra and $Z_4$
parafermion algebra.  Such a $Z_4$ vertex algebra is different from
any $Z_4$ simple-current vertex algebras, featured by an extra
dimension-2 primary field. However, both the $Z_4|Z_2$ and
$Z_2\otimes Z_4$ states have the same pattern of zeros.  This is an
example showing that the same pattern of zeros
$\{m;\hsc{1}..\hsc{n-1}\}=\{6;\frac54\ 1\ \frac54 \}$ can correspond
to more than one FQH wave functions.

\subsection{The $Z_n$ parafermion vertex algebra: $Z_n$
parafermion states with $\{M_k=0;p=1;m=2\}$}

In this simplest case we have $p=1,~M_k=0$.  For example the $Z_3$
parafermion state is described by the following pattern of zeros:
\begin{align}
 &n =3:\ \ \ Z_3\text{ state}
\nonumber\\
 &\{m;\hsc{1}..\hsc{n-1}\}=\{2;\frac23\ \frac23\}
\nonumber\\
 &\{p;M_{1}..M_{n-1}\}=\{1; 0\ 0\}
\nonumber\\
 &\ \ \ \ \{n_0..n_{m-1}\} = \{3\ 0\} .
\end{align}
In general, we have (we don't specify $p=1$ until necessary, trying
to obtain some general conclusions on $Z_n|...|Z_n$ series):
\begin{align}
\hsc{a} &= p\frac{a(n-a)}{n},
\nonumber\\
\alpha_{a,b} &= \frac{2pab}{n}-2p(a+b-n)\theta(a+b-n)
\end{align}
As a result we have
\begin{align}\label{mu=1}
\mu_{a,b}=1,\ \ \ C_{a,b}=C_{b,a}
\end{align}
Besides, $d_{a,b}$ defined in \eqn{d_a,b} has a simple form in this
case:
\begin{align}\label{d_a,b:Z_n}
d_{a,b}\equiv\frac{1}{2}(1+\frac{\hsc{a}-\hsc{b}}{\hsc{a+b}})=d_{n-a,n-b}=\frac{a}{a+b}\nonumber\\
~\text{if}~a+b<n
\end{align}
In \eqn{Delta3-ijk} we have $\Delta M[a,b,c]=0$ and
$\Delta_3(a,b,c)$ can only be multiples of $2p$.

At first let's take a look at
$\{A,B,C\}=\{\psi_a,\psi_b,\psi_c\},~a+b,b+c,a+c\neq0\mod n$. Only
when $\Delta_3(a,b,c)=0$ there are extra consistent conditions in
section \ref{consistency:a,b,c}, \ie $a+b+c\leq n$ or $a+b+c\geq2n$
we have
\begin{align}\label{a+b+c<n;>2n}
C_{a,b}C_{a+b,c}=C_{b,c}C_{a,b+c}=C_{a,c}C_{b,a+c}
\end{align}
Particularly when $a+b+c=0\mod n$ we have
\begin{align}\label{a+b+c=n;2n}
C_{a,b}=C_{a,c}=C_{b,c}
\end{align}
The other consistent condition is satisfied by \eqn{d_a,b:Z_n}.

For $A=B=C=\psi_{n/2},~n=$~even we know that
$\Delta_3(n/2,n/2,n/2)=4\hsc{n/2}=pn$. Only when $np/2\leq2$ there
are extra consistent conditions in section
\ref{consistency:n/2,n/2,n/2}, \ie when $n\leq4$ for
$p=1$,~~$n\leq2$ for $p=2$.

The above conclusions hold for any $p\in\dN$. Now let's enforce
$p=1$ for this special series.

For $\{A,B,C\}=\{\psi_a,\psi_b,\psi_{-b}\},~~a\pm b\neq0\mod n$ we
know $\Delta_3(a,b,-b)\geq2p$. Only when $\Delta_3(a,b,-b)=2$ there
are extra consistent conditions in section \ref{consistency:a,b,-b},
\ie when $a=1<b,n-b<n$,~ $b=1<a<n-1$ or $1\leq n,n-b<a=n-1$.

For $\{A,B,C\}=\{\psi_a,\psi_a,\psi_{-a}\},~~a\neq n/2\mod n$,
similarly only when $\Delta_3(a,a,-a)=2,~4$ there are consistent
conditions in section \ref{consistency:a,a,-a}, \ie when $a=1,~2$ or
$a=n-1,~n-2$.

First since $\Delta_3(1,1,-1)=\Delta_3(-1,-1,1)=2$ and
$\hsc1=1-1/n,~~\al_{1,1}=2/n$, from section \ref{consistency:a,a,-a}
we have
\begin{align}
c&=\frac{2(n-1)}{n+2}
\nonumber \\
C_{1,1}C_{2,n-1}&=C_{n-1,n-1}C_{n-2,1}=\frac{2(n-1)}{n} \nonumber
\end{align}
With central charge $c$ in hand, from
$\Delta_3(2,2,-2)=\Delta_3(-2,-2,2)=4$ we have
\begin{align}
C_{2,2}C_{4,n-2}=C_{n-2,n-2}C_{n-4,2}=\frac{6(n-2)(n-3)}{n(n-1)}
\end{align}
Similarly from
$\Delta_3(1,b,-b)=\Delta_3(a,1,-1)=\Delta_3(-1,-b,b)=\Delta_3(-a,-1,1)=2$
we have
\begin{align}
C_{b,n-1}C_{n-b,b-1}=\frac{b(n+1-b)}{n}
\nonumber \\
C_{a,n-1}C_{1,a-1}=\frac{a(n+1-a)}{n}
\end{align}
These are all the extra consistent conditions. Using
\eqn{a+b+c<n;>2n} and \eqn{a+b+c=n;2n} repeatedly we find out that
the independent conditions besides
\eqn{a+b+c<n;>2n}-\eqn{a+b+c=n;2n} and $C_{a,b}=C_{b,a}$ are just
\begin{align}\label{1,a*n-1,n-a:p=1}
C_{1,a}C_{n-1,n-a}=\frac{(a+1)(n-a)}{n}
\end{align}
Other structure constants can be expressed as
\begin{align}
\label{C_a,b<n}
C_{a,b}&=\frac{C_{a,b-1}C_{1,a+b-1}}{C_{1,b-1}}=\cdots
=\frac{\prod_{i=0}^{b-1}C_{1,a+i}}{\prod_{j=1}^{b-1}C_{1,j}}
\nonumber\\
&=\frac{\prod_{i=1}^{a+b-1}C_{1,i}}
{\prod_{i=1}^{a-1}C_{1,i}\prod_{i=1}^{b-1}C_{1,i} }
\end{align}
if $a+b\leq n$;
\begin{align}
\label{C_a,b>n}
C_{a,b}&=\frac{C_{a,b+1}C_{n-1,a+b+1}}{C_{n-1,b+1}}=\cdots
=\frac{\prod_{i=0}^{n-b-1}C_{n-1,a-i}}{\prod_{j=1}^{n-b-1}C_{n-1,n-j}}
\nonumber\\
&=\frac{\prod_{i=1}^{n-a+n-b-1}C_{n-1,n-i}}
{ \prod_{i=1}^{n-a-1}C_{n-1,n-i} \prod_{i=1}^{n-b-1}C_{n-1,n-i} }
\end{align}
if $a+b>n$. Notice that the above two equations are compatible with
\eqn{a+b+c<n;>2n}! Using \eqn{1,a*n-1,n-a:p=1} we immediately have
\begin{align}
&\nonumber C_{a,b}C_{n-a,n-b}=\frac{\prod_{i=0}^{b-1}C_{1,a+i}C_{n-1,n-a-i}}{\prod_{j=1}^{b-1}C_{1,j}C_{n-1,n-j}}\\
&=\frac{\Ga(a+b+1)\Ga(n-a+1)\Ga(n-b+1)}{\Ga(n+1)\Ga(a+1)\Ga(b+1)\Ga(n-a-b+1)},\\
&\forall~~~1\leq a,b<a+b\leq n
\end{align}
To summarize, the consistent conditions in section \ref{concon}
determine the central charge and fix the structure constants to the
following form:
\begin{align}
 C_{a,b}& =\frac{\prod_{i=0}^{b-1}\lambda_{a+i}}{\prod_{j=1}^{b-1}\lambda_{j}}
\times
\\
&
\sqrt{\frac{\Ga(a+b+1)\Ga(n-a+1)\Ga(n-b+1)}{\Ga(n+1)\Ga(a+1)\Ga(b+1)\Ga(n-a-b+1)}}
\nonumber
\end{align}
if $1\leq a,b<a+b\leq n$. Free parameters
$\{\lambda_a|a=1,\cdots,n-1\}$ are nonzero complex numbers, defined
by $C_{1,a}=\lambda_a^2C_{n-1,n-a}$. Moreover, the condition
\eq{a+b+c=n;2n} requires that $C_{a,n-a}=1$, so from \eqn{C_a,b<n}
we have the following ``reflection'' condition on $\{\lambda_a\}$
\begin{align}\label{normalization const:reflection}
\lambda_{n-1}=\frac{\lambda_{n-2}}{\lambda_1}=\cdots=\frac{\lambda_{n-1-k}}{\lambda_k}=\cdots=1
\end{align}

We point out that the above conclusions are all obtained from
conditions in section \ref{concon}, \ie we haven't introduced the
subleading order OPE \eq{OPE:i,j:1st} and new conditions in Appendix
\ref{app:subleading} yet. Now we apply conditions in Appendix
\ref{app:subleading} to see whether the normalization constants
$\{\lambda_a|a=1,\cdots,[\frac{n-1}{2}]\}$ can be determined or not.

According to Appendix \ref{consistency:a,b,c:1st}, choosing those
$\Delta_3(a,b,c)=2p=2$ with $a+b,b+c,a+c\neq0\mod n,~a+b+c=n+1$
leads to the following new constraints:
\begin{align}
\label{lala}
\lambda_{a-1}\lambda_{n-a}=1
\end{align}
which means that $\lambda_{a}=\pm1$. Other useful GJI's like
$\Delta_3(a,b,c)=4p=4$ with $a+b,b+c,a+c\neq0\mod n$ doesn't result
in any new constraints. So finally we can conclude that considering
the subleading order OPE \eq{OPE:i,j:1st}, the structure of such a
$Z_n$ simple-current vertex algebra is determined self-consistently
as:
\begin{align}
&c=\frac{2(n-1)}{n+2};~~\lambda_{a}=\pm1;\\
&\nonumber C_{a,b}=\frac{\prod_{i=0}^{b-1}\lambda_{a+i}}{\prod_{j=1}^{b-1}\lambda_{j}}\times\\
&\sqrt{\frac{\Ga(a+b+1)\Ga(n-a+1)\Ga(n-b+1)}{\Ga(n+1)\Ga(a+1)\Ga(b+1)\Ga(n-a-b+1)}},\\
&\nonumber
C_{n-a,n-b}=\frac{\prod_{j=1}^{b-1}\lambda_{j}}{\prod_{i=0}^{b-1}\lambda_{a+i}}\times\\
&\sqrt{\frac{\Ga(a+b+1)\Ga(n-a+1)\Ga(n-b+1)}{\Ga(n+1)\Ga(a+1)\Ga(b+1)\Ga(n-a-b+1)}},\\
&\nonumber\forall~~a+b\leq n;\\
&\lambda_{a-1}=\lambda_{n-a},~~\la_0=\lambda_{n-1}=1,\ \la_{a+n}=\la_a.
\end{align}

It is interesting to see that the $Z_n$-parafermion pattern of zeros
does not completely fix the structure constants $C_{a,b}$.  Do those
different structure constants $C_{a,b}$ corresponding to different
choices of $\la_a$ give rise to different FQH wave functions, even through
they all have the same pattern of zeros? In general, the different
structure constants (even with the same pattern of zeros) will give rise to
different FQH wave functions.  But in this particular case, all the
above different structure constants for different choices of
$\la_a=\pm 1$ give rise to the same FQH wave function.  So those
different structure constants describe the same FQH state.

To see this, let us introduce
\begin{align}
 \t\psi_a =\chi_a\psi_a,\ \ \
\chi_a=\pm 1,\ \chi_{a+n}=\chi_a .
\end{align}
Those modified simple current operators will generate the same
FQH state. But the structure constants of $ \t\psi_a$ is changed
\begin{align}
 \t C_{a,b}= C_{a,b} \frac{\chi_a\chi_b}{\chi_{a+b}}
\end{align}
So such kind of change in structure constants
\begin{align}
\label{CtC}
 C_{a,b}\to  \t C_{a,b}= C_{a,b} \frac{\chi_a\chi_b}{\chi_{a+b}}
\end{align}
does not generate new FQH wave function.  Therefore $C_{a,b}$
and $\t C_{a,b}= C_{a,b} \frac{\chi_a\chi_b}{\chi_{a+b}}$ describe the
same FQH state. We will call the transformation \eq{CtC} an
equivalence transformation.

Note that the factors can be rewritten as
\begin{align}
 \frac{\prod_{i=0}^{b-1}\lambda_{a+i}}{\prod_{j=1}^{b-1}\lambda_{j}}
&=
 \frac{\prod_{i=0}^{a+b-1}\lambda_{i}}{
\prod_{i=0}^{a-1}\lambda_{i} \prod_{i=0}^{b-1}\lambda_{i} }
\\
 \frac{ \prod_{j=1}^{b-1}\lambda_{j} }{ \prod_{i=0}^{b-1}\lambda_{a+i} }
&=
 \frac{\prod_{i=0}^{n-b-1}\lambda_{n-a+i}}{\prod_{j=1}^{n-b-1}\lambda_{j}}
 =
\frac{\prod_{i=0}^{n-a+n-b-1}\lambda_{i}}{
\prod_{i=0}^{n-a-1}\lambda_{i} \prod_{i=0}^{n-b-1}\lambda_{i} }
,
\nonumber
\end{align}
where we have used \eqn{lala}.  So if we choose $
\chi_a=\prod_{i=0}^{a-1}\lambda_{i} $, the equivalence transformation
\eq{CtC} will remove the $\la_a$ dependent factors in the structure
constants.  This completes our proof.  We see that the $Z_n$
parafermion patterns of zeros, $\{M_k=0;p=1;m=2\}$, completely
determine the structure constants $C_{ab}$ and the central charge $c$.

It is interesting to note that we can use the equivalence
transformation \eq{CtC} to make $C_{a,b}=1$ for all $a+b\leq n$, as
one can see from \eq{C_a,b<n}.  We can also use the transformation
\eq{CtC} to make $C_{a,b}=1$ for all $a+b\geq n$, as one can see from
\eq{C_a,b>n}.  But we cannot make all $C_{a,b}=1$.

\subsection{Quasiparticles in the $Z_2$ parafermion state }

\begin{table}[tb]
\begin{tabular}{|c|c||c|rr|c|c|}
\hline $I$ & $I_\text{na}$ & $n_{\ga;0..m-1}$ &
\multicolumn{2}{|c|}{$n\ksc{1..n}$} & $Q$ & $h^\text{sc}+h^\text{ga}$\\
\hline
0 & 0$_\text{na}$ & 2 0 & $\ 1$ & $-1$ & 0   & $0+0$\\
1 & 0$_\text{na}$ & 0 2 & $\ 1$ & $-1$ & 1   & $0+\frac{1}{2}$ \\
\hline
2 & 1$_\text{na}$ & 1 1 & $\ 0$ & $0$  & 1/2 & $\frac1{16}+\frac{1}{8}$ \\
\hline
\end{tabular}
\caption{ \label{Z2qppoz} The pattern of zeros and the charges $Q$
for the quasiparticles in the $Z_2$ parafermion state.
$n_{\ga;1}...n_{\ga,m-1}$ is the occupation sequence characterizing
the quasiparticle $\ga$ (defined below \eqn{lgaSga}).  The quasiparticles are labeled by the
index $I$.  The scaling dimensions of the quasiparticle operators
are sums of the contributions from the simple-current vertex algebra
and the Gaussian model: $h_\ga=h^\text{sc}+h^\text{ga}$. }
\end{table}

In this section, we will study the quasiparticles in the $Z_2$
parafermion state. For the $Z_2$ parafermion state, we have a simple
current with scaling dimension $\hsc{1}=1/2$ and central charge
$c=1/2$.  According to \Ref{poz3}, the patterns of zeros
$\{\ksc{a};Q_\ga\}$ for the quasiparticles are obtained by solving the
conditions \eq{concave1qp tlsc} and \eq{concave2qp tlsc}. The result
is listed in table \ref{Z2qppoz}.  There are three types of
quasiparticles.  In fact, these three quasiparticles belong to two
different families.  The quasiparticles in the same family can change
into each other by combining an Abelian quasiparticle.
The two quasiparticles in the first family
differ from each other merely by a
magnetic translation\cite{poz2,poz3} (\ie by an insertion of an
Abelian magnetic flux tube), while the 3rd quasiparticle differs
from the first two in their non-Abelian content. For a family of
quasiparticles differ by magnetic translations, we only need to
study one of them to obtain all the information of simple current
part (the difference between different quasiparticles in such a
family comes solely from a $U(1)$ factor).

\begin{table}[tb]
\begin{tabular}{|c|c||c|rrr|c|c|}
\hline $I$ & $I_\text{na}$ & $n_{\ga;0..m-1}$ &
\multicolumn{3}{|c|}{$n\ksc{1..n}$} & $Q$ & $h^\text{sc}+h^\text{ga}$\\
\hline
0 & 0$_\text{na}$ & 3 0 & $\ 2$ & $0$ & $-2$ & 0   & $0+0$\\
1 & 0$_\text{na}$ & 0 3 & $\ 2$ & $0$ & $-2$ & 3/2 & $0+\frac{3}{4}$ \\
\hline
2 & 1$_\text{na}$ & 2 1 & $\ 1$ & $-1$ & $0$ & 1/2 & $\frac{1}{15}+\frac{1}{12}$ \\
3 & 1$_\text{na}$ & 1 2 & $\ 0$ & $1$ & $-1$ & 1   & $\frac{1}{15}+\frac{1}{3}$ \\
\hline
\end{tabular}
\caption{ \label{Z3qppoz} The pattern of zeros and the charges $Q$
for the quasiparticles in the $Z_3$ parafermion state. The
quasiparticles are labeled by the index $I$. The scaling dimensions
of the quasiparticle operators are sums of the contributions from
the simple-current vertex algebra and the Gaussian model:
$h_\ga=h^\text{sc}+h^\text{ga}$. }
\end{table}

First let us study the 3rd quasiparticle
$\{\ksc1,\cdots,\ksc{n};Q_\ga\}=\{0,0;\frac12\}$. Using the
following relations derived from \eqn{hgaahga} (which will be used
frequently in the quasiparticle calculation)
\begin{align}
\al_{a,\ga+b}=\hsc a-\sum_{k=1}^a\ksc{b+k}
\nonumber \\
\Delta_3(a,b,\ga+c)=\al_{a,\ga+c}+\al_{b,\ga+c}-\al_{a+b,\ga+c} ,
\end{align}
we find $\Delta_3(1,1,\ga)=\Delta_3(1,1,\ga+1)=1$,
$\al_{1,\ga}=\al_{1,\ga+1}=1/2$.  So from section
\ref{consistency:a,-a,ga+b}, we see that
\begin{align}
\hsc{\ga}=\hsc{\ga+1}=1/16
\nonumber \\
C_{1,\ga+1}=1/2
\end{align}
This is the famous disorder operator of a $Z_2$ or Ising vertex
algebra.

Next we study the 1st quasiparticle
$\{\ksc1,\cdots,\ksc{n};Q_\ga\}=\{\frac12,-\frac12;0\}$ from its
family. With $\Delta_3(1,1,\ga)=0$ and $\Delta_3(1,1,\ga+1)=2$ we
have from section \ref{consistency:a,-a,ga+b} that
\begin{align}
\hsc{\ga}=0,~~~\hsc{\ga+1}=1/2
\nonumber \\
C_{1,\ga+1}=1,~~~\partial\si_\ga\equiv0
\end{align}
suggesting that $\si_\ga$ is proportional to the identity operator.
This means that this quasiparticle is simply the trivial vacuum
modulo electrons.

We like to stress that for the above two quasiparticles, the structure
constants $C_{a,\ga+b}$ are uniquely determined by the quasiparticles
pattern of zeros $\{\ksc{a}\}$.  So the quasiparticles in the
$Z_2$ parafermion theory are uniquely described by the quasiparticles
pattern of zeros $\{\ksc{a};Q_\ga\}$.
Each index $I$ in the table \ref{Z2qppoz}
label a unique quasiparticle pattern of zeros, and so the index $I$
also label a unique quasiparticle for the $Z_2$ parafermion state.

We can also obtain the fusion algebra using the method in
\Ref{poz3}. We find that
\begin{align}
\label{faZ2} 0 \times 0 &= 0, & 0 \times 1 &= 1, & 0 \times 2 &= 2,
\nonumber\\
1 \times 1 &= 0, & 1 \times 2 &= 2, & 2 \times 2 &= 0+1,
\end{align}
where we have used the index $I$ to label different quasiparticles
(see table \ref{Z2qppoz}).  We have regarded two quasiparticles to be
equivalent if they differ by some electrons. The index $I$ really
label the above equivalent classes of quasiparticles.  We may also
define a different equivalent class of quasiparticles by regarding two
quasiparticles to be equivalent if they differ by some electrons or by
some Abelian magnetic flux tubes.  Such classes of quasiparticles are
characterized by $\{\ksc1,\cdots,\ksc{n}\}$ up to a cyclic
permutation.  We introduce an index $I_\text{na}$ to label those
non-Abelian classes of quasiparticles. From the relation between the
two sets of indices $I$ and $I_\text{na}$ as shown in table
\ref{Z2qppoz}, we can reduce the fusion algebra \eq{faZ2} to a simpler
fusion algebra between the non-Abelian classes of quasiparticles
\begin{align}
0_\text{na} \times 0_\text{na} = 0_\text{na} , \ 
0_\text{na} \times 1_\text{na} = 1_\text{na} , \ 
1_\text{na} \times 1_\text{na} = 0_\text{na}+0_\text{na} .
\end{align}

\subsection{Quasiparticles in the $Z_3$ parafermion state}

The $Z_3$ simple-current vertex algebra is characterized by
\begin{align}
\hsc{1}=\hsc{2}=\frac23,~~~c=\frac45,
\nonumber \\
C_{1,1}=C_{2,2}=\frac2{\sqrt3},~~~C_{1,2}=C_{2,1}=1
\end{align}
where we have fixed the normalization factors to be $\la_a=1$.

There are two families of quasiparticles obtained from
\eq{concave1qp tlsc} and \eq{concave2qp tlsc} (see table
\ref{Z3qppoz}).  The 1st family is represented by quasiparticle
$\{\ksc1,\cdots,\ksc{n};Q_\ga\}=\{\frac23,0,-\frac23;0\}$. With
$\Delta_3(1,1,\ga)=\Delta_3(2,2,\ga)=
\Delta_3(1,1,\ga+1)=\Delta_3(2,2,\ga+2)=0$,
we find that (see section \ref{consistency:a,b,ga+c} or Appendix
\ref{consistency:a,b,ga+c:1st})
\begin{align}
C_{1,\ga+1}=C_{2,\ga+2}=C_{1,1}=C_{2,2}
\nonumber \\
C_{2,\ga+1}=C_{1,\ga+2}
\end{align}
Then with $\Delta_3(1,2,\ga)=0$ and
$\Delta_3(1,2,\ga+1)=\Delta_3(1,2,\ga+2)=2$ we have from section
\ref{consistency:a,-a,ga+b} that
\begin{align}
\hsc\ga=0,~~\hsc{\ga+1}=\hsc{\ga+2}=\frac23,~~~\partial\si_\ga\equiv0
\nonumber \\
C_{2,\ga+1}=C_{1,\ga+2}=1,~~~C_{1,\ga+1}C_{2,\ga+2}=\frac43
\end{align}
Therefore this quasiparticle is characterized by
\begin{align}
\hsc\ga=0,~~\hsc{\ga+1}=\hsc{\ga+2}=\frac23,~~~\partial\si_\ga=0
\nonumber \\
C_{2,\ga+1}=C_{1,\ga+2}=1,~~~C_{1,\ga+1}=C_{2,\ga+2}=\frac2{\sqrt3}
\end{align}
$\partial\si_\ga=0$ and $\hsc\ga=0$ imply that the quasiparticle
operator $\si_\ga$ is a constant operator with scaling dimension 0.
Such an operator is the trivial identity operator.

The 2nd family is represented by a quasiparticle with
$\{\ksc1,\cdots,\ksc{n};Q_\ga\}=\{\frac13,-\frac13,0;\frac12\}$.  With
$\Delta_3(1,1,\ga)=\Delta_3(2,2,\ga+2)=0$ and
$\Delta_3(1,1,\ga+1)=\Delta_3(1,1,\ga+2)=\Delta_3(2,2,\ga)
=\Delta_3(2,2,\ga+1)=1$,
we find that (see section \ref{consistency:a,b,ga+c} or Appendix
\ref{consistency:a,b,ga+c:1st})
\begin{align}
C_{1,\ga+1}=C_{1,1},~~~C_{2,\ga+2}=C_{2,2}/2
\nonumber \\
C_{1,\ga+2}=C_{1,1}C_{2,2}/4,~~~C_{2,\ga+1}=C_{1,1}C_{2,2}/2
\end{align}
Then with $\Delta_3(1,2,\ga)=\Delta_3(1,2,\ga+2)=1$ and
$\Delta_3(1,2,\ga+1)=2$, we find that (see section
\ref{consistency:a,-a,ga+b})
\begin{align}
\hsc\ga=\hsc{\ga+2}=\frac1{15},~~\hsc{\ga+1}=\frac25,
\nonumber \\
C_{2,\ga+1}=C_{1,\ga+1}C_{2,\ga+2}=\frac23,~~~C_{1,\ga+2}=\frac13
\end{align}
This nontrivial quasiparticle is characterized by
\begin{align}
\hsc\ga=\hsc{\ga+2}=\frac1{15},~~\hsc{\ga+1}=\frac25,
\nonumber \\
C_{2,\ga+1}=\frac23,~~~C_{1,\ga+2}=\frac13,
\nonumber \\
C_{1,\ga+1}=C_{1,1}=\frac2{\sqrt3},~C_{2,\ga+2}=C_{2,2}/2=\frac1{\sqrt3}
\end{align}

Again, for the above two quasiparticles, the structure constants
$C_{a,\ga+b}$ are uniquely determined by the quasiparticles pattern of
zeros $\{\ksc{a}\}$.  So the quasiparticles in the $Z_3$ parafermion
theory are uniquely described by the quasiparticles pattern of zeros
$\{\ksc{a};Q_\ga\}$.  Each index $I$ in the table \ref{Z3qppoz} label
a unique quasiparticle pattern of zeros, and so the index $I$ also
label a unique quasiparticle for the $Z_3$ parafermion state.

The full fusion algebra between the quasiparticles is\cite{poz3}
\begin{align}
0 \times 0 &= 0 , & 0 \times 1 &= 1 , & 0 \times 2 &= 2 ,
\nonumber\\
0 \times 3 &= 3 , & 1 \times 1 &= 0 , & 1 \times 2 &= 3 ,
\nonumber\\
1 \times 3 &= 2 , & 2 \times 2 &= 0+3 , & 2 \times 3 &= 1+2 ,
\nonumber\\
3 \times 3 &= 0+3 .
\end{align}
The fusion algebra between the non-Abelian classes of quasiparticles
is
\begin{align}
0_\text{na} \times 0_\text{na} &= 0_\text{na} , & 0_\text{na} \times
1_\text{na} &= 1_\text{na} ,
\nonumber \\
1_\text{na} \times 1_\text{na} &= 0_\text{na}+1_\text{na} .
\end{align}

\subsection{$Z_n|Z_n$ series: $\{M_k=0;p=2\}$}
\label{Z_n^2series}

The $Z_n|Z_n$ vertex algebra is called the ``second generation''
of $Z_n$ parafermion algebra and is studied in 
\Ref{DJS0359,DJS0377,DJS0464,DJS0486}.  In this case we have
\begin{align}
\hsc{a}=2\frac{a(n-a)}{n}
\end{align}
As a result we still have
\eqn{mu=1},~\eqn{d_a,b:Z_n},~\eqn{a+b+c<n;>2n} and \eqn{a+b+c=n;2n},
therefore \eqn{C_a,b<n} and \eqn{C_a,b>n} still hold.

Apparently in this case the extra conditions in section \ref{concon}
are not enough to determine the full structure of this vertex
algebra, since now $\Delta_3(a,b,c)$ are multiples of $2p=4$! So we
introduce the subleading order OPE \eq{OPE:i,j:1st} and resort to
new conditions in Appendix \ref{app:subleading}.

Since now we have
$\Delta_3(1,b,-b)=\Delta_3(-1,b,-b)=\Delta_3(1,a,-1)=2p=4$, from
Appendix \ref{consistency:a,b,-b:1st} we have:
\begin{align}
\nonumber&C_{1,a}C_{n-1,n-a}=\frac{(a+1)(n-a)}{n^2(n-2)}\times \\
&\Big[\frac{4a(n-a-1)(n-1)}{c}+(n-2a)(n-2a-2)\Big]
\end{align}
Representing the central charge $c$ in terms of a continuous
variable $g$ in the following way
\begin{align}
c=\frac{4(n-1)g(n+g-1)}{(n+2g-2)(n+2g)}
\end{align}
yields an expression of structure constants in terms of $g$
and normalization constants
$\{\lambda_a=\lambda_{n-1-a};a=1,\cdots,[\frac{n-1}{2}]\}$, which is
totally similar with $Z_n$ parafermion states:
\begin{align}
&C_{1,a}=\lambda_a\sqrt{\frac{(a+1)(n-a)}{n}
\frac{(a+g)(n+g-a-1)}{g(n+g-1)}}
\\
&C_{a,b}=\frac{\prod_{i=1}^{a+b-1}C_{1,i}}{\prod_{j=1}^{b-1}
C_{1,j}\prod_{k=1}^{a-1}C_{1,k}}=\frac{\prod_{i=0}^{b-1}
\lambda_{a+i}}{\prod_{j=1}^{b-1}\lambda_{j}}\times
\nonumber \\
&\sqrt{\frac{\Gamma(a+b+1)\Gamma(n-a+1)\Gamma(n-b+1)}
{\Gamma(n+1)\Gamma(a+1)\Gamma(b+1)\Gamma(n-a-b+1)}}\times
\nonumber \\
&\sqrt{\frac{\Gamma(a+b+g)\Gamma(n-a+g)\Gamma(n-b+g)
\Gamma(g)}{\Gamma(n+g)\Gamma(a+g)\Gamma(b+g)
\Gamma(n-a-b+g)}},
\\
&C_{n-a,n-b}=(\frac{\prod_{j=1}^{b-1}\lambda_{j}}
{\prod_{i=0}^{b-1}\lambda_{a+i}})^2C_{a,b}, \ \ \ \ a+b\leq n
\end{align}
where reflection condition \eq{normalization const:reflection} should
also be satisfied for the normalization constants $\la_a$.  It's easy
to verify that $\Delta_3(a,b,c)=2p=4$ doesn't result in any new
constraints on free parameters
$\{g;\lambda_a|a=1,\cdots,[\frac{n-1}{2}]\}$. Therefore the
above are all conditions on this $Z_n|Z_n$ series of vertex algebra.

Using the equivalence transformation \eq{CtC}, we can change the
normalization constants to $\la_a=1$. So only different $g$ in the
structure constants give rise to different FQH states.  All those
different FQH states have that same pattern of zeros, and we need an
additional parameter $g$ to completely characterize the FQH state.
For the simple ideal Hamiltonian introduced in \Ref{poz1,poz2}, all
those different FQH states have a zero energy.  In \Ref{S3},
additional terms are introduced in the Hamiltonian so that only the
$Z_3|Z_3$ state with a particular $g$ can be the zero energy states.

\subsection{Quasiparticles in the $Z_2|Z_2$ state}

\begin{table}[tb]
\begin{tabular}{|c|c||c|rr|c|c|}
\hline $I$ & $I_\text{na}$ & $n_{\ga;0..m-1}$ &
\multicolumn{2}{|c|}{$n\ksc{1..n}$} & $Q$ & $h^\text{sc}+h^\text{ga}$\\
\hline
0 & 0$_\text{na}$ & 2 0 0 0 & $2$ & $-2$ & 0 &    $0+0$\\
1 & 0$_\text{na}$ & 0 2 0 0 & $2$ & $-2$ & 1/2 &  $0+\frac14$ \\
2 & 0$_\text{na}$ & 0 0 2 0 & $2$ & $-2$  & 1 &   $0+1$ \\
3 & 0$_\text{na}$ & 0 0 0 2 & $2$ & $-2$  & 3/2 & $0+\frac94$ \\
\hline
4 & 1$_\text{na}$ & 1 1 0 0 & $1$ & $-1$  & 1/4 & $\frac{1}{16}+\frac1{16}$ \\
5 & 1$_\text{na}$ & 0 1 1 0 & $1$ & $-1$  & 3/4 & $\frac{1}{16}+\frac9{16}$ \\
6 & 1$_\text{na}$ & 0 0 1 1 & $1$ & $-1$  & 5/4 & $\frac{1}{16}+\frac{25}{16}$ \\
7 & 1$_\text{na}$ & 1 0 0 1 & $-1$ & $1$  & 3/4 & $\frac{9}{16}+\frac9{16}$ \\
\hline
8 & 2$_\text{na}$ & 1 0 1 0 & $0$ & $0$  & 1/2 &  $\eta+\frac14$ \\
9 & 2$_\text{na}$ & 0 1 0 1 & $0$ & $0$  & 1 &    $\eta+1$ \\
\hline
\end{tabular}
\caption{ \label{Z2Z2qppoz} The pattern of zeros and the charges $Q$
for the quasiparticles in the $Z_2|Z_2$ parafermion state.  The
quasiparticles are labeled by the index $I$.  The scaling dimensions
of the quasiparticle operators are given by
$h_\ga=h^\text{sc}+h^\text{ga}$, where $\eta=C_{1,\ga+1}/2$ is a
free real parameter.  Note the index $I=$ 8, 9 each actually
corresponds to a class of quasiparticles parameterized by a
continuous parameter $\eta$. }
\end{table}

The $Z_2|Z_2$ state is described by the following pattern of zeros:
\begin{align}
 &n =2:\ \ \ c=1\ \ \ (Z_2|Z_2\text{ state})
\nonumber\\
 &\{m;\hsc{1}..\hsc{n-1}\}=\{4;1 \}
\nonumber\\
 &\ \ \ \ \{n_0..n_{m-1}\} = \{2\ 0\ 0\ 0\} .
\end{align}
Here we have $n=2,~p=2,~M_1=0$ and thus $\hsc{1}=1$. Since
$\Delta_3(1,1,1)=4$ we have according to section
\ref{consistency:n/2,n/2,n/2}:
\begin{align}
c=1
\end{align}
There is no free parameter in such a $Z_2$ simple current vertex
algebra.

Now let's turn to the quasiparticles of this state. There are three
families of different quasiparticles, and we will discuss them one
by one.

The 1st family has $\{\ksc1,\cdots,\ksc{n};Q_\ga\}=\{1,-1;0\}$ as
its representative. With $\Delta_3(1,1,\ga)=0$ and
$\Delta_3(1,1,\ga+1)=4$ we have
\begin{align}
\hsc\ga=0,~~~\partial\si_\ga\equiv0,~~~C_{1,\ga+1}=1
\end{align}
indicating this quasiparticle is trivial. We also have
$\hsc{\ga+1}=\hsc\ga+\ksc1=1$ from \eqn{hgaahga}.

The 2nd family is represented by
$\{\ksc1,\cdots,\ksc{n};Q_\ga\}=\{\frac12,-\frac12;\frac14\}$. With
$\Delta_3(1,1,\ga)=0$ and $\Delta_3(1,1,\ga+1)=3$, we find that (see
section \ref{consistency:a,-a,ga+b})
\begin{align}
\hsc{\ga}=1/16,~~~\hsc{\ga+1}=9/16,~~~C_{1,\ga+1}=1/2
\end{align}
This is a nontrivial quasiparticle, resembling the one in an Ising
vertex algebra, except for the charge being $Q_\ga=1/4$ rather than
$1/2$ in the Ising case.

The 3rd family is represented by
$\{\ksc1,\cdots,\ksc{n};Q_\ga\}=\{0,0;\frac12\}$. With
$\Delta_3(1,1,\ga)=\Delta_3(1,1,\ga+1)=2$, we find that
(see section \ref{consistency:a,-a,ga+b})
\begin{align}
\hsc\ga=\hsc{\ga+1}=C_{1,\ga+1}/2 .
\end{align}

Remember that the quasiparticles are described by the data
$\{\ksc{a};Q_\ga;C_{a,\ga+b}\}$.  For the first two family of the
quasiparticles, the quasiparticle structure constants $C_{a,\ga+b}$
are uniquely determined by the quasiparticle pattern of zeros
$\{\ksc{a};Q_\ga\}$. In this case, a pattern of zeros correspond to
a single type of quasiparticle.  For the 3rd family, the pattern of
zeros does not fix $C_{a,\ga+b}$.  Therefore the quasiparticles in
the third family are labeled by the pattern of zeros
$\{\ksc{a};Q_\ga\}$ and a free parameter $\eta=C_{1,\ga+1}/2$. So
there are infinite types of quasiparticles in the 3rd family. The
energy gap for such kind of quasiparticles must vanish at least in
the $\eta\to 0$ limit.

We want to mention here that even introducing subleading order OPE
terms, $[\psi_a\si_{\ga+b}]_{\al_{a,\ga+b}-1}
=C_{a,\ga+b}d_{a,\ga+b}\partial\si_{\ga+a+b}$, like we did in Appendix
\ref{app:subleading} cannot fix this free parameter here.  There are
indeed infinite types of quasiparticles in the $Z_2|Z_2$
simple-current FQH state.  This suggests that the $Z_2|Z_2$
simple-current FQH state is gapless for the ideal Hamiltonian introduced
in \Ref{poz1}.

Using the method in \Ref{poz3}, we obtain the full fusion algebra
between the quasiparticles (expressed in terms of the index $I$ in
table \ref{Z2Z2qppoz}):
\begin{align}
0 \times 0 &= 0 & 0 \times 1 &= 1 & 0 \times 2 &= 2
\nonumber \\
0 \times 3 &= 3 & 0 \times 4 &= 4 & 0 \times 5 &= 5
\nonumber \\
0 \times 6 &= 6 & 0 \times 7 &= 7 & 0 \times 8 &= 8
\nonumber \\
0 \times 9 &= 9 & 1 \times 1 &= 2 & 1 \times 2 &= 3
\nonumber \\
1 \times 3 &= 0 & 1 \times 4 &= 5 & 1 \times 5 &= 6
\nonumber \\
1 \times 6 &= 7 & 1 \times 7 &= 4 & 1 \times 8 &= 9
\nonumber \\
1 \times 9 &= 8 & 2 \times 2 &= 0 & 2 \times 3 &= 1
\nonumber \\
2 \times 4 &= 6 & 2 \times 5 &= 7 & 2 \times 6 &= 4
\nonumber \\
2 \times 7 &= 5 & 2 \times 8 &= 8 & 2 \times 9 &= 9
\nonumber \\
3 \times 3 &= 2 & 3 \times 4 &= 7 & 3 \times 5 &= 4
\nonumber \\
3 \times 6 &= 5 & 3 \times 7 &= 6 & 3 \times 8 &= 9
\nonumber \\
3 \times 9 &= 8 & 4 \times 4 &= 1+8 & 4 \times 5 &= 2+9
\nonumber \\
4 \times 6 &= 3+8 & 4 \times 7 &= 0+9 & 4 \times 8 &= 5+7
\nonumber \\
4 \times 9 &= 4+6 & 5 \times 5 &= 3+8 & 5 \times 6 &= 0+9
\nonumber \\
5 \times 7 &= 1+8 & 5 \times 8 &= 4+6 & 5 \times 9 &= 5+7
\nonumber \\
6 \times 6 &= 1+8 & 6 \times 7 &= 2+9 & 6 \times 8 &= 5+7
\nonumber \\
6 \times 9 &= 4+6 & 7 \times 7 &= 3+8 & 7 \times 8 &= 4+6
\nonumber \\
7 \times 9 &= 5+7 & 8 \times 8 &= 0+2+9 & 8 \times 9 &= 1+3+8
\nonumber \\
& & 9 \times 9 &= 0+2+9
\end{align}
Here index $I=$ 8 or $I=$ 9 does not correspond to a single quasiparticle.
They each actually corresponds to a class of quasiparticles
parameterized by a continuous parameter $\eta$.  We can use $(8,\eta)$
and $(9,\eta)$ to uniquely label those quasiparticles.  Thus, for
example, the fusion rule $8 \times 9 = 1+3+8$ should to interpreted as
$(8,\eta) \times (9,\eta') = 1+3+(8,\eta'')$, for some $\eta$,
$\eta'$, and $\eta''$.

The fusion algebra between the non-Abelian classes of quasiparticles
is
\begin{align}
0_\text{na} \times 0_\text{na} &= 0_\text{na} & 0_\text{na} \times
1_\text{na} &= 1_\text{na}
\\
0_\text{na} \times 2_\text{na} &= 2_\text{na} & 1_\text{na} \times
1_\text{na} &= 0_\text{na}+2_\text{na}
\nonumber\\
1_\text{na} \times 2_\text{na} &= 1_\text{na}+1_\text{na} &
2_\text{na} \times 2_\text{na} &=
0_\text{na}+0_\text{na}+2_\text{na} \nonumber
\end{align}
where the relation between $I$ and $I_\text{na}$ is given in table
\ref{Z2Z2qppoz}.

\subsection{Quasiparticles in the $Z_3|Z_3$ state}

The $Z_3|Z_3$ state is described by the following pattern of zeros:
\begin{align}
 &n =3:\ \ \ \ \ \ (Z_3|Z_3\text{ state})
\nonumber\\
 &\{m;\hsc{1}..\hsc{n-1}\}=\{4;\frac43\ \frac43 \}
\nonumber\\
 &\ \ \ \ \{n_0..n_{m-1}\} = \{3\ 0\ 0\ 0\} .
\end{align}
Here we have $n=3,~p=2,~M_1=M_2=0$ and thus $\hsc{1}=\hsc2=\frac43$.
As shown in section \ref{Z_n^2series} we have
\begin{align}
C_{1,1}C_{2,2}=\frac49(\frac8c-1)
\end{align}
as the only extra consistent condition of this vertex algebra from
useful GJI's. We can use two free parameters $\{c,\la\}$ to express
the structure constants:
\begin{align}\label{Z3Z3Cstr}
C_{1,1}=\la,~~~C_{2,2}=\frac{4}{9\la}(\frac8c-1).
\end{align}
However, using the equivalence transformation (see \eqn{CtC})
\begin{align}
 \psi_1 \to \chi \psi_1,\ \ \
 \psi_2 \to \chi^{-1} \psi_2,\ \ \
\la \to \la /\chi^3,
\end{align}
we can set $\la=1$.  So the infinite $Z_3|Z_3$ simple-current vertex
algebras are parameterized by only a single real number $c$.

There are 5 classes of non-Abelian quasiparticles as
shown in TABLE \ref{Z3Z3qppoz}. We shall study these 5 classes one
by one in this section.
The 1st class is the trivial one, represented by the data
\begin{align}
\{\ksc1,\cdots,\ksc{n};Q_\ga\}=\{\frac43,0,-\frac43;0\}.
\end{align}
With
$\Delta_3(1,1,\ga)=\Delta_3(1,1,\ga+1)=\Delta_3(2,2,\ga)=\Delta_3(2,2,\ga+2)=0$
we have for the structure constants:
\begin{align}
C_{1,\ga+1}=C_{1,1},~C_{2,\ga+2}=C_{2,2},~C_{1,\ga+2}=C_{2,\ga+1}.
\end{align}
Then with $\Delta_3(1,2,\ga)=0$ we have
\begin{align}
C_{1,\ga+2}=C_{2,\ga+1}=1,~~\hsc{\ga}=0,~~\partial\si_{\ga}=0.
\end{align}
which dictates that this is a trivial quasiparticle, proportional to
the identity operator.

The 2nd class is represented by the data
\begin{align}
\{\ksc1,\cdots,\ksc{n};Q_\ga\}=\{1,-\frac13,-\frac23;\frac14\}.
\end{align}
With $\Delta_3(1,1,\ga)=\Delta_3(2,2,\ga+2)=0$ and
$\Delta_3(1,1,\ga+1)=\Delta_3(2,2,\ga)=1$ we have for the structure
constants:
\begin{align}
C_{1,\ga+1}=C_{1,1},~C_{2,\ga+2}=C_{2,2}/2,~C_{2,\ga+1}=2C_{1,\ga+2}.
\end{align}
Then with $\Delta_3(1,2,\ga)=1$ and $\Delta_3(1,2,\ga+2)=3$ we have
\begin{align}
&C_{1,\ga+2}=\frac13,~~~C_{2,\ga+1}=\frac23,
\nonumber \\
&\hsc{\ga}=\frac{c}{24},~~~\hsc{\ga+2}=c\frac{1+3C_{1,\ga+1}C_{2,\ga+2}}{8}=\frac{c}{24}+\frac23.
\end{align}
after using the structure constants (\ref{Z3Z3Cstr}). The above
results from GJI's are consistent with (\ref{hgaahga}).

The 3rd class is represented by the data
\begin{align}
\{\ksc1,\cdots,\ksc{n};Q_\ga\}=\{\frac23,\frac13,-1;\frac12\}.
\end{align}
With $\Delta_3(1,1,\ga)=\Delta_3(2,2,\ga+2)=1$ and
$\Delta_3(1,1,\ga+1)=\Delta_3(2,2,\ga)=0$ we have for the structure
constants:
\begin{align}
C_{1,\ga+1}=C_{1,1}/2,~C_{2,\ga+2}=C_{2,2},~C_{1,\ga+2}=2C_{2,\ga+1}.
\end{align}
Then with $\Delta_3(1,2,\ga)=1$ and $\Delta_3(1,2,\ga+1)=3$ we have
\begin{align}
C_{1,\ga+2}=\frac23,~~~C_{2,\ga+1}=\frac13,
\nonumber \\
\hsc{\ga}=\frac{c}{24},~~~\hsc{\ga+1}=\frac{c}{24}+\frac23.
\end{align}
where we have used (\ref{Z3Z3Cstr}) in calculating $\hsc{\ga+1}$ as
well.

The 4th class is represented by the data
\begin{align}
\{\ksc1,\cdots,\ksc{n};Q_\ga\}=\{\frac23,-\frac23,0;\frac12\}.
\end{align}
With $\Delta_3(1,1,\ga)=\Delta_3(2,2,\ga+2)=0$ we have for the
structure constants:
\begin{align}
C_{1,\ga+1}=C_{1,1},~~C_{2,\ga+2}C_{2,\ga+1}=C_{2,2}C_{1,\ga+2}.
\end{align}
Then with $\Delta_3(1,2,\ga)=\Delta_3(1,2,\ga+2)=2$ we have the
following consistent conditions:
\begin{align}
C_{2,\ga+1}=C_{1,\ga+2}+\frac13=C_{1,\ga+1}C_{2,\ga+2}
\nonumber \\
\hsc\ga=\hsc{\ga+2}=-\frac{c}{12}+\frac{3c}{8}C_{2,\ga+1}
\end{align}
Solving the above nonlinear equations gives us the structure
constants and quasiparticle scaling dimensions:
\begin{align}
&C_{1,\ga+1}=C_{1,1},~~~C_{1,\ga+2}=C_{2,\ga+1}-\frac13,
\nonumber \\
&C_{2,\ga+1}=-\frac29+\frac{16}{9c}\pm\frac{4\sqrt{(c-2)(c-8)}}{9c},
\nonumber \\
&C_{2,\ga+2}=C_{2,2}(C_{2,\ga+1}-\frac13)/C_{2,\ga+1},
\nonumber \\
\label{etapm} &\hsc{\ga}=\hsc{\ga+2}=\frac23-\frac c6 \pm
\frac{\sqrt{(c-2)(c-8)}}{6} \equiv \eta_\pm
\end{align}
$\pm$ corresponds to two different branches of solutions. Here
$c\leq2$ or $c\geq8$ is required to guarantee the scaling dimension
$\hsc\ga$ to be a real number.

For a $Z_3|Z_3$ simple-current algebra described by a fixed $c$ and
a quasiparticle pattern of zeros indexed by $I=$12, 13, 14, or 15,
there are two sets of quasiparticle structure constants that satisfy
all the consistent conditions for the GJI.  This implies that the
index $I=$12, 13, 14, 15 in the table \ref{Z3Z3qppoz} each actually
corresponds to two types of quasiparticles parameterized by the two
sets of structure constants.  Those quasiparticles are uniquely
labeled by $(I,+)$ and $(I,-)$, $I=$12, 13, 14, 15.  When $c=2$ or
$c=8$, then there is only one type of quasiparticle for each  $I=$12,
13, 14, 15.

\begin{table}[tb]
\begin{tabular}{|c|c||c|rrr|c|c|}
\hline $I$ & $I_\text{na}$ & $n_{\ga;0..m-1}$ &
\multicolumn{3}{|c|}{$n\ksc{1..n}$} & $Q$ & $h^\text{sc}+h^\text{ga}$\\
\hline
0 & 0$_\text{na}$ & 3 0 0 0 & $4$ & $0$ & $-4$  & 0   & $0+0$\\
1 & 0$_\text{na}$ & 0 3 0 0 & $4$ & $0$ & $-4$  & 3/4 & $0+\frac{3}{8}$ \\
2 & 0$_\text{na}$ & 0 0 3 0 & $4$ & $0$ & $-4$  & 3/2 & $0+\frac{3}{2}$ \\
3 & 0$_\text{na}$ & 0 0 0 3 & $4$ & $0$ & $-4$  & 9/4 & $0+\frac{27}{8}$ \\
\hline
4 & 1$_\text{na}$ & 2 1 0 0 & $3$ & $-1$ & $-2$ & 1/4 & $\frac{c}{24}+\frac{1}{24}$ \\
5 & 1$_\text{na}$ & 0 2 1 0 & $3$ & $-1$ & $-2$ & 1   & $\frac{c}{24}+\frac{2}{3}$ \\
6 & 1$_\text{na}$ & 0 0 2 1 & $3$ & $-1$ & $-2$ & 7/4 & $\frac{c}{24}+\frac{49}{24}$ \\
7 & 1$_\text{na}$ & 1 0 0 2 & $-2$ & $3$ & $-1$ & 3/2 & $(\frac{c}{24}+\frac23)+\frac{3}{2}$ \\
\hline
8 & 2$_\text{na}$ & 1 2 0 0 & $2$ & $1$ & $-3$  & 1/2 & $\frac{c}{24}+\frac{1}{6}$ \\
9 & 2$_\text{na}$ & 0 1 2 0 & $2$ & $1$ & $-3$  & 5/4 & $\frac{c}{24}+\frac{25}{24}$ \\
10& 2$_\text{na}$ & 0 0 1 2 & $2$ & $1$ & $-3$  & 2   & $\frac{c}{24}+\frac{8}{3}$ \\
11& 2$_\text{na}$ & 2 0 0 1 & $1$ & $-3$ & $2$  & 3/4 & $(\frac{c}{24}+\frac23)+\frac{3}{8}$ \\
\hline
12& 3$_\text{na}$ & 2 0 1 0 & $2$ & $-2$ & $0$  & 1/2 & $\eta_\pm+\frac{1}{6}$ \\
13& 3$_\text{na}$ & 0 2 0 1 & $2$ & $-2$ & $0$  & 5/4 & $\eta_\pm+\frac{25}{24}$ \\
14& 3$_\text{na}$ & 1 0 2 0 & $0$ & $2$ & $-2$  & 1   & $\eta_\pm+\frac{2}{3}$ \\
15& 3$_\text{na}$ & 0 1 0 2 & $0$ & $2$ & $-2$  & 7/4 & $\eta_\pm+\frac{49}{24}$ \\
\hline
16& 4$_\text{na}$ & 1 1 1 0 & $1$ & $0$ & $-1$  & 3/4 & $\eta+\frac{3}{8}$ \\
17& 4$_\text{na}$ & 0 1 1 1 & $1$ & $0$ & $-1$  & 3/2 & $\eta+\frac{3}{2}$ \\
18& 4$_\text{na}$ & 1 0 1 1 & $-1$ & $1$ & $0$  & 5/4 & $(\eta+\frac13)+\frac{25}{24}$ \\
19& 4$_\text{na}$ & 1 1 0 1 & $0$ & $-1$ & $1$  & 1   & $(\eta+\frac13)+\frac{2}{3}$ \\
\hline
\end{tabular}
\caption{ \label{Z3Z3qppoz} The pattern of zeros and the charges $Q$
for the quasiparticles in the $Z_3|Z_3$ parafermion state
parameterized by $\{c,\la\}$.  Note that the quasiparticle quantum
numbers do not depend on the second parameter $\la$.  The
quasiparticles are labeled by the index $I$. The scaling dimensions
of the quasiparticle operators are sums of the contributions from
the simple-current vertex algebra and the Gaussian model:
$h_\ga=h^\text{sc}+h^\text{ga}$, where $\eta_\pm$ is given by
\eqn{etapm}.  Note the index $I=$16, 17, 18, 19 each actually
corresponds to a class of quasiparticles parameterized by a
continuous parameter $\eta$.  Similarly the index $I=$12, 13, 14, 15
each corresponds to two types of quasiparticles parameterized by
$\pm$.}
\end{table}

The 5th class is represented by the data
\begin{align}
\{\ksc1,\cdots,\ksc{n};Q_\ga\}=\{\frac13,0,-\frac13;\frac34\}.
\end{align}
With
$\Delta_3(1,1,\ga)=\Delta_3(1,1,\ga+1)=\Delta_3(2,2,\ga)=\Delta_3(2,2,\ga+2)=1$
we have for the structure constants:
\begin{align}
C_{1,\ga+1}=C_{1,1}/2,~C_{2,\ga+2}=C_{2,2}/2,~C_{1,\ga+2}=C_{2,\ga+1}.
\end{align}
Then with $\Delta_3(1,2,\ga)=2$ and
$\Delta_3(1,2,\ga+1)=\Delta_3(1,2,\ga+2)=3$ we have
\begin{align}
&\hsc\ga=\frac{3c}{8}C_{1,\ga+2}\equiv \eta,
\nonumber \\
&\hsc{\ga+1}=\hsc{\ga+2}=\frac{3c}{8}(C_{1,\ga+2}+C_{1,\ga+1}C_{2,\ga+2})+\frac{c}{24}\nonumber\\
&=\frac{3c}{8}C_{1,\ga+2}+\frac13.
\end{align}
where we have used (\ref{Z3Z3Cstr}).  Just like the $Z_2|Z_2$ states,
there are infinite sets of quasiparticles structure constants the
satisfy the consistent conditions.  Those sets of structure constants
is parameterized by a single real number
$\eta=\frac{3c}{8}C_{1,\ga+2}=\frac{3c}{8}C_{2,\ga+1}$.  This implies
that the index $I=$16, 17, 18, 19 in table \ref{Z3Z3qppoz} each
corresponds to a class of quasiparticles parameterized by a continuous
parameter $\eta$.  Those quasiparticles are uniquely labeled by
$(I,\eta)$, $I=$16, 17, 18, 19. We see that there are infinite types
of quasiparticles in the $Z_3|Z_3$ state, suggesting that the
$Z_3|Z_3$ state is gapless for the ideal Hamiltonian introduced
in \Ref{poz1,S3}.

\subsection{The $Z_2|Z_2|Z_2$ state}

This $Z_2$ simple-current state is described by the pattern of
zeros:
\begin{align}
 &n =2:\ \ \ (Z_2|Z_2|Z_2\text{ state})
\nonumber\\
 &\{m;\hsc{1}..\hsc{n-1}\}=\{6;3/2 \}
\nonumber\\
 &\{p;M_{1}..M_{n-1}\}=\{3; 0\}
\nonumber\\
 &\ \ \ \ \{n_0..n_{m-1}\} = \{2\ 0\ 0\ 0\ 0\ 0\} .
\end{align}

Since there are no structure constants for a $Z_2$ vertex algebra
after choosing the proper normalization, the only free parameter in
this simple-current vertex algebra is the central charge $c$. However,
since $\Delta_3(1,1,1)=6$ in this case, consistent conditions from
GJI's cannot fix the central charge according to section
\ref{consistency:n/2,n/2,n/2}.

Explicit calculations of simple currents correlation functions suggest
that the electron wave functions uniquely depends on the central
charge $c$. We like to stress that the $Z_2|Z_2|Z_2$ state provides an
interesting example that the vertex algebra is not determined by the
structure constants $C_{ab}$ of the leading terms, but by a structure
constant of a subleading term.

\begin{table}[tb]
\begin{tabular}{|c|c||c|rr|c|}
\hline $I$ & $I_\text{na}$ & $n_{\ga;0..m-1}$ &
\multicolumn{2}{|c|}{$n\ksc{1..n}$} & $Q$ \\
\hline
  0 &$0_\text{na}$ & 2 0 0 0 0 0 & $3$ & $-3$ & 0   \\
  1 &$0_\text{na}$ & 0 2 0 0 0 0 & $3$ & $-3$ & 1/3   \\
  2 &$0_\text{na}$ & 0 0 2 0 0 0 & $3$ & $-3$ & 2/3   \\
  3 &$0_\text{na}$ & 0 0 0 2 0 0 & $3$ & $-3$ & 1   \\
  4 &$0_\text{na}$ & 0 0 0 0 2 0 & $3$ & $-3$ & 4/3   \\
  5 &$0_\text{na}$ & 0 0 0 0 0 2 & $3$ & $-3$ & 5/3   \\
\hline
  6 &$1_\text{na}$ & 1 1 0 0 0 0 & $2$ & $-2$ & 1/6   \\
  7 &$1_\text{na}$ & 0 1 1 0 0 0 & $2$ & $-2$ & 1/2   \\
  8 &$1_\text{na}$ & 0 0 1 1 0 0 & $2$ & $-2$ & 5/6   \\
  9 &$1_\text{na}$ & 0 0 0 1 1 0 & $2$ & $-2$ & 7/6   \\
 10 &$1_\text{na}$ & 0 0 0 0 1 1 & $2$ & $-2$ & 3/2   \\
 11 &$1_\text{na}$ & 1 0 0 0 0 1 & $-2$ & $2$ & 5/6   \\
\hline
 12 &$2_\text{na}$ & 1 0 1 0 0 0 & $1$ & $-1$ & 1/3   \\
 13 &$2_\text{na}$ & 0 1 0 1 0 0 & $1$ & $-1$ & 2/3   \\
 14 &$2_\text{na}$ & 0 0 1 0 1 0 & $1$ & $-1$ & 1   \\
 15 &$2_\text{na}$ & 0 0 0 1 0 1 & $1$ & $-1$ & 4/3   \\
 16 &$2_\text{na}$ & 1 0 0 0 1 0 & $-1$ & $1$ & 2/3   \\
 17 &$2_\text{na}$ & 0 1 0 0 0 1 & $-1$ & $1$ & 1   \\
\hline
 18 &$3_\text{na}$ & 1 0 0 1 0 0 & $0$ & $0$ & 1/2   \\
 19 &$3_\text{na}$ & 0 1 0 0 1 0 & $0$ & $0$ & 5/6   \\
 20 &$3_\text{na}$ & 0 0 1 0 0 1 & $0$ & $0$ & 7/6   \\
\hline
\end{tabular}
\caption{
\label{Z2Z2Z2qppoz}
The pattern of zeros and the charges $Q$
for the quasiparticles in the $Z_2|Z_2|Z_2$ state. The quasiparticles are
labeled by the index $I$.
}
\end{table}

In table \ref{Z2Z2Z2qppoz}, we list 21 distinct quasiparticle patterns of
zeros which give rise to at least 21 different quasiparticles.
Those quasiparticles group into 4 classes of non-Abelian quasiparticles.

\subsection{Gaffnian: a non unitary $Z_4$ example}

A $Z_4$ solution
$\{m;\hsc{1},\cdots,\hsc{n-1}\}=\{6;\frac34,0,\frac34\}$ is called
Gaffnian in literature\cite{Gaffnian}. It has the following
commutation factors:
\begin{align}
\mu_{1,2}=\mu_{1,3}=\mu_{2,3}=-1
\end{align}
Therefore it is a generalized $Z_4$ simple-current vertex algebra.

With $\Delta_3(2,2,2)=0$ we know from section
\ref{consistency:n/2,n/2,n/2} that
\begin{align}
\partial\psi_2\equiv0
\end{align}
Since $\Delta_3(1,1,2)=\Delta_3(2,3,3)=0$  we know from section
\ref{consistency:a,b,c} that
\begin{align}
C_{1,1}=C_{1,2}=-C_{2,1}
\nonumber \\
C_{3,3}=C_{2,3}=-C_{3,2}
\nonumber \\
C_{3,1}=-1,~~~C_{1,3}=1
\end{align}
With $\Delta_3(1,1,3)=\Delta_3(1,3,3)=3$ we know from section
\ref{consistency:a,b,-b} that
\begin{align}
c=\frac{-2(\hsc1)^2}{(2\hsc1-3)(\hsc1-2)}=-\frac{3}{5}
\nonumber \\
C_{1,1}C_{2,3}=C_{3,3}C_{1,2}=-1
\end{align}
Other useful GJI's with
$\Delta_3(1,2,3)=\Delta_3(2,2,3)=\Delta_3(1,2,2)=0$ in section
\ref{consistency:a,b,-b} and $\Delta_3(1,1,1)=\Delta_3(3,3,3)=3$ in
section \ref{consistency:a,b,c} don't produce any new conditions.
Further calculations show that even introducing the subleading order
OPE \eq{OPE:i,j:1st} and applying new conditions in Appendix
\ref{app:subleading} wouldn't not supply any extra conditions. In
summary we have
\begin{align}
c=-\frac35,~~~\partial\psi_2\equiv0
\nonumber \\
C_{1,1}=C_{1,2}=-C_{2,1}=\lambda\neq0
\nonumber \\
C_{3,3}=C_{2,3}=-C_{3,2}=-\lambda^{-1}
\nonumber \\
C_{1,3}=1,~~~C_{3,1}=-1
\end{align}
for this $Z_4$ simple-current vertex algebra, which corresponds to
the Gaffnian wave function.
Using the equivalence transformation (see \eqn{CtC})
\begin{align}
 (\psi_1,\psi_2,\psi_3) \to
 (\chi \psi_1,\psi_2,\chi^{-1}\psi_3),\ \ \ \
\la \to \la \chi^{-2}
\end{align}
we can set $\la=1$. So there is only a single Gaffnian wave function.

\begin{table}[tb]
\begin{tabular}{|c|c||c|rrrr|c|c|}
\hline $I$ & $I_\text{na}$ & $n_{\ga;0..m-1}$ &
\multicolumn{4}{|c|}{$n\ksc{1..n}$} & $Q$ & $h^\text{sc}+h^\text{ga}$\\
\hline
0 & 0$_\text{na}$ & 2 0 0 2 0 0 & $3$ & $-3$ & $3$ & $-3$ & 0 & $0+0$\\
1 & 0$_\text{na}$ & 0 2 0 0 2 0 & $3$ & $-3$ & $3$ & $-3$ & 2/3 &$0+\frac{1}{3}$ \\
2 & 0$_\text{na}$ & 0 0 2 0 0 2 & $3$ & $-3$ & $3$ & $-3$ & 4/3 &$0+\frac{4}{3}$ \\
\hline
3 & 1$_\text{na}$ & 1 1 0 1 1 0 & $1$ & $-1$ & $1$ & $-1$ & 1/3 &$-\frac1{20}+\frac{1}{12}$ \\
4 & 1$_\text{na}$ & 0 1 1 0 1 1 & $1$ & $-1$ & $1$ & $-1$ & 1   &$-\frac1{20}+\frac{3}{4}$ \\
5 & 1$_\text{na}$ & 1 0 1 1 0 1 & $-1$ & $1$ & $-1$ & $1$ & 2/3 &$\frac15+\frac{1}{3}$ \\
\hline
\end{tabular}
\caption{ \label{Gafqppoz} The pattern of zeros and the charges $Q$
for the quasiparticles in the Gaffnian state. The quasiparticles are
labeled by the index $I$. The scaling dimensions of the
quasiparticle operators are sums of the contributions from the
simple-current vertex algebra and the Gaussian model:
$h_\ga=h^\text{sc}+h^\text{ga}$. }
\end{table}

Gaffnian state
$\{m;\hsc{1},\cdots,\hsc{n-1}\}=\{6;\frac34,0,\frac34\}$
has two families of different quasiparticles according to conditions
\eq{concave1qp tlsc} and \eq{concave2qp tlsc} (see table
\ref{Gafqppoz}). The 1st family has the following representative:
$\{\ksc1,\cdots,\ksc{n};Q_\ga\}=\{\frac34,-\frac34,\frac34,-\frac34;0\}$.
With $\Delta_3(1,1,\ga)=\Delta_3(1,1,\ga+2)=\Delta_3(1,2,\ga+a)
=\Delta_3(2,3,\ga+a)=\Delta_3(3,3,\ga)=\Delta_3(3,3,\ga+2)=0$ we
obtain all the structure constants from section
\ref{consistency:a,b,ga+c} or Appendix
\ref{consistency:a,b,ga+c:1st}
\begin{align}
C_{1,\ga+1}=C_{1,\ga+2}=-C_{2,\ga+1}=\lambda,
\nonumber \\
C_{1,\ga+3}=C_{2,\ga+2}=-C_{3,\ga+1}=1,
\nonumber \\
C_{2,\ga+3}=C_{3,\ga+3}=-C_{3,\ga+2}=-\lambda^{-1}.
\end{align}

With
$\Delta_3(1,3,\ga)=1=\Delta_3(1,3,\ga+2),~\Delta_3(1,3,\ga+1)=\Delta_3(1,3,\ga+3)=2$
and $\Delta_3(2,2,\ga+a)=0$ we have the quasiparticle scaling
dimensions from section \ref{consistency:a,-a,ga+b}
\begin{align}
\hsc\ga=\hsc{\ga+2}=0,~~\hsc{\ga+1}=\hsc{\ga+3}=\frac34,
\nonumber \\
\partial\si_\ga=\partial\si_{\ga+2}\equiv0.
\end{align}
Since $C_{a,b}=C_{a,\ga+b}$ and $\partial\si_\ga=0$ here, we know
this quasiparticle $\si_\ga$ must be proportional to the identity
operator $1$.

The 2nd family has the following representative:
$\{\ksc1,\cdots,\ksc{n};Q_\ga\}=\{\frac14,-\frac14,\frac14,-\frac14;\frac13\}$.
With
$\Delta_3(1,1,\ga)=\Delta_3(1,1,\ga+2)=\Delta_3(3,3,\ga)=\Delta_3(3,3,\ga+2)=1$
and $\Delta_3(1,2,\ga+a)=\Delta_3(2,3,\ga+a)=0$ we obtain all the
structure constants from section \ref{consistency:a,b,ga+c} or
Appendix \ref{consistency:a,b,ga+c:1st}
\begin{align}
C_{1,\ga+1}=\lambda/2,~~C_{3,\ga+3}=-\lambda^{-1}/2,
\nonumber \\
C_{1,\ga+3}=-C_{3,\ga+1}=1/2,~~C_{2,\ga+2}=1,
\nonumber \\
C_{1,\ga+2}=-C_{2,\ga+1}=\lambda,
\nonumber \\
C_{3,\ga+2}=-C_{2,\ga+3}=\lambda^{-1}.
\end{align}
With $\Delta_3(1,3,\ga)=\Delta_3(1,3,\ga+2)=\Delta_3(2,2,\ga+a)=0$
and $\Delta_3(1,3,\ga+1)=\Delta_3(1,3,\ga+3)=3$ we have the
quasiparticle scaling dimensions from section
\ref{consistency:a,-a,ga+b}
\begin{align}
\hsc\ga=\hsc{\ga+2}=-\frac1{20},~~\hsc{\ga+1}=\hsc{\ga+3}=\frac15
\end{align}
and the structure constants are consistent with all the useful
GJI's. Apparently this quasiparticle is a nontrivial one.

Using the method in \Ref{poz3}, we obtain the full fusion algebra
between the quasiparticles (expressed in terms of the index $I$ in
table \ref{Gafqppoz}):
\begin{align}
0 \times 0 &= 0 & 0 \times 1 &= 1 & 0 \times 2 &= 2
\nonumber\\
0 \times 3 &= 3 & 0 \times 4 &= 4 & 0 \times 5 &= 5
\nonumber\\
1 \times 1 &= 2 & 1 \times 2 &= 0 & 1 \times 3 &= 4
\nonumber\\
1 \times 4 &= 5 & 1 \times 5 &= 3 & 2 \times 2 &= 1
\nonumber\\
2 \times 3 &= 5 & 2 \times 4 &= 3 & 2 \times 5 &= 4
\nonumber\\
3 \times 3 &= 1+5 & 3 \times 4 &= 2+3 & 3 \times 5 &= 0+4
\nonumber\\
4 \times 4 &= 0+4 & 4 \times 5 &= 1+5 & 5 \times 5 &= 2+3
\end{align}
The fusion algebra between the non-Abelian classes of quasiparticles
is
\begin{align}
0_\text{na} \times 0_\text{na} &= 0_\text{na} & 0_\text{na} \times
1_\text{na} &= 1_\text{na}
\nonumber\\
1_\text{na} \times 1_\text{na} &= 0_\text{na}+1_\text{na} .
\end{align}

\subsection{The $Z_4|Z_2$ state}

This solution
$\{m;\hsc{1},\cdots,\hsc{n-1}\}=\{6;\frac54,1,\frac54\}$ is a direct
product of a $n=4$ Pfaffian state
$\{m;\hsc{1},\cdots,\hsc{n-1}\}=\{4;\frac12,0,\frac12\}$ and a $Z_4$
parafermion state
$\{m;\hsc{1},\cdots,\hsc{n-1}\}=\{2;\frac32,1,\frac32\}$.

In this case we have
\begin{align}
p=3,\ \ \ M_1=M_3=1,\ \ \ M_2=2
\end{align}
It's easy to verify that $\mu_{i,j}=1$ and thus $C_{i,j}=C_{j,i}$.

{}From section \ref{consistency:n/2,n/2,n/2} we see that
$\Delta_3(2,2,2)=4$ determines the central charge
\begin{align}
c=1
\end{align}
Then with $\Delta_3(1,1,2)=\Delta_3(2,3,3)=0$ we know from section
\ref{consistency:a,b,c} that
\begin{align}
C_{1,1}=C_{1,2}=C_{2,1}
\nonumber \\
C_{3,3}=C_{2,3}=C_{3,2}
\end{align}
$\Delta_3(1,1,3)=\Delta_3(1,3,3)=4$ in section
\ref{consistency:a,a,-a} and
$\Delta_3(1,2,3)=\Delta_3(1,2,2)=\Delta_3(2,2,3)=2$ in section
\ref{consistency:a,b,-b} both lead to the following conclusions:
\begin{align}
C_{1,1}C_{2,3}=C_{1,2}C_{3,3}=\frac{5}{2c}=-\frac58+\frac{25}{8c}=\frac52
\end{align}
Note that $\Delta_3(1,1,1)=\Delta_3(2,2,2)=2$ doesn't bring us any
new constraints. Further studies after introducing subleading order
OPE \eq{OPE:i,j:1st} show that there are no new constraints on the
structure constants, so we conclude that:
\begin{align}
c=1,~~~C_{1,3}=C_{3,1}=1
\nonumber \\
C_{1,1}=C_{1,2}=C_{2,1}=\lambda\neq0
\nonumber \\
C_{2,3}=C_{3,3}=C_{3,2}=\frac{5}{2\lambda}
\end{align}
characterizes this $Z_4$ simple-current vertex algebra.
Using the equivalence transformation (see \eqn{CtC})
\begin{align}
 (\psi_1,\psi_2,\psi_3) \to
 (\chi \psi_1,\psi_2,\chi^{-1}\psi_3),\ \ \ \
\la \to \la \chi^{-2}
\end{align}
we can set $\la=1$. So there is only a single $Z_4|Z_2$ simple-current
vertex algebra which correspond to a single FQH wave function.

\begin{table}[t]
\begin{tabular}{|c|c|c|c||c|c|c|c|}
\hline
$I$ & $I_\text{na}$ & $n_{\ga;0..m-1}$ &  $Q$ &
$I$ & $I_\text{na}$ & $n_{\ga;0..m-1}$ &  $Q$ \\
\hline
  0 & 0$_\text{na}$ & 2 0 2 0 0 0 & 0   &
  1 & 0$_\text{na}$ & 0 2 0 2 0 0 & $\frac{2}{3}$   \\
  2 & 0$_\text{na}$ & 0 0 2 0 2 0 & $\frac{4}{3}$  &
  3 & 0$_\text{na}$ & 0 0 0 2 0 2 & 2   \\
  4 & 0$_\text{na}$ & 2 0 0 0 2 0 & $\frac{2}{3}$  &
  5 & 0$_\text{na}$ & 0 2 0 0 0 2 & $\frac{4}{3}$   \\
\hline
  6 & 1$_\text{na}$ & 2 0 1 1 0 0 & $\frac{1}{6}$  &
  7 & 1$_\text{na}$ & 0 2 0 1 1 0 & $\frac{5}{6}$   \\
  8 & 1$_\text{na}$ & 0 0 2 0 1 1 & $\frac{3}{2}$  &
  9 & 1$_\text{na}$ & 1 0 0 2 0 1 & $\frac{7}{6}$   \\
 10 & 1$_\text{na}$ & 1 1 0 0 2 0 & $\frac{5}{6}$  &
 11 & 1$_\text{na}$ & 0 1 1 0 0 2 & $\frac{3}{2}$   \\
\hline
 12 & 2$_\text{na}$ & 1 1 1 1 0 0 & $\frac{1}{3}$   &
 13 & 2$_\text{na}$ & 0 1 1 1 1 0 & 1   \\
 14 & 2$_\text{na}$ & 0 0 1 1 1 1 & $\frac{5}{3}$   &
 15 & 2$_\text{na}$ & 1 0 0 1 1 1 & $\frac{4}{3}$   \\
 16 & 2$_\text{na}$ & 1 1 0 0 1 1 & 1   &
 17 & 2$_\text{na}$ & 1 1 1 0 0 1 & $\frac{2}{3}$   \\
\hline
 18 & 3$_\text{na}$ & 2 0 0 2 0 0 & $\frac{1}{3}$   &
 19 & 3$_\text{na}$ & 0 2 0 0 2 0 & 1   \\
 20 & 3$_\text{na}$ & 0 0 2 0 0 2 & $\frac{5}{3}$
&&&&\\
\hline
 21 & 4$_\text{na}$ & 1 1 0 2 0 0 & $\frac{1}{2}$   &
 22 & 4$_\text{na}$ & 0 1 1 0 2 0 & $\frac{7}{6}$   \\
 23 & 4$_\text{na}$ & 0 0 1 1 0 2 & $\frac{11}{6}$   &
 24 & 4$_\text{na}$ & 2 0 0 1 1 0 & $\frac{1}{2}$   \\
 25 & 4$_\text{na}$ & 0 2 0 0 1 1 & $\frac{7}{6}$   &
 26 & 4$_\text{na}$ & 1 0 2 0 0 1 & $\frac{5}{6}$   \\
\hline
 27 & 5$_\text{na}$ & 2 0 1 0 1 0 & $\frac{1}{3}$   &
 28 & 5$_\text{na}$ & 0 2 0 1 0 1 & 1   \\
 29 & 5$_\text{na}$ & 1 0 2 0 1 0 & $\frac{2}{3}$   &
 30 & 5$_\text{na}$ & 0 1 0 2 0 1 & $\frac{4}{3}$   \\
 31 & 5$_\text{na}$ & 1 0 1 0 2 0 & 1   &
 32 & 5$_\text{na}$ & 0 1 0 1 0 2 & $\frac{5}{3}$   \\
\hline
 33 & 6$_\text{na}$ & 1 1 1 0 1 0 & $\frac{1}{2}$   &
 34 & 6$_\text{na}$ & 0 1 1 1 0 1 & $\frac{7}{6}$   \\
 35 & 6$_\text{na}$ & 1 0 1 1 1 0 & $\frac{5}{6}$   &
 36 & 6$_\text{na}$ & 0 1 0 1 1 1 & $\frac{3}{2}$   \\
 37 & 6$_\text{na}$ & 1 0 1 0 1 1 & $\frac{7}{6}$   &
 38 & 6$_\text{na}$ & 1 1 0 1 0 1 & $\frac{5}{6}$  \\
\hline
 39 & 7$_\text{na}$ & 1 1 0 1 1 0 & $\frac{2}{3}$  &
 40 & 7$_\text{na}$ & 0 1 1 0 1 1 & $\frac{4}{3}$  \\
 41 & 7$_\text{na}$ & 1 0 1 1 0 1 & 1
&&&& \\
\hline
\end{tabular}
\caption{
\label{Z2Z4qppoz}
The pattern of zeros and the charges $Q$
for the quasiparticles in the $Z_4|Z_2$ state. The quasiparticles are
labeled by the index $I$.
}
\end{table}

In table \ref{Z2Z4qppoz}, we list 42 distinct quasiparticle patterns of
zeros which give rise to at least 42 different quasiparticles.
Those quasiparticles group into 8 classes of non-Abelian quasiparticles.

\subsection{$C_n|C_n$ series with
$\{m;\hsc1,\cdots,\hsc{n-1}\}=\{2n;2,\cdots,2\}$}

\begin{table}[t]
\begin{tabular}{|c|c|c|c||c|c|c|c|}
\hline
$I$ & $I_\text{na}$ & $n_{\ga;0..m-1}$ &  $Q$ &
$I$ & $I_\text{na}$ & $n_{\ga;0..m-1}$ &  $Q$ \\
\hline
  0 & 0$_\text{na}$ & 3 0 0 0 0 0 & 0 &
  1 & 0$_\text{na}$ & 0 3 0 0 0 0 & 1/2 \\
  2 & 0$_\text{na}$ & 0 0 3 0 0 0 & 1 &
  3 & 0$_\text{na}$ & 0 0 0 3 0 0 & 3/2 \\
  4 & 0$_\text{na}$ & 0 0 0 0 3 0 & 2 &
  5 & 0$_\text{na}$ & 0 0 0 0 0 3 & 5/2 \\
\hline
  6 & 1$_\text{na}$ & 2 1 0 0 0 0 & 1/6 &
  7 & 1$_\text{na}$ & 0 2 1 0 0 0 & 2/3 \\
  8 & 1$_\text{na}$ & 0 0 2 1 0 0 & 7/6 &
  9 & 1$_\text{na}$ & 0 0 0 2 1 0 & 5/3 \\
 10 & 1$_\text{na}$ & 0 0 0 0 2 1 & 13/6 &
 11 & 1$_\text{na}$ & 1 0 0 0 0 2 & 5/3 \\
\hline
 12 & 2$_\text{na}$ & 1 2 0 0 0 0 & 1/3 &
 13 & 2$_\text{na}$ & 0 1 2 0 0 0 & 5/6 \\
 14 & 2$_\text{na}$ & 0 0 1 2 0 0 & 4/3 &
 15 & 2$_\text{na}$ & 0 0 0 1 2 0 & 11/6 \\
 16 & 2$_\text{na}$ & 0 0 0 0 1 2 & 7/3 &
 17 & 2$_\text{na}$ & 2 0 0 0 0 1 & 5/6 \\
\hline
 18 & 3$_\text{na}$ & 2 0 1 0 0 0 & 1/3 &
 19 & 3$_\text{na}$ & 0 2 0 1 0 0 & 5/6 \\
 20 & 3$_\text{na}$ & 0 0 2 0 1 0 & 4/3 &
 21 & 3$_\text{na}$ & 0 0 0 2 0 1 & 11/6 \\
 22 & 3$_\text{na}$ & 1 0 0 0 2 0 & 4/3 &
 23 & 3$_\text{na}$ & 0 1 0 0 0 2 & 11/6 \\
\hline
 24 & 4$_\text{na}$ & 1 1 1 0 0 0 & 1/2 &
 25 & 4$_\text{na}$ & 0 1 1 1 0 0 & 1 \\
 26 & 4$_\text{na}$ & 0 0 1 1 1 0 & 3/2 &
 27 & 4$_\text{na}$ & 0 0 0 1 1 1 & 2 \\
 28 & 4$_\text{na}$ & 1 0 0 0 1 1 & 3/2 &
 29 & 4$_\text{na}$ & 1 1 0 0 0 1 & 1 \\
\hline
 30 & 5$_\text{na}$ & 1 0 2 0 0 0 & 2/3 &
 31 & 5$_\text{na}$ & 0 1 0 2 0 0 & 7/6 \\
 32 & 5$_\text{na}$ & 0 0 1 0 2 0 & 5/3 &
 33 & 5$_\text{na}$ & 0 0 0 1 0 2 & 13/6 \\
 34 & 5$_\text{na}$ & 2 0 0 0 1 0 & 2/3 &
 35 & 5$_\text{na}$ & 0 2 0 0 0 1 & 7/6 \\
\hline
 36 & 6$_\text{na}$ & 2 0 0 1 0 0 & 1/2 &
 37 & 6$_\text{na}$ & 0 2 0 0 1 0 & 1 \\
 38 & 6$_\text{na}$ & 0 0 2 0 0 1 & 3/2 &
 39 & 6$_\text{na}$ & 1 0 0 2 0 0 & 1 \\
 40 & 6$_\text{na}$ & 0 1 0 0 2 0 & 3/2 &
 41 & 6$_\text{na}$ & 0 0 1 0 0 2 & 2 \\
\hline
 42 & 7$_\text{na}$ & 1 1 0 1 0 0 & 2/3 &
 43 & 7$_\text{na}$ & 0 1 1 0 1 0 & 7/6 \\
 44 & 7$_\text{na}$ & 0 0 1 1 0 1 & 5/3 &
 45 & 7$_\text{na}$ & 1 0 0 1 1 0 & 7/6 \\
 46 & 7$_\text{na}$ & 0 1 0 0 1 1 & 5/3 &
 47 & 7$_\text{na}$ & 1 0 1 0 0 1 & 7/6 \\
\hline
 48 & 8$_\text{na}$ & 1 0 1 1 0 0 & 5/6 &
 49 & 8$_\text{na}$ & 0 1 0 1 1 0 & 4/3 \\
 50 & 8$_\text{na}$ & 0 0 1 0 1 1 & 11/6 &
 51 & 8$_\text{na}$ & 1 0 0 1 0 1 & 4/3 \\
 52 & 8$_\text{na}$ & 1 1 0 0 1 0 & 5/6 &
 53 & 8$_\text{na}$ & 0 1 1 0 0 1 & 4/3 \\
\hline
 54 & 9$_\text{na}$ & 1 0 1 0 1 0 & 1 &
 55 & 9$_\text{na}$ & 0 1 0 1 0 1 & 3/2 \\
\hline
\end{tabular}
\caption{
\label{C32qppoz}
The pattern of zeros and the charges $Q$
for the quasiparticles in the $C_3|C_3$ state
(which also the $Z_3|Z_3|Z_3$ state). The quasiparticles are
labeled by the index $I$.
}
\end{table}

This corresponds to a series of FQH states with filling fraction
$\nu=1/2$ for bosonic electrons (and $\nu=1/3$ for fermionic
electrons). A $C_4|C_4$ example is given in \eqn{C42}.

First, from \eqn{muij} we know that $\mu_{a,b}=1$
for such a $C_n|C_n$ simple-current vertex algebra, since all the simple
current scaling dimensions are even integers and so are all
$\al_{a,b},~\forall a,b$. As a result we have
\begin{align}
C_{a,b}=C_{b,a},\hspace{1cm}\forall a,b\in\dZ
\end{align}

It's straightforward to check that if we don't have the subleading
term \eq{OPE:i,j:1st} in OPE, this solution only has the following
extra consistent conditions shown in section \ref{consistency:a,b,c}
with $\Delta_3(a,b,-a-b)=0,~~a,b,a+b\neq0\mod n$:
\begin{align}
C_{a,b}=C_{a,-a-b}=C_{b,-a-b}
\end{align}
which for sure can be satisfied for all $a,b\in\dZ$.

Now we introduce the subleading OPE term \eq{OPE:i,j:1st} and the
new consistent conditions in Appendix \ref{app:subleading} to see
whether they are satisfied for this vertex algebra. Note that here
we have
\begin{align}
\al_{a,b}=\left \{\begin{aligned}2,\hspace{1cm}&a+b\neq0\mod
n\\4,\hspace{1cm}&a+b=0\mod n\end{aligned}\right.
\end{align}
for any $a,b\neq0\mod n$, and also $d_{a,b}=1/2,~a+b\neq0\mod n$
from \eqn{d_a,b}.

Taking any integers $a,b,c\neq0\mod n$, for this such a $Z_n$ simple
current vertex algebra we have:

$\Delta_3(a,b,c)=2$ for $a+b,b+c,a+c,a+b+c\neq0\mod n$, so according
to Appendix \ref{consistency:a,b,c:1st} we have:
\begin{align}
C_{a,b}C_{a+b,c}=C_{b,c}C_{a,b+c}=C_{a,c}C_{b,a+c}
\end{align}
Then all consistent conditions are satisfied without requiring that
$\partial\psi_a=0$.

$\Delta_3(a,b,c=-a-b)=0$ for $a+b\neq0\mod n$, so according to
Appendix \ref{consistency:a,b,-a-b:1st} we have:
\begin{align}
C_{a,b}=C_{a,-a-b}=C_{b,-a-b}
\end{align}

$\Delta_3(a,b,c=-b)=4$ for $a\pm b\neq0\mod n$, so according to
Appendix \ref{consistency:a,b,-b:1st} we have:
\begin{align}
C_{a,b}C_{a+b,-b}=C_{a,-b}C_{b,a-b}=\frac{8}{c}
\end{align}

$\Delta_3(a,b=a,c=-a)=6$ for $2a\neq0\mod n$, so according to
Appendix \ref{consistency:a,a,-a:1st} we have:
\begin{align}
d_{a,a}=1/2
\end{align}
which is consistent with \eqn{d_a,b}.

$\Delta_3(a=n/2,b=n/2,c=n/2)=8$ for $n=$~even, so according to
section \ref{consistency:n/2,n/2,n/2} there are no extra consistent
conditions.

In summary, this series of solutions
$\{m;\hsc1,\cdots,\hsc{n-1}\}=\{2n;2,\cdots,2\}$ corresponds to a
$Z_n$ simple-current vertex algebra 
satisfying the following consistent conditions:
\begin{align}
&C_{a,b}=C_{b,a},~~~\forall~a,b\in\dZ;\\
&\nonumber C_{a,b}C_{a+b,c}=C_{a,c}C_{b,a+c}=C_{b,c}C_{a,b+c},\\
&\text{if}~~a+b,~b+c,~a+c\neq0\mod n~;\\
&\nonumber C_{a,b}C_{a+b,-b}=C_{a,-b}C_{b,a-b}=\frac{8}{c},\\
&\text{if}~~a\pm b\neq0\mod n~;
\end{align}

By solving the above conditions in the similarly way as with the $Z_n$ and the
$Z_n|Z_n$
series, we find that
\begin{align}
C_{a,b}&=\frac{\prod_{i=1}^{a+b-1}\lambda_i}
{\prod_{i=1}^{a-1}\lambda_i\prod_{j=1}^{b-1}\lambda_j}\sqrt{\frac8c},~~~a+b<n
\\
C_{n-a,n-b}&=\frac{\prod_{i=1}^{a-1}\lambda_i\prod_{j=1}^{b-1}\lambda_j}
{\prod_{i=1}^{a+b-1}\lambda_i}\sqrt{\frac8c},~~~a+b<n
\\
C_{a,n-a}&=1.
\end{align}
where nonzero complex parameters $\{\lambda_i|i=1,2,\cdots,n-2\}$
satisfy the following constraint:
\begin{align}
\lambda_{a-1}=\lambda_{n-a},~~~1\leq a\leq n-1,\ \ \ \ \ \
\la_0=1.
\end{align}
If we choose $ \chi_a=\prod_{i=0}^{a-1}\lambda_{i} $,
the equivalence transformation \eq{CtC} will remove the
$\la_a$ dependent factors in the structure constants.
We find that the $C_n|C_n$ series is characterized by the following
data:
\begin{align}
C_{a,b}&= \sqrt{\frac8c},~~~a+b<n \text{ or } a+b>n
\\
C_{a,n-a}&=1.
\end{align}
Therefore this theory has one free parameters $c$
if $n>2$.

In table \ref{C32qppoz}, we list 56 distinct quasiparticle patterns of zeros
which give rise to at least 56 different quasiparticles for the $C_3|C_3$ state
(or $Z_3|Z_3|Z_3$ state).  Those quasiparticles group into 10 classes of
non-Abelian quasiparticles. 

\section{Summary}

The pattern-of-zeros is a powerful way to characterize FQH
states.\cite{poz1,poz2,poz3} However, the pattern-of-zeros approach is
not quite complete. It is known that some patterns of zeros do not
uniquely describe the FQH states.  As a result, we cannot obtain all
the topological properties of FQH states from the data of pattern of
zeros $\{n;m;S_a\}$.

In this paper, we combine the pattern-of-zero approach with the vertex
algebra approach. We find that we can generalize the data of pattern
of zeros $\{n;m;S_a\}$ to $\{n;m;S_a;c\}$ to completely describe a FQH
state, at least for the many examples discussed in this paper.  Many
consistent conditions on the new set of data $\{n;m;S_a;c\}$ are
obtained from the GJI of the simple-current vertex algebra. Those
consistent conditions are sufficient: \ie if the data $\{n;m;S_a;c\}$
satisfy those conditions, then the data will define a $Z_n$
simple-current vertex algebra and a FQH wave function. Using the new
characterization scheme and the $Z_n$ simple-current vertex algebra,
we can calculate quasiparticle scaling dimensions, fractional
statistics, the central charge of the edge states, as well as many
other properties, from the data $\{n;m;S_a;c\}$.

For example, for the $Z_2$ parafermion state characterized by
pattern of zeros $\{n;m;\hsc{1}..\hsc{n-1}\}=\{2;2;\frac12 \}$, we
find the well known scaling dimensions (the non-Abelian part) $0$,
$\frac12$, and $\frac{1}{16}$ for the three kind of quasiparticles.
For the $Z_2|Z_2$ state characterized by pattern of zeros
$\{n;m;\hsc{1}..\hsc{n-1}\}=\{2;4;1 \}$, we find the scaling
dimensions and the charges for all its quasiparticles (see table
\ref{Z2Z2qppoz}).  We find that the FQH state described by the
$Z_2|Z_2$ simple-current vertex algebra contains infinite types of
quasiparticles and two classes of them are parameterized by a real
parameter. This indicates that the $Z_2|Z_2$ state is gapless for
the ideal Hamiltonian introduced in \Ref{poz1}.

We also studied the
$Z_3|Z_3$ state described by the $Z_3|Z_3$ simple-current vertex
algebra, with the pattern of zeros
$\{n;m;\hsc{1}..\hsc{n-1}\}=\{3;4;\frac43\ \frac43\}$.  Such a state
is also studied in \Ref{S3}.  We show that the $Z_3|Z_3$ state
cannot be completely characterized by the pattern-of-zeros data
$\{n;m;\hsc{1}..\hsc{n-1}\}=\{3;4;\frac43\ \frac43\}$.  We need to
add one more parameter $c$ and use the expanded data
$\{n;m;\hsc{1}..\hsc{n-1};c\}=\{3;4;\frac43\ \frac43;c\}$ to
completely characterize the $Z_3|Z_3$ state.  We find the scaling
dimensions and the charges for all its quasiparticles
(see table \ref{Z3Z3qppoz}). Again there are
infinite types of quasiparticles and four classes of them are
parameterized by a real parameter.  This again suggests that the
$Z_3|Z_3$ state is gapless for the ideal Hamiltonian introduced in
\Ref{poz1,S3}.

The study in this paper is based on the $Z_n$ simple-current vertex
algebra. But the $Z_n$ simple-current vertex algebra makes some
unnecessary assumptions.  It is much more natural to study FQH state
based on the more general $Z_n$ vertex algebra.  This will be a
direction of future exploration.

\begin{acknowledgments}
YML is grateful to Boris Noyvert for many helpful discussions on the
algebraic approach to conformal field theory. This research is
supported by DOE Grant DE-FG02-99ER45747 (YML,ZQW), NSF Grant No.
DMR-0706078 (XGW), and by NSF Grant No. DMS-034772 (ZHW).  XGW is
also supported by Perimeter Institute for Theoretical Physics.
Research at Perimeter Institute is supported by the Government of
Canada through Industry Canada and by the Province of Ontario
through the Ministry of Research \& Innovation.
\end{acknowledgments}

\appendix

\section{Other ways to label the pattern of zeros}

In section \ref{POZ}, we have discussed two ways to label the pattern of
zeros, one in terms of $\{n;m;S_a\}$ and the other in terms of
$\{n;m;h^\text{sc}_a\}$.  In this section, we will introduce two other
more efficient ways to label pattern of zeros.  The new ways of
labeling automatically satisfy more self consistent conditions.

\subsection{Label the pattern of zeros by a set of non-negative
integers $\{a_j\}$}

\label{label:a_j}

Since $\Delta_3(a,b,c)$ in \eqn{hcond3} is just a linear combination
of the $\hsc a$, there are only $n-1$ independent equations among
all the possible $n^3$ choices of $(i,j,k)$ in \eqn{del3con}. A
convenient choice would be $(i,j,k)=(1,1,a),~a=1,2,\cdots,n-1$.
These equations are
\begin{widetext}
\begin{align}
\nonumber\Delta_3(1,1,1)=(2\hsc1-\hsc2)+\hsc1-2\hsc2+\hsc3=a_1\in\dN\\
\nonumber\cdots\\
\nonumber\Delta_3(1,1,j)=(2\hsc1-\hsc2)+\hsc j-2\hsc{j+1}+\hsc{j+2}=a_j\in\dN\\
\nonumber\cdots\\
\nonumber\Delta_3(1,1,n-3)=(2\hsc 1-\hsc 2)+\hsc{n-3}-2\hsc{n-2}+\hsc{n-1}=a_{n-3}\in\dN\\
\nonumber\Delta_3(1,1,n-2)=(2\hsc1-\hsc2)+\hsc{n-2}-2\hsc{n-1}=a_{n-2}\in\dN\\
\nonumber\Delta_3(1,1,n-1)=(2\hsc1-\hsc2)+\hsc{n-1}+\hsc{1}=a_{n-1}\in\dN
\end{align}
\end{widetext} Here we only used the $\hsc{n}=\hsc{0}=0$ condition.
Adding up these equations together we immediately obtain the
following equation:
\begin{equation}
n(2\hsc1-\hsc2)=\sum_{i=1}^{n-1}a_i
\end{equation}
By defining another vector $\{A_j\}$:
\begin{equation}
A_j=a_j-(2\hsc1-\hsc2)=a_j-\frac{1}{n}\sum_{i=1}^{n-1}a_i
\end{equation}
we have a simple relation:
\begin{equation}
\textbf{X}\cdot\textbf{h}^{sc}=\textbf{A}
\end{equation}
where $\textbf{h}^{sc}=(\hsc1,\cdots,\hsc{n-1})^T$ and
$\textbf{A}=(A_1,\cdots,A_{n-1})^T$ are column vectors and the
matrix
\begin{equation}
\textbf{X}=\begin{bmatrix}1&-2&1&0&0&\cdots\\0&1&-2&1&0&\cdots\\\cdots\\\cdots&0&0&1&-2&1\\
\cdots&0&0&0&1&-2\\
1&\cdots&0&0&0&1\end{bmatrix}_{(n-1)\times(n-1)}
\end{equation}
It would be straightforward to check that the above matrix is not
singular and its inverse equals
\begin{equation}
n\cdot\textbf{X}^{-1}_{i,j}=\{\begin{aligned}ij-jn\hspace{1cm}j<i\\ij-(i-1)n\hspace{1cm}j\geq
i\end{aligned}
\end{equation}
So we can express $(\hsc{1},...,\hsc{n-1})$ in terms of the
$(a_1,...,a_{n-1})$:
\begin{align}
&\nonumber\hsc{a}=\sum_{b=1}^{n-1}\textbf{X}^{-1}_{a,b}A_b=\frac{a(n-a)}{2n}\sum_{j=1}^{n-1}a_j\\
&+\frac{a-n}{n}\sum_{j=1}^{n-1}(j+1)a_j+\sum_{j=a}^{n-1}(j+1-a)a_j
\end{align}
Since the date $\{ n;m;h^\text{sc}_1,...,h^\text{sc}_{n-1}\}$ and
$\{ n;m;a_1,...,a_{n-1}\}$ have an one-to-one correspondence, we can
also use $\{ n;m;a_1,...,a_{n-1}\}$ to label the pattern of zeros.
The $\{ n;m;a_1,...,a_{n-1}\}$ labeling scheme is more efficient:
once we choose non-negative integers $\{a_1,...,a_{n-1}\}$, we
generate $\hsc{a}$ that already satisfy a part of \eqn{hcond3}.

{}From the reflection conditions \eq{symmetric4scaling dimension} of
$\{\hsc{a}\}$, we can obtain similar reflection conditions for
$a_j$:
\begin{align}\label{symmetric4aj}
a_{j}=a_{n-2-j}\ \ \ (1\leq j\leq n-3);\ \ \ a_{n-2}=0
\end{align}
then the independent sequence of non-negative integers is actually
$\{a_j,1\leq j\leq[\frac{n}{2}]-1;~a_{n-1}\}$, which contain
$[\frac{n}{2}]$ integers. (We use $[x]$ to denote the biggest
integer no larger than $x$.) The $\{a_j\}$ label of the patterns of
zeros provides us an efficient way to numerically find the solutions
of \eqn{hscperi}, \eqn{hcond1}, \eqn{hcond2}, and \eqn{hcond3}.

\subsection{Label patterns of zeros by $\{M_k;p;m\}$}
\label{pMk}

\subsubsection{The $\{M_k;p;m\}$ labeling scheme}

Using reflection conditions \eq{symmetric4aj} we can define:
\begin{align}\label{p-def}
p & \equiv\frac{1}{2}\sum_{j=1}^{n-1}a_j
\\
&=\frac{a_{n-1}}{2}+\sum_{j=1}^{[\frac{n}{2}]-2}a_j +\left\{
\begin{aligned}
a_{[\frac{n}{2}]-1},&\ \ n=\text{odd} \\
{a_{[\frac{n}{2}]-1}}/{2},&\ \ n=\text{even}
\end{aligned}
\right. \nonumber
\end{align}
and
\begin{align}\label{mk-def}
M_k & \equiv \frac{n-k}{n}\sum_{j=1}^{n-1}(j+1)a_j -
\sum_{j=k}^{n-1}(j+1-k)a_j
\end{align}
It's easy to verify the following reflection condition for $\{M_k\}$
\begin{align}\label{symmetric4mk}
&\nonumber\ \text{for}~~k=1,2,\cdots,[\frac{n}{2}]:\\
&M_k=M_{n-k}=\sum_{j=1}^{k-2}(j+1)a_j
+k\sum_{j=k-1}^{[\frac{n}{2}]-2}a_j\nonumber\\
&\ \ \ \ \ \ \ \ \ \ \  \ \ \ \ +k\cdot\left\{
\begin{aligned}
&a_{[\frac{n}{2}]-1},\ \ &n=\text{odd}\\
&{a_{[\frac{n}{2}]-1}}/{2},\ \ &n=\text{even}
\end{aligned}
\right.
\end{align}
and another important relation
\begin{align}
M_2=2M_1
\end{align}

Then we can express $\{\hsc{a}\}$ in terms of this new set of
independent variables $\{M_k,k=1,3,4,\cdots,[\frac{n}{2}];~p\}$
(also $[\frac{n}{2}]$-dimensional)
\begin{align}\label{hsc-pmk}
\hsc{a}=p\frac{a(n-a)}{n}-M_a
\end{align}

{}From definitions we see that both $p$ and $\{M_k,~k=\text{odd}\}$
can be half integers, while $\{M_k,~k=\text{even}\}$ must be
integers. When $n=\text{odd}$, $\{M_k,~k=\text{odd}\}$ must be
integers too. In fact, the simplest parafermion vertex
algebra\cite{ZF0} (which describes the $Z_n$ parafermion
states\cite{RR99} in a FQH context) corresponds to the case in which
$p=1,~M_k=0$.

Certainly, not all possible choices of $\{M_k;p;m\}$ correspond to
valid patterns of zeros.  Only those that satisfy the conditions
\eqn{hcond1}, \eqn{hcond2}, and \eqn{hcond3} are valid. But
$\{M_k;p;m\}$ labeling scheme is an efficient way to  generate the
valid patterns of zeros.

Now we have two $[\frac{n}{2}]$-dimensional vectors describing the
$\{\hsc{a}\}$: $\{a_j,j=1,2,\cdots,[\frac{n}{2}]-1;~a_{n-1}\}$ and
$\{M_k,k=1,3,4,\cdots,[\frac{n}{2}];~p\}$. The latter is expressed
in terms of the former in \eqn{p-def} and \eqn{symmetric4mk}.
Conversely, we can express the former in terms of the latter in the
following way
\begin{align}
&a_j=-M_j+2M_{j+1}-M_{j+2}\ \ \ j=1,2,\cdots,[\frac{n}{2}]-2
\nonumber \\
&a_{[\frac{n}{2}]-1}=( M_{[\frac{n}{2}]} -M_{[\frac{n}{2}]-1}
)\cdot\left\{
\begin{aligned}
1,\ \ \ &n=\text{odd}\\
2,\ \ \ &n=\text{even}
\end{aligned}\right.
\nonumber \\
&a_{n-1}=2p - 2M_1
\end{align}
The $\{M_k;p;m\}$ label of the pattern of zeros has a close tie to
simple parafermion CFT.

\subsubsection{Consistent conditions on $\{M_k;p;m\}$}

Now we use this new labeling scheme in (\ref{hsc-pmk}) to see what
are the constraints on $\{M_k;p;m\}$, from all the consistent
conditions \eq{hcond1}, \eq{hcond2} and \eq{hcond3} on
$\{\hsc{a};m\}$.

At first, with $M_2=2M_1$ \eq{hcond1} leads to
\begin{eqnarray}
S_2=D_{1,1}=\frac{m-2p}{n}=\text{even}
\end{eqnarray}
therefore we have
\begin{eqnarray}\label{m-cond}
m=2p+2nk_m,~~~k_m\in\dN.
\end{eqnarray}
This determines the electron filling fraction
$\nu_e=(1+m/n)^{-1}=\frac{n}{2p+(2k_m+1)n}$. 
\begin{widetext}
To guarantee
$mn=\text{even}$ with \eq{m-cond} we have another condition on $p$~:
\begin{eqnarray}\label{p-cond}
p\in\dN,~~~\text{if}~n=\text{odd}.
\end{eqnarray}
Since $S_a=\sum_{i=2}^{a-1}D_{i,1}$, \eq{hcond2} naturally
guarantees $S_a\in\dN$. Moreover we have $D_{a+n,b}=D_{a,b}+mb$,
thus we only need to satisfy the following conditions for
\eq{hcond2}:
\begin{eqnarray}\label{Dab-cond}
D_{a,b}=\left\{\begin{aligned}M_a+M_b-M_{a+b}+2k_mab,~~~&0\leq
a,b\leq a+b\leq n\\
M_a+M_b-M_{a+b}+2k_mab+2p(a+b-n),~~~&0<a,b\leq
n<a+b\end{aligned}\right.~~~\in\dN
\end{eqnarray}

Since $\Delta_3(i,j,k),~0\leq i\leq j\leq k<n$ should be
non-negative integers, $M_k$ and $p$ satisfy some additional
conditions as shown in \eq{hcond3}:
\begin{align}
\label{Delta3-ijk} &\ \ \
\Delta_3(i,j,k)
=\hsc i+\hsc j+\hsc k+\hsc{i+j+k}-\hsc{i+j}-\hsc{i+k}-\hsc{j+k}
\nonumber \\
&=\left\{\begin{aligned}
-\Delta{M}[i,j,k]\ \ \ &i+j+k\leq n\\
-\Delta{M}[i,j,k]+2p(i+j+k-n)\ \ \ &j+k\leq n<i+j+k\leq2n\\
-\Delta{M}[i,j,k]+2pi\ \ \ &i+k\leq n<j+k\leq i+j+k\leq2n\\
-\Delta{M}[i,j,k]+2p(n-k)\ \ \ &i+j\leq n<i+k\leq j+k\leq i+j+k\leq2n\\
-\Delta{M}[i,j,k]+2p(2n-i-j-k)\ \ \ &n<i+j\leq i+k\leq j+k\leq i+j+k\leq2n\\
-\Delta{M}[i,j,k]\ \ \ &n<i+j\leq i+k\leq j+k\leq2n<i+j+k
\end{aligned}\right.
\nonumber \\
&\in \dN
\end{align}
where we defined
\begin{equation}
\Delta{M}[i,j,k]=M_i+M_j+M_k+M_{(i+j+k\mod n)}-M_{(i+j\mod
n)}-M_{(i+k\mod n)}-M_{(j+k\mod n)}
\end{equation}
\end{widetext}
By partially solving the consistent conditions \eq{hcond1},
\eq{hcond2} and \eq{hcond3} on $\{\hsc{a};m\}$, we obtain a finite
set of conditions (\ref{m-cond})-(\ref{Delta3-ijk}). They are the
consistent conditions to be satisfied by the pattern of zeros
$\{M_k;p;m\}$, a sequence of integers and half-integers. For
instance, the simplest $Z_n$ parafermion states\cite{RR99,ZF0}
correspond to the pattern of zeros $\{M_k=0;p=1;m=2p=2\}$, by
choosing non-negative integer $k_m=0$.

\section{Consistent conditions on the commutation factor $\mu_{AB}$}
\label{app:muABcon}

To introduce some useful notations, let us write the OPE between two
generic operators $A(z)$ and $B(w)$ as the follwoing\cite{va0}:
\begin{align}
\label{OPE-general}
A(z)B(w)&=\frac{1}{(z-w)^{\alpha_{AB}}}\Big([AB]_{\alpha_{AB}}(w)
\nonumber\\
&\ \ \ \ \ +(z-w)[AB]_{\alpha_{AB}-1}(w)
\nonumber\\
&\ \ \ \ \ +(z-w)^2[AB]_{\alpha_{AB}-2}(w)+\cdots\Big)
\end{align}
where
\begin{equation}
\alpha_{AB}=h_A+h_B-h_{[AB]_{\alpha_{AB}}} ,
\end{equation}
and $h_A$ ia the scaling dimension of operator $A$. $A$ and $B$
satisfy the following commutation relation
\begin{equation}
\label{muAB} (z-w)^{\alpha_{AB}}A(z)B(w) =
\mu_{AB}(w-z)^{\alpha_{AB}}B(w)A(z) .
\end{equation}

Let us derive some conditions on $\mu_{AB}$ from the associativity
of the vertex algebra. By exchanging $A$ and  $B$ twice we have
\begin{equation}\label{cf0}
\mu_{AB}\mu_{BA}=1
\end{equation}
which immediately leads to
\begin{equation}\label{cf00}
\mu_{AA}=1
\end{equation}
$\mu_{AA}\neq-1$ because the leading term in the OPE of two $A$
fields would vanish otherwise.

Let
\begin{equation}
B(z)C(w)=\frac{D(w)}{(z-w)^{\alpha_{BC}}}+\cdots
\end{equation}
To exchange $A$ with $B$ and then with $C$ is equivalent to exchange
$A$ with $D=[BC]_{\alpha_{BC}}$, so we have
\begin{align}
\label{cf1} & \mu_{AB}(-1)^{\alpha_{AB}}\mu_{AC}(-1)^{\alpha_{AC}}
=\mu_{AD}(-1)^{\alpha_{AD}}
\nonumber\\
\Rightarrow\ \ \ & \mu_{AD}=\mu_{AB}\mu_{AC}r_{ABC}
\end{align}
in which
\begin{equation}\label{cf2}
r_{ABC}\equiv(-1)^{\alpha_{AB}+\alpha_{AC}-\alpha_{AD}}=\pm1
\end{equation}
and
\begin{align}
\label{consistency0}
\nonumber&\Delta_3(A,B,C)\equiv\alpha_{AB}+\alpha_{AC}-\alpha_{AD}\\
\nonumber&=\hsc A+\hsc B+\hsc C
+\hsc{[AD]_{\alpha_{AD}}}\\
&-\hsc{[AB]_{\alpha_{AB}}}-\hsc{[AC]_{\alpha_{AC}}}-\hsc{[BC]_{\alpha_{BC}}}\in\mathds{N}
\end{align}

In a vertex algebra the identity operator $1$ (e.g. $\psi_0$ in a
$Z_n$ simple-current vertex algebra) is a zero-scaling-dimension
operator with the following OPEs:
\begin{align}\label{OPE:identity}
[1,A]_{-j}=A~\delta_{j,0};~~~[A,1]_{-j}=\frac{1}{j!}\partial^j A
\end{align}
and $\al_{1,A}=\al_{A,1}=0,~~\mu_{1,A}=\mu_{A,1}=1$ for any operator
$A$ in the vertex algebra. $\partial^j1=0,~j\geq1$ is understood as
usual.

One thing needs to be pointed out here: the derivative of an
operator $A$ (let's suppose that $A$ is not the identity operator:
$A\neq1$), \ie $\partial A$ could be zero or not, depending on the
definition of this operator $A$. For example, simple current
$\psi_2$ in a $Z_4$ Gaffnian vertex algebra obeys
$\partial\psi_2=0$. However, this simple current $\psi_2$ is not the
identity operator $1=\psi_0$ since it has nontrivial commutations
factors $\mu_{2,1}=\mu_{2,3}=-1\neq1$.

\section{Determine the quasiparticle commutation factor $\mu_{\ga,a}$
from the quasiparticle pattern of zeros $\{k^\text{sc}_{\ga;a}\}$
}
\label{mugakga}

The quasiparticle commutation factors $\mu_{\ga,a}$
are not fully independent of the patter-of-zero data
$\{\ksc{a};Q_\ga\}$.  In this section, we try to determine
$\mu_{\ga+b,a}$ from $\ksc{a}$.  Note that
\begin{align}
 C_{\ga+b,a} = \mu_{\ga+b,a} C_{a,\ga+b}
\end{align}

By choosing $A=\si_{\ga+a},~B=\psi_b,~C=\psi_c,~D=\psi_{b+c}$ we see
from \eqn{cf1} that:
\begin{align}\label{mu:ga+a,b+c}
\mu_{\ga+a,b+c}=\mu_{\ga+a,b}\mu_{\ga+a,c}\e^{\imth\pi(\al_{\ga+a,b}+\al_{\ga+a,c}-\al_{\ga+a,b+c})}
\end{align}
where we have defined
\begin{align}
\al_{\ga+a,b}\equiv\hsc{\ga+a}+\hsc b-\hsc{\ga+a+b}
\end{align}
Repeatedly using \eqn{mu:ga+a,b+c} we immediately have
\begin{align}
\nonumber
\mu_{\ga+a,b}&=\mu_{\ga+a,b-1}\mu_{\ga+a,1}\e^{\imth\pi(\al_{\ga+a,1}+\al_{\ga+a,b-1}-\al_{\ga+a,b})}\\
\nonumber
&=\mu_{\ga+a,b-2}\mu^2_{\ga+a,1}\e^{\imth\pi(2\al_{\ga+a,1}+\al_{\ga+1,b-2}-\al_{\ga+a,b})}\\
\nonumber
&=\cdots\\
&=\mu^b_{\ga+a,1}\e^{\imth\pi(b\al_{\ga+a,1}-\al_{\ga+a,b})}
\end{align}
By requiring that $\mu_{\ga+a,n}=1$ since $\psi_n$ is the identity
operator, we have
\begin{align}
\mu_{\ga+a,1}=\e^{-\imth\pi\al_{\ga+a,1}+2\pi\imth\frac{\ka_{\ga+a}}{n}}
\end{align}
where $\ka_{\ga+a}$ is an $Z_n$ integer. As a result we can obtain all
commutation factors:
\begin{align}\label{mu:ga+a,b}
\mu_{\ga+a,b}=\e^{-\imth\pi\al_{\ga+a,b}+2\pi\imth\frac{b\ka_{\ga+a}}{n}}
\end{align}
Due to the consistency condition \eq{cf0} we also have
\begin{align}\label{mu:b,ga+a}
\mu_{b,\ga+a}=\e^{\imth\pi\al_{\ga+a,b}-2\pi\imth\frac{b\ka_{\ga+a}}{n}}
\end{align}

Now we implement \eqn{cf1} again with
$A=\psi_b,~B=\si_{\ga+a},~C=\psi_c,~D=\si_{\ga+a+c}$ to see whether
there are any new consistency conditions for quasiparticle scaling
dimensions $\hsc{\ga+a}$
\begin{align}
\nonumber&1=\mu_{b,\ga+a}\mu_{b,c}\mu_{\ga+a+c,b}\e^{\imth\pi(\al_{\ga+a,b}+\al_{b,c}-\al_{b,\ga+a+c})}\\
\nonumber&=\e^{2\pi\imth(\al_{\ga+a,b}+\al_{b,c}-\al_{\ga+a+c,b})}\mu_{b,c}\e^{-\imth\pi\al_{b,c}+2\pi\imth\frac{b(\ka_{\ga+a+c}-\ka_{\ga+a})}{n}}\\
\nonumber&=\e^{-\imth\pi
bc\al_{1,1}+2\pi\imth\frac{b(\ka_{\ga+a+c}-\ka_{\ga+a})}{n}}\\
&=\e^{2\pi\imth b\frac{\ka_{\ga+a+c}-pc-\ka_{\ga+a}}{n}}
\end{align}
where we used \eqn{muij}, $q\equiv n \al_{1,1}/2\in \dZ$, and that
$\Delta_3(\ga+a,b,c)=\al_{\ga+a,b}+\al_{b,c}-\al_{\ga+a+c,b}\in\mathds{N}$.
\Eqn{cf1} with $A=\si_{\ga+a},~B=\psi_b,~C=\psi_c,~D=\psi_{b+c}$
does not produce any new conditions. We can see that all the
consistency conditions on $\{\mu_{\ga+a,b}\}$, i.e. \eqn{cf0} and
\eqn{cf1} can be guaranteed by choosing
\eqn{mu:ga+a,b},~\eqn{mu:b,ga+a} and the integer $\ka_{\ga+a}$ as
\begin{align}
\ka_{\ga+a}=\ka_\ga+qa\mod n
\end{align}

We find that $\mu_{\ga+a,b}$ and
$\mu_{b,\ga+a}$ can almost be determined from $\ksc{a}$:
\begin{align}\label{mu_ga+a,b}
\mu_{\ga+a,b}&=\e^{-\imth\pi\al_{\ga+a,b}+2\pi\imth\frac{b(\ka_{\ga}+aq)}{n}}
,
\nonumber\\
\mu_{b,\ga+a}&=\e^{\imth\pi\al_{\ga+a,b}-2\pi\imth\frac{b(\ka_{\ga}+aq)}{n}}
,
\nonumber\\
q &= n \al_{1,1}/2\in \dZ \
.
\end{align}
However, we need to supply a $Z_n$ integer $\ka_\ga$ to fully fix
$\mu_{\ga+a,b}$ and $\mu_{b,\ga+a}$ from $\ksc{a}$.  (Note that $
\al_{\ga+a,b}=\hsc{b}-\sum_{i=a}^{a+b} \ksc{i}$.)

\section{Generalized Jacobi Identity}
\label{app:GJI}

\subsection{GJI's of an associative vertex algebra}

In the above, we only considered the associativity of the vertex
algebra through the commutation factor $\mu_{AB}$.  Although some
new conditions on pattern of zeros and some relations between the
quasiparticle scaling dimensions are obtained, the associativity of
the algebra is not fully utilized.  To fully use the associativity
condition of the vertex algebra, we need to derive the generalized
Jacobi Identity.

\begin{widetext}
Choose $f(z,w)$ in the following relation
\begin{equation}
\oint_{|z|>|w|}\text{d}z\oint_0\text{d}wf(z,w)-\oint_{|w|>|z|}\text{d}w\oint_0\text{d}zf(z,w)=\oint_0\text{d}w\oint_w\text{d}zf(z,w)
\end{equation}
to be the operator function
\begin{equation}\label{gf}
f(z,w)=A(z)B(w)C(0)(z-w)^{\gamma_{AB}}w^{\gamma_{BC}}z^{\gamma_{AC}}
\end{equation}
with
$\alpha_{AB}-\gamma_{AB},~\alpha_{AC}-\gamma_{AC},~\alpha_{BC}-\gamma_{BC}\in\dZ$,
we obtain the generalized Jacobi Identity (GJI)
\begin{multline}\label{mul:gJacob}
\sum_{j=0}^{\alpha_{BC}-\gamma_{BC}-1}(-1)^j
\binom{\gamma_{AB}}{j}[A[BC]_{\gamma_{BC}+j+1}]_{\gamma_{AB}
+\gamma_{AC}+1-j}
\\
-\mu_{AB}(-1)^{\alpha_{AB}-\gamma_{AB}}\sum_{j=0}^{\alpha_{AC}
-\gamma_{AC}-1}(-1)^j \binom{\gamma_{AB}}{j}
[B[AC]_{\gamma_{AC}+j+1}]_{\gamma_{AB}+\gamma_{BC}+1-j}
\\
=\sum_{j=0}^{\alpha_{AB}-\gamma_{AB}-1} \binom{\gamma_{AC}}{j}
[[AB]_{\gamma_{AB}+j+1}C]_{\gamma_{BC}+\gamma_{AC}+1-j}
\end{multline}
where $\binom{n}{m}$ is the binomial function.

When we choose
$\gamma_{AB}-\alpha_{AB}=\gamma_{AC}-\alpha_{AC}=\gamma_{BC}-\alpha_{BC}=0$,
\eqn{gf} is a regular function with the asymptotic behavior
\begin{equation}
\lim_{z\rightarrow0}\lim_{w\rightarrow0}f(z,w)=\lim_{z\rightarrow0}z^{\alpha_{AB}+\alpha_{AC}}A(z)D(0)=\lim_{z\rightarrow0}z^{\alpha_{AB}+\alpha_{AC}-\alpha_{AD}}[AD]_{\alpha_{AD}}(0)
\end{equation}
Since $f(z,w))$ is an analytic function of both $z$ and $w$,
$z^{\alpha_{AB}+\alpha_{AC}-\alpha_{AD}}$ should still be an
analytic function of $z$. Thus $\alpha_{AB}+\alpha_{AC}-\alpha_{AD}$
should be a non-negative integer, allowing us to obtain the
consistency condition \eq{consistency0}.

For clarity we introduce three integers $n_{AB},~n_{AC},~n_{BC}$ as
\begin{align}
\gamma_{AB}&=\alpha_{AB}-1-n_{AB} &
\gamma_{AC}&=\alpha_{AC}-1-n_{AC} &
\gamma_{BC}&=\alpha_{BC}-1-n_{BC}
\end{align}
and the GJI \eq{mul:gJacob} can be rewritten as
\begin{multline}\label{GJI}
(-1)^{n_{BC}}\sum_{j=0}^{n_{BC}}(-1)^j
\binom{\alpha_{AB}-1-n_{AB}}{n_{BC}-j}
[A[BC]_{\alpha_{BC}-j}]_{\alpha_{AB}+\alpha_{AC}-1-(n_{AB}+n_{AC}+n_{BC})+j}
\\
+\mu_{AB}(-1)^{n_{AB}+n_{AC}}\sum_{j=0}^{n_{AC}}(-1)^j
\binom{\alpha_{AB}-1-n_{AB}}{n_{AC}-j}
[B[AC]_{\alpha_{AC}-j}]_{\alpha_{AB}+\alpha_{BC}-1-(n_{AB}+n_{AC}+n_{BC})+j}
\\
=\sum_{j=0}^{n_{AB}}\binom{\alpha_{AC}-n_{AC}-1}{n_{AB}-j}
[[AB]_{\alpha_{AB}-j}C]_{\alpha_{BC}+\alpha_{AC}-1-(n_{AB}+n_{AC}+n_{BC})+j}
\end{multline}
\end{widetext}

The GJI \eq{mul:gJacob} or \eq{GJI} is the associativity condition
of a vertex algebra. It generalizes the usual Jacobi identity of a
Lie algebra to the case of an infinite-dimensional Lie algebra (the
vertex algebra here), with the usual Lie bracket (the commutator)
defined by OPE in \eqn{OPE-general} We say that the theory is
associative up to a certain order if all the GJI's are satisfied up
to this order in OPE.  Applying the GJI, more conditions on the
patterns of zeros can be found.  More importantly, those conditions
are likely to be the necessary and the sufficient conditions.

For example, by choosing $C$ in GJI \eq{GJI} to be the identity
operator $1$ (note that we have
$\Delta_3(A,B,1)=0$),~$\nx=k\geq0,~\ny=-1,~\nz=0$ and making use of
OPE \eq{OPE:identity}, we immediately reach the following relation
\begin{align}\label{OPE:AB->BA}
\mu_{AB}[BA]_{\al_{AB}-k}=\sum_{j=0}^k\frac{(-1)^{j}}{(k-j)!}\partial^{k-j}[AB]_{\al_{AB}-j}
\end{align}
This allows us to obtain the OPE of $[B,A]$ to the same order with
the OPE of $[A,B]$ to a certain order in hand. For example, we have
\begin{align}
\nonumber[\psi_iT]_1=\partial[T\psi_i]_2-[T\psi_i]_1=(\hsc{i}-1)\partial\psi_i
\end{align}
since $\mu_{T,\psi_i}=1$ and $\al_{T,\psi_i}=2$. As a special case
of \eqn{OPE:AB->BA} we have
\begin{align}
[AA]_{\al_{AA}-2k-1}=
\frac{1}{2}\sum_{j=0}^{2k}\frac{(-1)^j}{(2k+1-j)!}\partial^{2k+1-j}[AA]_{\al_{AA}-j}
\end{align}

This relation is actually an example, showing how we ``derive'' (or
more precisely, obtain the consistent conditions of) higher order
OPE's from the known OPE's up to a certain order based on GJI's.

\subsection{``Useful'' GJI's of a vertex algebra up to a certain order}
\label{usefulGJI}

In practice we need to extract the consistent conditions of a vertex
algebra from a set of ``useful'' GJI's concerning only the OPE's up
to a certain order. The OPE's up to this order are determined
already except for some structure constants (usually complex
numbers). Other GJI's involving higher order OPE's do not serve as
constraints to the vertex algebra up to this order, since we can
always introduce new operators into this infinite-dimensional Lie
algebra in higher order OPE's. For example, a generic $Z_n$ vertex
algebra is defined by OPE's between currents $\{\psi_a\}$ up to
leading order with
$[\psi_i,\psi_j]_{\alpha_{i,j}}=C_{i,j}\psi_{i+j}$.

Let's now consider the GJI \eq{GJI} of three operators $(A,B,C)$,
with the corresponding vertex algebra defined up to
$(N_{AB},N_{BC},N_{AC})$ order, i.e. $[AB]_{\alpha_{AB}-i}$ is known
up to structure constants for all $0\leq i\leq N_{AB}$ in the OPE
\eq{OPE-general} of operators $A$ and $B$. For example, in a special
$Z_n$ simple-current vertex algebra defined by OPE's
(\ref{OPE:i,j})-(\ref{OPE:T,T}), with $(A=\psi_i,B=\psi_j)$ we have
$N_{AB}=0$ if $i+j\neq0\mod n$ or $N_{AB}=2$ if $i+j=0\mod n$. Let's
further assume the following relation
\begin{align}\label{GJI:assumption0}
\alpha_{[AB]_{\alpha_{AB}-j},C}=\alpha_{[AB]_{\alpha_{AB}},C}+j,~\text{if}~[AB]_{\alpha_{AB}-j}\neq0
\end{align}
is satisfied by any operators $(A,B,C)$ of this vertex algebra. It's
straightforward to verify that $Z_n$ simple-current vertex algebra
indeed obeys the above relation: e.g.
$(A=\psi_i,B=\psi_{n-i},C=\psi_j)$ we have
$1=[AB]_{\alpha_{AB}},~T=[AB]_{\alpha_{AB}-2}$ and
$\alpha_{T,\psi_j}=2,~\alpha_{1,\psi_j}=0$. Defining the following
quantity:
\begin{align}\label{N_ABC}
\nonumber
&N_{ABC}(\nx,\ny,\nz)\equiv\\
\nonumber\min\{&N_{[AB]_{\alpha_{AB}-j},C},~0\leq j\leq
N_{AB};\\
\nonumber&N_{B,[AC]_{\alpha_{AC}-j}},~0\leq j\leq N_{AC};\\
&N_{A,[BC]_{\alpha_{BC}-j}},~0\leq j\leq N_{BC}\}
\end{align}
then we can obtain all the ``useful'' consistent conditions of the
vertex algebra from the GJI \eq{GJI}, by choosing
\begin{align}\label{GJI:useful}
\nonumber&n_{AB}\leq N_{AB},~n_{BC}\leq N_{BC},~n_{AC}\leq N_{AC};\\
&\nonumber\Delta_3(A,B,C)-1\leq
n_{AB}+n_{BC}+n_{AC}\\
&\leq\Delta_3(A,B,C)-1+N_{ABC}(\nx,\ny,\nz)
\end{align}
Any other choice with larger $(n_{AB},n_{AC},n_{BC})$ will involve
higher order OPE's. Generally speaking, the set of ``useful'' GJI's
satisfying \eqn{GJI:useful} will be translated into a set of
nonlinear equations of structure constants. (here ``structure
constants'' have a broader meaning than usual, e.g.
$\frac{2\hsc{i}}{c}$ in \eqn{OPE:i,n-i} and $\hsc{i}$ in
\eqn{OPE:T,i} should also be considered as structure constants.)
Some of these equations become consistent conditions of this vertex
algebra, while others help define this vertex algebra e.g. by
determining the structure constants $\{C_{a,b},C_{a,\gamma+b}\}$ and
central charge $c$ of a $Z_n$ simple-current vertex algebra, as is
shown in section \ref{GJIVA} and \ref{EXM}.

\section{Associativity of $Z_n$ vertex algebra
and new conditions on $\hsc{a}$ and $C_{ab}$} \label{gVAhC}

In this section, we apply the consistency conditions of commutation
factor discussed in Appendix \ref{app:muABcon} and GJI discussed in
Appendix \ref{app:GJI} to a $Z_n$  vertex algebra.  This allows us
to derive additional conditions on the scaling dimension $\hsc{a}$
from the associativity of $Z_n$  vertex algebra.

As mentioned earlier, a generic FQH wave function can be expressed
as a correlation function of an associative vertex algebra obeying
the following OPE:
\begin{align}\label{OPE:i,j:generic}
\psi_a(z)\psi_b(w)=\frac{C_{a,b}\psi_{a+b}(w)+O(z-w)}{(z-w)^{\al_{a,b}}}
\end{align}
where we have $\al_{a,b}\equiv\hsc{a}+\hsc{b}-\hsc{a+b\mod n}$. This
guarantees the quasi-Abelian fusion rule
$\psi_a\psi_b\sim\psi_{a+b}$ (see \ref{psiapsibG}).  Moreover, we
choose the normalization of simple currents $\psi_a$ to be
\eqn{Cstr:normalization}.

\subsection{New conditions from the commutation factors}
\label{nconcom}

If we use the radial order \eq{rorder} to calculate correlation
function, then the continuity of the correlation function requires
that
\begin{equation}
(z-w)^{\alpha_{V_aV_b}}V_a(z)V_b(w)
=\mu_{V_a,V_b}(w-z)^{\alpha_{V_aV_b}}V_b(w)V_a(z)
\end{equation}
Since the operators $\e^{\imth a\phi(z)/\sqrt\nu}$ in the Gaussian
model satisfy
\begin{align}
&\ \ \ (z-w)^{ \frac{a^2}{2\nu} +\frac{b^2}{2\nu}
-\frac{(a+b)^2}{2\nu} } \e^{\imth a\phi(z)/\sqrt\nu} \e^{\imth
b\phi(w)/\sqrt\nu}
\nonumber\\
&= (w-z)^{ \frac{a^2}{2\nu} +\frac{b^2}{2\nu} -\frac{(a+b)^2}{2\nu}
} \e^{\imth b\phi(w)/\sqrt\nu} \e^{\imth a\phi(z)/\sqrt\nu} ,
\end{align}
the simple current operator satisfy the following commutation
relation
\begin{equation}
\label{mupsiapsib} (z-w)^{\alpha_{ab}}\psi_a(z)\psi_b(w) =
\mu_{ab}(w-z)^{\alpha_{ab}}\psi_b(w)\psi_a(z),
\end{equation}
where
\begin{equation}
\alpha_{ab}=\hsc a+\hsc b-\hsc {a+b} .
\end{equation}
We stress that the relation \eq{mupsiapsib} is required by the
continuity of the correlation function of the electron operators.

The commutation factors $\mu_{ij}$ satisfy some consistency
relations, which is discussed in the appendix \ref{app:muABcon}
under a more general setting.  The conditions \eq{cf0}, \eq{cf00},
and \eq{cf1} are the conditions on the commutation factors
$\mu_{ij}$ that were obtained from the associativity of the vertex
algebra. From the $n$-cluster condition $\psi^n=\psi_n=1$, we also
have
\begin{equation}
\label{muAB1}
 \mu_{ij}=1,\ \ \ \ \text{if } i=0 \text{ mod } n
\text{ or } j=0 \text{ mod } n  .
\end{equation}
due to the definition of the identity operator $\psi_0=1$ shown in
Appendix \ref{app:muABcon}.

Those conditions, \eq{cf0}, \eq{cf00}, \eq{cf1}, and \eq{muAB1}, can
be expressed as the extra condition on the scaling dimensions
$\hsc{a}$.
We note that according to \eqn{cf1} and \eqn{cf2}
\begin{align}
\mu_{ij} & =\mu_{i,1}\mu_{i,j-1}(-1)^{\alpha_{i,1}
+\alpha_{i,j-1}-\alpha_{i,j}}
\nonumber\\
&=\mu_{i,1}^2\mu_{i,j-2}(-1)^{2\alpha_{i,1}+\alpha_{i,j-2}-\alpha_{i,j}}
\nonumber\\
&=\cdots
\nonumber\\
&=\mu_{i,1}^j(-1)^{j\alpha_{i,1}-\alpha_{ij}}
\end{align}
A similar manipulation leads to
$\mu_{1,i}=\mu_{1,1}^i(-1)^{i\alpha_{1,1}-\alpha_{1,i}}$, and we can
write the commutation factor in a symmetric way
\begin{equation}
\label{muij} {\mu_{ij}}{=(-1)^{ij\alpha_{1,1}-\alpha_{i,j}}=\pm1} .
\end{equation}
We see that $\mu_{ij}$ can be expressed in terms of $\hsc{i}$. Eq.
\eq{muij} also implies that
\begin{equation}
\label{alijint}
 ij\alpha_{1,1}-\alpha_{i,j} =\text{ integer} ,
\end{equation}
which is actually guaranteed by \eqn{hcond3}. 
The condition \eq{muAB1} becomes
\begin{equation}
\label{alijeven}
 ij\alpha_{1,1}-\alpha_{i,j} =\text{ even },\
\text{ if } i = 0 \text{ mod } n \ \text{ or }
 j = 0 \text{ mod } n .
\end{equation}

Now we use $\{M_k,k=1,3,4,\cdots,[\frac{n}{2}];~p\}$ to describe
$\hsc{a}$ (see section \ref{pMk}). So the consistent conditions on
$\{\hsc{a}\}$ can be translated into the consistent conditions on
$\{M_k~;~p\}$. We note that
\begin{align}
\label{alphaijpMk}
 \alpha_{i,j} &  \equiv\alpha_{\psi_i,\psi_j} =\hsc{i}+\hsc{j}-\hsc{i+j}
\nonumber\\
& =p\frac{2ij}{n}-M_i-M_j+M_{i+j\text{ mod } n}
\nonumber\\
& \ \ \ \ \ \ \  -2p(i+j-n)\theta(i+j-n)
\end{align}
for $1\leq i,j\leq n-1$.

By choosing $j=n$ in (\ref{alijeven}) we have
\begin{align}
\label{cchsc1}
{n\alpha_{1,1}=2p=\text{even}}
\end{align}
As a result
\begin{equation}
\label{pdN}
 p\in \dN .
\end{equation}
Besides, $\mu_{i,i}=1$ becomes another constraint
\begin{equation}\label{cchsc2}
{i^2\alpha_{1,1}-\alpha_{i,i}=\text{even}\ \ \ \forall i=1,2,\cdots
n-1}
\end{equation}
What's more, from OPE \eq{OPE:i,n-i} of a special $Z_n$ simple
current vertex algebra and the definition of commutation factor
\eq{mupsiapsib} we immediately have
\begin{align}
\mu_{a,b}=1,~~~\text{if}~~a+b=0\mod n
\end{align}
which becomes an extra constraint
\begin{align}
\label{cchsc3} {i(n-i)\alpha_{1,1}-\alpha_{i,n-i}=\text{even}\ \ \
\forall i=1,2,\cdots n-1}
\end{align}

If we require OPE \eq{OPE:i,n-i} to be satisfied, combining
\eq{alphaijpMk}, \eq{cchsc2} and \eq{cchsc3} we find
\begin{align}\label{mu:constraint}
2M_i-M_{2i\mod n}=2M_i=\text{even}
\end{align}
This determines the parity of $\{M_k\}$ as summarized in Table
\ref{parity-mk}. Notice that we always have $M_k\in\dZ$ for such a
special $Z_n$ vertex algebra.

\begin{table}[tb]
\begin{tabular}{|c||c|c|}
\hline
$n=\text{odd}$ & \multicolumn{2}{|c|}{$M_k\in2\dZ$}\\
\hline $n=\text{even}$ & $M_{2\dZ+1}\in\dZ$ & $M_{2\dZ}\in2\dZ$ \\
\hline 
\end{tabular}
\caption{\label{parity-mk} The parity of $M_k$ with different parity
of $n$ for a special $Z_n$ simple-current vertex algebra with OPE
\eq{OPE:i,n-i}, according to constraint \eq{mu:constraint}. The
parafermion scaling dimension $\{\hsc{a}\}$ is given by
$\hsc{a}=pa(n-a)/n-M_a$, with $p$ being a non-negative integer.}
\end{table}

However, the constraint \eq{cchsc3} is too strong and is not
necessary (this is why we use "special" as a description here). For
example, a $4$-cluster state called Gaffnian explicitly violates it
since we have $\mu_{1,3}=-1\neq1$ for a Gaffnian vertex algebra. To
remove the constraint \eq{cchsc3}, we need to
modify the normalization $C_{a,-a}=1$ in OPE \eq{psiapsibOPE} to:\\
for $a\leq n/2\mod n$
\begin{align}
\label{OPE:i,n-i:i<=n/2} \psi_a(z)\psi_{-a}(w)
&=\frac{1}{(z-w)^{2h^\text{sc}_a} } + ...
\end{align}
and for $a>n/2\mod n$
\begin{align}
\label{OPE:i,n-i:i>n/2} \psi_a(z) \psi_{-a}(w)
&=\frac{\mu_{a,-a}}{(z-w)^{2h^\text{sc}_a} } + ...
\end{align}

If we adopt the more general OPE \eq{OPE:i,n-i:i<=n/2} and
\eq{OPE:i,n-i:i>n/2}, the corresponding consistent conditions on
$\{M_k\}$ from \eqn{alphaijpMk} and \eqn{cchsc2} become
\begin{align}\label{mu:constraint:general}
2M_i-M_{2i\mod n}=\text{even}
\end{align}
This leads to some conclusions on the parity of $\{M_k\}$ as
summarized in Table \ref{parity-mk:general}. Generally we don't have
$M_k\in\dZ$ for such a generic $Z_n$ vertex algebra.

\begin{table}[tb]
\begin{tabular}{|c||c|c|}
\hline
$n=\text{odd}$ & \multicolumn{2}{|c|}{$M_k\in2\dZ$}\\
\hline $n=2\mod4$ & $M_{2\dZ+1}\in\dZ$ & $M_{2\dZ}\in2\dZ$ \\
\hline $n=0\mod4$ & $2M_i-M_{2i\mod n}\in2\dZ$ & $M_{4\dZ}\in2\dZ$\\
\hline
\end{tabular}
\caption{\label{parity-mk:general} The parity of $M_k$ with
different parity of $n$ for a $Z_n$ simple current vertex algebra
with OPE \eq{OPE:i,n-i:general}, according to constraint
\eq{mu:constraint:general}. The parafermion scaling dimension
$\{\hsc{a}\}$ is given by $\hsc{a}=pa(n-a)/n-M_a$, with $p$ being a
non-negative integer.}
\end{table}

To summarize, (\ref{cchsc1}) and (\ref{cchsc2}) are the extra
conditions from commutation factors for a generic $Z_n$ vertex
algebra. They become (\ref{pdN}) and (\ref{mu:constraint:general})
when translated into consistent conditions of $\{M_k;p;m\}$, as a
supplement to conditions (\ref{m-cond})-(\ref{Delta3-ijk}) in
Appendix \ref{pMk}.

\subsection{New conditions from GJI}
\label{nconGJI}

As shown in Appendix \ref{app:GJI}, all GJI's must be satisfied for
the associativity of the vertex algebra. With the OPE
(\ref{OPE:i,j:generic}), we have $N_{a,b}\equiv N_{\psi_a,\psi_b}=0$
and the useful GJI's are very limited. We list the consistent
conditions from useful GJI's of this generic $Z_n$ vertex algebra
below. Then we summarize the new consistent conditions on
$\{\hsc{a}\}$ and $\{c\}$.

\subsubsection{A list of useful GJI's:
$(A=\psi_a,B=\psi_b,C=\psi_c)$}\label{app:list of useful
GJIs:example}

Using the notations in Appendix \ref{app:GJI}, here we have $\Nx=\Ny=\Nz=0$.\\

If $\Delta_3(a,b,c)=0$, all the 3 useful GJI's satisfying
\eqn{GJI:useful} are
\begin{align}
\nonumber & \ngji{-1}{0}{0} &
(C_{b,c}C_{a,b+c}-\mu_{a,b}C_{a,c}C_{b,a+c})\psi_{a+b+c}=0
\end{align}
\begin{align}
\nonumber & \ngji{0}{-1}{0} &
(\mu_{a,b}C_{a,c}C_{b,a+c}-C_{a,b}C_{a+b,c})\psi_{a+b+c}=0
\end{align}
\begin{align}
\nonumber & \ngji{0}{0}{-1} &
(C_{b,c}C_{a,b+c}-C_{a,b}C_{a+b,c})\psi_{a+b+c}=0
\end{align}

If $\Delta_3(a,b,c)=1$, the only 1 useful GJI satisfying
\eqn{GJI:useful} is
\begin{align}
\nonumber & \ngji{0}{0}{0} &
(C_{a,b}C_{a+b,c}-C_{b,c}C_{a,b+c}-\mu_{a,b}C_{a,c}C_{b,a+c})\psi_{a+b+c}=0
\end{align}

For $\Delta_3(a,b,c)\geq2$ there are no useful GJI's satisfying
\eqn{GJI:useful} and thus no new conditions on $C_{i,j}$.

\subsubsection{Summary of new consistent conditions from GJI}

As shown in Appendix \ref{app:list of useful GJIs:example}, for
$\Delta_3(a,b,c)\equiv\Delta_3(\psi_a,\psi_b,\psi_c)=0$ the extra
consistent conditions are
\begin{align}
\label{GJIC1}
C_{a,b}C_{a+b,c}=C_{b,c}C_{a,b+c}=\mu_{a,b}C_{a,c}C_{b,a+c}
\end{align}

For $\Delta_3(a,b,c)=1$ the corresponding consistent condition is
\begin{align}
\label{GJIC2}
C_{a,b}C_{a+b,c}=C_{b,c}C_{a,b+c}+\mu_{a,b}C_{a,c}C_{b,a+c}
\end{align}
For $\Delta_3(a,b,c)\geq2$ there are no useful GJI's and no extra
consistent conditions.

The above conditions should be satisfied no matter what $(a,b,c)$
are. Now let's further specify $(a,b,c)$ and use the normalization
(\ref{Cstr:normalization}) of structure constants to obtain new
conditions.

If $a+b,~a+c,~b+c\neq0\mod n$:\\
For $\Delta_3(a,b,c)=0$ we have
\begin{align}
C_{a,b}C_{a+b,c}=C_{b,c}C_{a,b+c}=\mu_{a,b}C_{a,c}C_{b,a+c}
\end{align}
For $\Delta_3(a,b,c)=1$ we have
\begin{align}
C_{a,b}C_{a+b,c}=C_{b,c}C_{a,b+c}+\mu_{a,b}C_{a,c}C_{b,a+c}
\end{align}
If $a\pm b\neq0\mod n$:\\
For $\Delta_3(a,b,-b)=0$ we have
\begin{align}
C_{a,b}C_{a+b,-b}=C_{b,-b}=\mu_{a,b}C_{a,-b}C_{b,a-b}
\end{align}
For $\Delta_3(a,b,-b)=1$ we have
\begin{align}
C_{a,b}C_{a+b,-b}=C_{b,-b}+\mu_{a,b}C_{a,-b}C_{b,a-b}
\end{align}
If $a\neq n/2\mod n$:\\
For $\Delta_3(a,a,-a)=0$ we have
\begin{align}
\label{GJIC3} \mu_{a,-a}=C_{a,-a}=C_{a,a}C_{2a,-a}=1
\end{align}
For $\Delta_3(a,a,-a)=1$ we have
\begin{align}
\label{GJIC4}
\mu_{a,-a}=\mu_{-a,a}=-1,\\
C_{a,a}C_{2a,-a}=2C_{a,-a}.
\end{align}
If $n=$~even,
we require $\Delta_3(\frac n2,\frac n2,\frac n2)\neq1$ since
otherwise
\begin{align}
\psi_{n/2}\equiv0
\end{align}
must be required to satisfy GJI's.

Among the above consistent conditions, some are just conditions on
the structure constants $\{C_{a,b}\}$, while others serve as the new
conditions on the pattern of zeros $\{\hsc a\}$ or $\{S_a\}$. As a
summary, the extra consistent conditions for the pattern of zeros
from GJI's are
\begin{align}
\label{pozGJI} &\mu_{a,-a}=1~\text{if}~\Delta_3(a,a,-a)=0,
\nonumber \\
&\mu_{a,-a}=-1,~\text{if}~\Delta_3(a,a,-a)=1,
\nonumber \\
&\Delta_3(\frac n2,\frac n2,\frac n2)=4\hsc{\frac
n2}\neq1,~~~n=~\text{even}.
\end{align}
where we need (\ref{muij}) to relate commutation factor $\mu_{a,b}$
with the pattern of zeros.  Note that the first two conditions in
the above can be rewritten as $a^2\alpha_{1,1}+\alpha_{a,-a}=$even
if $\Delta_3(a,a,-a)=0$ and $a^2\alpha_{1,1}+\alpha_{a,-a}=$odd if
$\Delta_3(a,a,-a)=1$.  Since $a^2\alpha_{1,1}-\alpha_{a,a} =$even
and $\Del_3(a,a,-a)=\al_{a,a}+\al_{a,-a}$, the two conditions are
always satisfied.

Obviously these extra conditions, based on the most generic OPE
(\ref{OPE:i,j:generic}) of a $Z_n$ vertex algebra, are not enough to
determine the structure constants $\{C_{a,b}\}$ of this vertex
algebra. In order to have more consistent conditions and to
determine the structure constants, we need to specify higher order
terms in the OPE (\ref{OPE:i,j:generic}). This is done through
defining $Z_n$ simple current vertex algebra in section \ref{GJIVA},
essentially by introducing the energy momentum tensor $T$ and
Virasoro algebra into the vertex algebra. The corresponding extra
consistent conditions are summarized in section \ref{concon} and
\ref{ConsistentConditionsQP}. We can obtain even more extra
conditions from GJI's when we fix the subleading term of OPE's
between simple currents, as shown in  Appendix \ref{app:subleading}.

\section{Subleading terms in OPE of a $Z_n$ simple-current vertex
algebra and more consistent conditions} \label{app:subleading}

\subsection{``Deriving'' subleading terms in OPE from GJI's}

In this section we show how to ``derive'' the subleading term in OPE
\eq{OPE:i,j} of a $Z_n$ simple-current vertex algebra as an example.

First we notice that the subleading term
$[\psi_a\psi_b]_{\al_{a,b}-1}$ should have a scaling dimension of
$\hsc{a+b}+1$, thus we propose the following conclusion:
\begin{align}\label{OPE:i,j:1st}
[\psi_a\psi_b]_{\al_{a,b}-1}=C_{a,b}d_{a,b}\partial\psi_{a+b}
\end{align}
Then we can use GJI's to determine the expression of $d_{a,b}$ in
terms of scaling dimensions $\{\hsc{i}\}$.

First we choose $(A=T,~B=\psi_a,~C=1)$ and $(\nx=-1~\ny=1,~\nz=-1)$
in GJI \eq{GJI}. Since $\Delta_3(T,\psi_a,1)=0$ and
$[\psi_a,1]_1=\partial\psi_a$ we have the following consistent
conditions from this GJI:
\begin{align}
[T\partial\psi_a]_3=2[T\psi_a]_2=2\hsc{a}\psi_a
\end{align}

Then we choose $(A=T,~B=\psi_a,~C=\psi_b),~~a+b\neq0\mod n$ and
$(\nx=0,~\ny=1,~\nz=0)$ in GJI \eq{GJI}. It's easy to verify that
$\Delta_3(T,\psi_a,\psi_b)=2$ since we have
$\al_{T,\psi_i}=2,~\forall i$. Plugging in \eqn{OPE:i,j:1st} and
this GJI yields a consistent condition on $d_{a,b}$:
\begin{align}
2d_{a,b}\hsc{a+b}=(\hsc{a+b}+\hsc{a}-\hsc{b})
\end{align}
Thus we conclude that as long as $\hsc{a+b}\neq0$ (this should hold
in most cases except for ``strange'' examples like Gaffnian) we have
\begin{align}\label{d_a,b}
d_{a,b}=\frac{1}{2}(1+\frac{\hsc{a}-\hsc{b}}{\hsc{a+b}})
\end{align}

\subsection{More consistent conditions due to subleading terms in OPE}

Now with the subleading terms we have more useful GJI's and
therefore more consistent conditions on the data
$\{n;m;\hsc{a};c\}$ characterizing a $Z_n$ simple current
vertex algebra. In this section we shall show the extra consistent
conditions accompanied with the introduction of the subleading term
\eq{OPE:i,j:1st},~\eq{d_a,b} in OPE of $[\psi_a,~\psi_b]$.

It turns out that there are many more useful GJI's considering the
subleading order OPE \eq{OPE:i,j:1st} with \eqn{d_a,b}. In many
cases the new consistent conditions are extremely complicated, so we
will only show the complete consistent conditions in several cases
(which will be utilized in section \ref{EXM} for some examples).

\subsubsection{$\{A,B,C\}=\{\psi_a,\psi_b,\psi_c\}$,
$a+b$, $b+c$, $a+c$, $a+b+c$ $\neq0\mod n$}
\label{consistency:a,b,c:1st} Right now we have $\Nx=\Ny=\Nz=1>0$
thus there are more useful GJI's and more conditions compared with
in section \ref{consistency:a,b,c}.

For $\Delta_3(a,b,c)=0$ 
the complete consistent conditions are summarized as:
\begin{align}
&C_{a,b}C_{a+b,c}=C_{b,c}C_{a,b+c}=\mu_{a,b}C_{a,c}C_{b,a+c}
\nonumber \\
&\al_{a,b+c}d_{b,c}=\al_{b,a+c}d_{a,c}=\al_{a,b}
\nonumber \\
&(d_{a,b+c}+d_{b,a+c}-d_{a+b,c})\partial\psi_{a+b+c}\equiv0
\nonumber \\
&[d_{a,b+c}-d_{a,c}(1-d_{b,a+c})]\partial\psi_{a+b+c}\equiv0
\nonumber \\
&(d_{a,b+c}-d_{a,c}d_{a+c,b})\partial\psi_{a+b+c}\equiv0
\end{align}
The above conditions should also be satisfied with respect to any
permutations of $(a,b,c)$.

For $\Delta_3(a,b,c)=1$ 
some of the new consistent conditions are:
\begin{align}
&\frac{C_{b,c}C_{a,b+c}}{C_{a,b}C_{a+b,c}}=(1-\al_{a,b}+d_{b,c}\al_{a,b+c})^{-1}
\nonumber \\
&\frac{\mu_{a,b}C_{a,c}C_{b,a+c}}{C_{a,b}C_{a+b,c}}=\frac{\al_{a,b}-d_{b,c}\al_{a,b+c}}{\al_{a,b}-d_{b,c}\al_{a,b+c}-1}
\nonumber \\
&(\al_{a,b}+d_{a,c}\al_{b,a+c})(\al_{a,b}-d_{b,c}\al_{a,b+c})=1
\nonumber \\
&(1-\al_{a,b}+d_{b,c}\al_{a,b+c})(\al_{a,c}-d_{a,b}\al_{a+b,c})=1
\end{align}
we didn't show those lengthy consistent conditions with the form of
$(\cdots)\partial\psi_{a+b+c}=0$ here.

For $\Delta_3(a,b,c)=2$ 
the new consistent condition are:
\begin{align}
&(\al_{a,b}-d_{b,c}\al_{a,b+c})C_{b,c}C_{a,b+c}\\
&\nonumber=\mu_{a,b}C_{a,c}C_{b,a+c}(\al_{a,b}-\al_{b,a+c}d_{a,c}),
\end{align}
\begin{align}
&(\al_{a,b}-d_{b,c}\al_{a,b+c}-2)C_{b,c}C_{a,b+c}\\
&\nonumber=-C_{a,b}C_{a+b,c}(\al_{a,c}-\al_{a+b,c}d_{a,b}),
\end{align}
\begin{align}
&(\al_{a,b}-d_{a,c}\al_{b,a+c}-2)C_{a,b}C_{a+b,c}\\
&\nonumber=\mu_{a,b}C_{a,c}C_{b,a+c}(\al_{a,b}-\al_{b,a+c}d_{a,c}-2),
\end{align}
\begin{align}
&0\equiv\partial\psi_{a+b+c}\{C_{a,b}C_{a+b,c}d_{a+b,c}\\
&\nonumber+C_{b,c}C_{a,b+c}[d_{a,b+c}(\al_{a,b}+d_{b,c}-1-d_{b,c}\al_{a,b+c})-d_{b,c}]\\
&\nonumber+\mu_{a,b}C_{a,c}C_{b,a+c}[d_{b,a+c}(\al_{a,b}+d_{a,c}-1-d_{a,c}\al_{b,a+c})-d_{a,c}]\},
\end{align}
\begin{align}
&0\equiv\partial\psi_{a+b+c}\{C_{b,c}C_{a,b+c}d_{a,b+c}\\
&\nonumber+C_{a,b}C_{a+b,c}d_{a+b,c}(-\al_{a,c}-d_{a,b}+1+d_{a,b}\al_{a+b,c})\\
&\nonumber+\mu_{a,b}C_{a,c}C_{b,a+c}[d_{b,a+c}(\al_{a,b}+d_{a,c}-2-d_{a,c}\al_{b,a+c})-d_{a,c}]\},
\end{align}
\begin{align}
&0\equiv\partial\psi_{a+b+c}\{\mu_{a,b}C_{a,c}C_{b,a+c}d_{b,a+c}\\
&\nonumber+C_{b,c}C_{a,b+c}[d_{a,b+c}(\al_{a,b}+d_{b,c}-2-d_{b,c}\al_{a,b+c})-d_{b,c}]\\
&\nonumber+C_{a,b}C_{a+b,c}d_{a+b,c}(\al_{a,c}+d_{a,b}-1-d_{a,b}\al_{a+b,c})\}.
\end{align}

For $\Delta_3(a,b,c)=4$ 
the new consistent conditions are:
\begin{align}
&\nonumber\ngji{1}{1}{1}
&\nonumber C_{a,b}C_{a+b,c}(2-\al_{a,c}+d_{a,b}\al_{a+b,c})\\
&\nonumber+C_{b,c}C_{a,b+c}(2-\al_{a,b}+d_{b,c}\al_{a,b+c})\\
&-\mu_{a,b}C_{a,c}C_{b,a+c}(2-\al_{a,b}+d_{a,c}\al_{b,a+c})=0
\end{align}

For $\Delta_3(a,b,c)\geq5$ there are no useful GJI's and thus no
extra consistent conditions.

\subsubsection{$\{A,B,C\}=\{\psi_a,\psi_b,\psi_{-a-b}\}$, $a,b,a+b\neq0\mod
n$}\label{consistency:a,b,-a-b:1st}

For $\Delta_3(a,b,-a-b)=0$ 
the new consistent condition are:
\begin{align}
&C_{a,b}C_{a+b,-a-b}=C_{b,-a-b}C_{a,-a} =\mu_{a,b}C_{a,-a-b}C_{b,-b}
\nonumber \\
&\al_{a,b}=\al_{a,-a}d_{b,-a-b}=\al_{b,-b}d_{a,-a-b}
\nonumber \\
&\al_{a,-a-b}=\al_{a,-a}d_{-a-b,b}=\al_{-a-b,a+b}d_{a,b}
\nonumber \\
&d_{b,-a-b}\hsc{a}=d_{a,-a-b}\hsc{b}=\al_{a,b}/2
\nonumber \\
&d_{-a-b,b}\hsc{a}=d_{a,b}\hsc{a+b}=\al_{a,-a-b}/2
\nonumber \\
&d_{a,b}\hsc{a}=\al_{a,-a-b}(\hsc{a}+1-\hsc{a+b})/2
\nonumber \\
&d_{-a-b,b}\hsc{a+b}=\al_{-a-b,a}(\hsc{a+b}+1-\hsc{a})/2
\nonumber \\
&d_{a,-a-b}\hsc{a}=\al_{a,b}(\hsc{a}+1-\hsc{b})/2
\end{align}
The above conditions should also be satisfied with $a\leftrightarrow
b$ exchange.

For $\Delta_3(a,b,-a-b)=4$ there is still only 1 useful GJI for
$(A,B,C)$ in a certain order and the new consistent condition are:
\begin{align}
&\nonumber\ngji{1}{1}{1}
&\nonumber C_{a,b}C_{a+b,-a-b}(2-\al_{a,-a-b}+d_{a,b}\al_{a+b,-a-b})\\
&\nonumber+C_{b,-a-b}C_{a,-a}(2-\al_{a,b}+d_{b,-a-b}\al_{a,-a})-\\
&\mu_{a,b}C_{a,-a-b}C_{b,-b}(2-\al_{a,b}+d_{a,-a-b}\al_{b,-b})=0
\end{align}

For $\Delta_3(a,b,-a-b)\geq5$ there are still no useful GJI's and
thus no extra consistent conditions.

\subsubsection{$\{A,B,C\}=\{\psi_a,\psi_b,\psi_{-b}\}$, $a\pm b\neq0\mod
n$} \label{consistency:a,b,-b:1st}

Now we have $N_{\psi_a,\psi_{\pm b}}=1,~~N_{\psi_{b},\psi_{-b}}=2$.

For $\Delta_3(a,b,-b)=0$ 
the consistent conditions are:
\begin{align}
&\hsc{b}\hsc{a}=\al_{a,\pm b}=0,~~~\hsc{b}\partial\psi_a\equiv0
\nonumber \\
&C_{a,b}C_{a+b,-b}=\mu_{a,b}C_{a,-b}C_{b,a-b}=C_{b,-b}
\nonumber \\
&\nonumber(d_{a\pm b,\mp b}-1)\partial\psi_a=(d_{a\pm b,\mp
b}-1)\partial\psi_a\\
&=d_{\pm b,a}\partial\psi_a=d_{\pm b,a\mp b}\partial\psi_a\equiv0
\end{align}

For $\Delta_3(a,b,-b)=2$ 
the consistent conditions are:
\begin{align}
&\nonumber\frac{\mu_{a,b}C_{a,-b}C_{b,a-b}}{C_{b,-b}}=\frac{\al_{a,b}}{\al_{a,b}-d_{a,-b}\al_{b,a-b}}\\
&=\frac{\al_{a,b}(\al_{a,b}-1)}{2}+\frac{2\hsc{a}\hsc{b}}{c}
\end{align}
\begin{align}
&\nonumber\frac{C_{a,b}C_{a+b,-b}}{C_{b,-b}}=\frac{2-\al_{a,b}}{\al_{a,-b}-d_{a,b}\al_{a+b,-b}}\\
&=\frac{(\al_{a,b}-1)(\al_{a,b}-2)}{2}+\frac{2\hsc{a}\hsc{b}}{c}
\end{align}
\begin{align}
&\frac{C_{a,b}C_{a+b,-b}}{\mu_{a,b}C_{a,-b}C_{b,a-b}}=\frac{\al_{a,-b}-2-d_{a,b}\al_{a+b,-b}}{\al_{a,b}-2-d_{a,-b}\al_{b,a-b}}
\end{align}
\begin{align}
&\nonumber\{d_{\pm b,a\mp b}[\frac{2\hsc{a}\hsc{b}}{c}+\frac{\al_{a,\pm b}(\al_{a,\pm b}-1)}{2}]\\
&\nonumber+(1-d_{a\pm b,\mp b})[\frac{2\hsc{a}\hsc{b}}{c}+\frac{(\al_{a,\pm b}-2)(\al_{a,\pm b}-1)}{2}]\\
&-\frac{2\hsc{b}}{c}\}\partial\psi_a\equiv0
\end{align}
\begin{align}
&\nonumber\{[\frac{2\hsc{a}\hsc{b}}{c}+\frac{\al_{a,\pm b}(\al_{a,\pm b}-1)}{2}][d_{a,-b}(d_{b,a-b}-1)+1]\\
&+\al_{a,b}d_{b,a-b}-\frac{2\hsc{b}}{c}\}\partial\psi_a\equiv0
\end{align}
\begin{align}
&\nonumber\{[\frac{(\al_{a,b}-1)(\al_{a,b}-2)}{2}+\frac{2\hsc{a}\hsc{b}}{c}](1-d_{a,b}d_{a+b,-b})\\
&+(\al_{a,b}-2)d_{a+b,-b}-\frac{2\hsc{b}}{c}\}\partial\psi_a\equiv0
\end{align}
\begin{align}
&\nonumber\{[\frac{(\al_{a,b}-1)(\al_{a,b}-2)}{2}+\frac{2\hsc{a}\hsc{b}}{c}]d_{a+b,-b}(d_{a,b}-1)\\
&\nonumber+[\frac{\al_{a,b}(\al_{a,b}-1)}{2}+\frac{2\hsc{a}\hsc{b}}{c}]d_{b,a-b}\\
&+(\al_{a,b}-2)(1-d_{a+b,-b})\}\partial\psi_a\equiv0
\end{align}
\begin{align}
&\nonumber\{[\frac{(\al_{a,b}-1)(\al_{a,b}-2)}{2}+\frac{2\hsc{a}\hsc{b}}{c}]d_{a+b,-b}\\
&\nonumber+[\frac{\al_{a,b}(\al_{a,b}-1)}{2}+\frac{2\hsc{a}\hsc{b}}{c}](d_{a,-b}d_{b,a-b}-d_{a,-b}-d_{b,a-b})\\
&+\al_{a,b}(d_{b,a-b}+1)-1\}\partial\psi_a\equiv0
\end{align}
\begin{align}
&\nonumber\{[\frac{(\al_{a,b}-1)(\al_{a,b}-2)}{2}+\frac{2\hsc{a}\hsc{b}}{c}](2-d_{a,b})d_{a+b,-b}\\
&\nonumber+[\frac{(\al_{a,b}-1)\al_{a,b}}{2}+\frac{2\hsc{a}\hsc{b}}{c}](d_{a,-b}d_{b,a-b}-d_{a,-b}-2d_{b,a-b})\\
&+1+\al_{a,b}d_{b,a-b}+(\al_{a,b}-2)d_{a+b,-b}\}\partial\psi_a\equiv0
\end{align}
We see that central charge $c$ can be determined consistently from
the first two conditions. Notice that after a $b\leftrightarrow-b$
exchange the above conditions should also be satisfied.

For $\Delta_3(a,b,-b)=4$ there are 4 useful GJI's for $(A,B,C)$ in a
certain order now and the consistent conditions are:
\begin{align}
&\nonumber
C_{a,b}C_{a+b,-b}=\mu_{a,b}C_{a,-b}C_{b,a-b}(\al_{b,a-b}d_{a,-b}-\al_{a,b}+1)\\
&+C_{b,-b}[\frac{(\al_{a,b}-1)(\al_{a,b}-2)}{2}+\frac{2\hsc{a}\hsc{b}}{c}]\\
&\nonumber
\mu_{a,b}C_{a,-b}C_{b,a-b}=C_{a,b}C_{a+b,-b}(d_{a,b}\al_{a+b,-b}-\al_{a,-b}+1)\\
&+C_{b,-b}[\frac{(\al_{a,b}-3)(\al_{a,b}-2)}{2}+\frac{2\hsc{a}\hsc{b}}{c}]\\
&\nonumber\{C_{a,b}C_{a+b,-b}d_{a+b,-b}(2-\al_{a,-b}-d_{a,b}+d_{a,b}\al_{a+b,-b})\\
&\nonumber+C_{b,-b}[\frac{(\al_{a,b}-3)(\al_{a,b}-2)}{2}+\frac{2(\hsc{a}-1)\hsc{b}}{c}]\\
&\nonumber+\mu_{a,b}C_{a,-b}C_{b,-a-b}[d_{b,a-b}(\al_{a,b}+d_{a,-b}-2-d_{a,-b}\al_{b,a-b})\\
&-d_{a,-b}]\}\partial\psi_a\equiv0
\end{align}
All the above conditions should also hold when we exchange
$b\leftrightarrow-b$.

For $\Delta_3(a,b,-b)\geq6$ there are no useful GJI's and no extra
consistent conditions.

\subsubsection{$\{A,B,C\}=\{\psi_a,\psi_a,\psi_{-a}\}$, $a\neq n/2\mod
n$}

\label{consistency:a,a,-a:1st}

Now we have $N_{\psi_a,\psi_a}=1,~N_{\psi_a,\psi_{-a}}=2$.

For $\Delta_3(a,a,-a)=\al_{a,a}+2\hsc{a}=0$ 
the consistent conditions are the same as in section \ref{concon}:
\begin{align}
& \hsc{a}=\hsc{2a}=\al_{a,a}=0,~~~\partial\psi_a\equiv0
\nonumber \\
& C_{a,a}C_{2a,-a}=C_{a,-a}=C_{-a,a}=\mu_{a,-a}=1
\end{align}

For $\Delta_3(a,a,-a)=\al_{a,a}+2\hsc{a}=1$ 
the consistent conditions are:
\begin{align}
& \hsc{2a}=3,~~\hsc{a}=1,~~c=-2,~~\mu_{a,-a}=-1
\nonumber \\
& C_{a,a}C_{2a,-a}=2C_{a,-a},~~~C_{a,-a}=-C_{-a,a}
\nonumber \\
& (d_{2a,-a}-\frac32)\partial\psi_a
=(d_{-a,2a}+\frac12)\partial\psi_a\equiv0
\nonumber \\
& d_{a,a}=1/2
\end{align}
Notice that $d_{-a,2a}=-1/2,~~d_{2a,-a}=3/2,~~d_{a,a}=1/2$ are
consistent with \eqn{d_a,b}) and $\hsc{a}=1,~\al_{a,a}=-1$.

For $\Delta_3(a,a,-a)=\al_{a,a}+2\hsc{a}=2$ 
the consistent conditions are
\begin{align}
& c=\frac{2\hsc{a}}{3-2\hsc{a}},~~\al_{a,a}=2-2\hsc{a},
\nonumber \\
& C_{a,a}C_{2a,-a}=2\hsc{a}\neq0
\nonumber \\
& C_{a,-a}=C_{-a,a}=\mu_{a,-a}=1
\nonumber \\
&(d_{2a,-a}-2+\frac{1}{\hsc{a}})\partial\psi_a
=(d_{-a,2a}+1-\frac{1}{\hsc{a}})\partial\psi_a\equiv0
\nonumber \\
&(2\hsc{a}-1)d_{a,a}=1/2
\end{align}
Again notice that \eqn{d_a,b} is consistent with
$d_{2a,-a}=2-(\hsc{a})^{-1},~~d_{-a,2a}=(\hsc{a})^{-1}-1$.

For $\Delta_3(a,a,-a)=\al_{a,a}+2\hsc{a}=4$ 
the consistent conditions are
\begin{align}
&\al_{a,a}=4-2\hsc{a},~~C_{a,-a}=C_{-a,a}=\mu_{a,-a}=1
\nonumber \\
&C_{a,a}C_{2a,-a}=\hsc{a}(2\hsc{a}-3+\frac{2\hsc{a}}{c})\neq0
\nonumber \\
&(d_{2a,-a}-2+\frac{2}{\hsc{a}})\partial\psi_a
=(d_{-a,2a}+1-\frac{2}{\hsc{a}})\partial\psi_a\equiv0
\nonumber \\
&(\hsc{a}-1)(d_{a,a}-1/2)=0
\end{align}

For $\Delta_3(a,a,-a)=\al_{a,a}+2\hsc{a}=6$ there is only 1 useful
GJI for $(A,B,C)$ in a certain order now, and the consistent
conditions are:
\begin{align}
& (2\hsc{a}-3)(d_{a,a}-1/2)=0
\nonumber \\
& (\mu_{a,-a}-1)[\frac{2(\hsc{a})^2}{c}+(\hsc{a}-2)(2\hsc{a}-3)]=0
\end{align}

For $\Delta_3(a,a,-a)=\al_{a,a}+2\hsc{a}\geq7$ we don't have any
useful GJI's and there are no consistent conditions.

\subsubsection{$\{A,B,C\}=\{\psi_{n/2},\psi_{n/2},\psi_{n/2}\}$, $n=$even}

This section is exactly the same as section
{\ref{consistency:n/2,n/2,n/2}} since we still have $\Nx=\Ny=\Nz=2$
if $A=B=C=\psi_{n/2}$. The subleading term in OPE \eq{OPE:i,j:1st}
has no effect on these GJI's.

\subsubsection{$\{A,B,C\}=\{\psi_{a},\psi_{b},\si_{\ga+c}\}$, $a+b\neq0\mod
n$}

\label{consistency:a,b,ga+c:1st} Now we have
$N_{\psi_a,\psi_b}=1>0$, so there are new useful GJI's in this case
than in section \ref{consistency:a,b,ga+c}. Therefore we have more
consistent conditions.

For $\Delta_3(a,b,\ga+c)=0$ 
the consistent conditions are:
\begin{align}
&\mu_{a,b}C_{a,\ga+c}C_{b,\ga+a+c}=C_{a,b}C_{a+b,\ga+c}
=C_{b,\ga+c}C_{a,\ga+b+c},
\nonumber \\
&\al_{a,\ga+c}=\al_{a+b,\ga+c}d_{a,b}=\al_{c,\ga+a+b}d_{a,b}.
\end{align}

For $\Delta_3(a,b,\ga+c)=1$ 
the consistent conditions are:
\begin{align}
&\mu_{a,b}C_{a,\ga+c}C_{b,\ga+a+c}
=C_{a,b}C_{a+b,\ga+c}(\al_{a,\ga+c}-d_{a,b}\al_{a+b,\ga+c}),
\nonumber \\
& C_{b,\ga+c}C_{a,\ga+b+c}
=C_{a,b}C_{a+b,\ga+c}(1-\al_{a,\ga+c}+d_{a,b}\al_{a+b,\ga+c}),
\nonumber \\
&d_{a,b}(\al_{a+b,\ga+c}-\al_{c,\ga+a+b})=0.
\end{align}

For $\Delta_3(a,b,\ga+c)=2$ 
the consistent conditions are:
\begin{align}
&\nonumber C_{b,\ga+c}C_{a,\ga+b+c}-\mu_{a,b}C_{a,\ga+c}C_{b,\ga+a+c}\\
&=C_{a,b}C_{a+b,\ga+c}(\al_{a,\ga+c}-1-d_{a,b}\al_{a+b,\ga+c}),
\nonumber \\
&d_{a,b}(\al_{a+b,\ga+c}-\al_{c,\ga+a+b})=0.
\end{align}

For $\Delta_3(a,b,\ga+c)\geq3$ there are no useful GJI's, and thus
no extra consistent conditions.

\subsubsection{$\{A,B,C\}=\{\psi_{a},\psi_{-a},\si_{\ga+c}\}$}
This section is exactly the same as section
\ref{consistency:a,-a,ga+b} since the subleading term in OPE
\eq{OPE:i,j:1st} has no effect on these GJI's.


\end{document}